\documentclass[fleqn,usenatbib]{mnras}

\usepackage{newtxtext,newtxmath}

\usepackage[T1]{fontenc}

\DeclareRobustCommand{\VAN}[3]{#2}
\let\VANthebibliography\thebibliography
\def\thebibliography{\DeclareRobustCommand{\VAN}[3]{##3}\VANthebibliography}

\usepackage{graphicx}	
\usepackage{amsmath}	
\usepackage{xcolor}

\usepackage{bm}		

\title[Excess radio models and 21-cm Bispectrum]{The Signature of Strong High-Redshift Radio Backgrounds on the Cosmic Dawn 21-cm Bispectrum}

\author[S. Sikder et al.]{
Sudipta Sikder,$^{1}$\thanks{E-mail: sudiptas@mail.tau.ac.il}
Rajesh Mondal,$^{2}$
Rennan Barkana$^{1}$
and Anastasia Fialkov$^{3,4}$
\\
$^{1}$School of Physics and Astronomy, Tel-Aviv University, Tel-Aviv, 69978, Israel
\\
$^{2}$Department of Physics, National Institute of Technology Calicut, Calicut 673601, Kerala, India
\\
$^{3}$Institute of Astronomy, University of Cambridge, Madingley Road, Cambridge, CB3 0HA, UK\\
$^{4}$Kavli Institute for Cosmology, Madingley Road, Cambridge, CB3 0HA, UK
}

\date{Accepted XXX. Received YYY; in original form ZZZ}

\pubyear{2025}

\begin{document}
\label{firstpage}
\pagerange{\pageref{firstpage}--\pageref{lastpage}}
\maketitle

\begin{abstract}
Measurements from the Absolute Radiometer for Cosmology, Astrophysics, and Diffuse Emission 2 (ARCADE-2) reveal a strong radio background in the GHz frequency range. Since the cosmological 21-cm signal is measured relative to the background radiation temperature, the presence of a radio excess can significantly alter its characteristics. Previous studies have explored the impact of an inhomogeneous radio background on the global 21-cm signal and 21-cm power spectrum. This non-uniform radio background is also expected to introduce substantial non-Gaussianity. In this work, using the bispectrum, we analyze the non-Gaussianity in the 21-cm signal in the presence of an excess galactic radio background and investigate how line-of-sight radio fluctuations from early galaxies influence its nature. We find that even a moderate enhancement in radio efficiency in early galaxies significantly affects the small-scale 21-cm bispectrum. Furthermore, the delayed heating transition caused by a galactic radio background shifts the sign change in the squeezed-limit bispectrum to lower redshifts ($z\sim11$), providing a potential observational signature for distinguishing different radio background models. These results demonstrate that the 21-cm bispectrum, particularly in the squeezed limit, is highly sensitive to radio background fluctuations, making it a powerful tool for probing high-redshift radio-loud sources and the physics of the early cosmic epoch.

\end{abstract}




\section{Introduction}

The Universe expanded and cooled after the Big Bang. This cooling led to the formation of neutral hydrogen atoms, as electrons and protons combined. Radiation and matter then decoupled, which marked the end of the Epoch of Recombination. The universe remained dark for a time after this. However, the first luminous structures eventually formed, during a phase known as the Cosmic Dawn\,(CD). These first luminous sources emitted ionizing radiation, which gradually ionized the neutral hydrogen in the intergalactic medium (IGM). This period is therefore known as the Epoch of Reionization\,(EoR). The limited number of observations has hindered our understanding of the CD-EoR. However, a considerable international observational effort is underway to learn more about this important period in the history of the universe.

The redshifted 21-cm signal is a powerful probe of astrophysical and cosmological information during the CD-EoR. This signal arises due to the hyperfine transition of the electron-proton system from parallel to anti-parallel spin in the ground state of neutral hydrogen\,(HI). Several low-frequency radio interferometers, such as GMRT\footnote{\url{http://www.gmrt.ncra.tifr.res.in}} \citep{swarup91}, LOFAR\footnote{\url{http://www.lofar.org/}} \citep{haarlem13}, MWA\footnote{\url{http://www.mwatelescope.org/}} \citep{tingay13}, and HERA\footnote{\url{https://reionization.org/}} \citep{deboer17}, have been dedicated to detecting the fluctuations in the 21-cm signal from the EoR. However, the signal is very weak, being approximately 4 to 5 orders of magnitude smaller than the foreground emission (see e.g., \citealt{ali08}). This has posed great challenges to these observational efforts. Despite these challenges, the first-generation radio interferometers have successfully put significant upper limits on the expected fluctuations in the EoR 21-cm signal \citep{barry19, li19, Mertens2020, trott20, HERA_limit22}. These upper limits have placed constraints on the properties of early galaxies and the IGM during the EoR (see e.g., \citealt{Mondal2020, Ghara2020, Greig2021, Abdurashidova2022}). The next generation telescope, the SKA\footnote{\url{http://www.skatelescope.org/}} \citep{koopmans15}, is expected to be much more sensitive than its predecessors. Therefore it is
likely to be able to detect the fluctuations in the 21-cm signal from the CD-EoR. The detection of the 21-cm signal from the EoR would be a major breakthrough in astronomy. It would allow us to study the formation of the first stars and galaxies, and to learn more about the evolution of the universe at this early epoch.

The CD-EoR 21-cm signal is highly non-Gaussian due to the underlying non-linear matter density field, non-uniform Ly-$\alpha$ coupling, non-uniform heating of HI gas, and the formation of ionized regions that should appear as holes in the 21-cm signal (see e.g., \citealt{Bharadwaj2005, Mondal2015, Kamran2022}). One-point statistics such as the probability distribution function (PDF) \citep{Ciardi_2003, Mellema2006, Ichikawa2010}, skewness and kurtosis \citep{Wyithe2007, Harker2009, Watkinson2014, Watkinson2015, Shimabukuro2015, Kubota2016, Kittiwisit2017}  can be used to quantify this non-Gaussianity, but they do not describe its scale dependence. 

The spherically averaged power spectrum (SAPS) measures the variance of the 21-cm signal at different length scales. It quantifies the amplitudes of the fluctuations in the signal at different wavenumbers $k$. The SAPS fully describes the statistical properties of a field that consists of Gaussian random fluctuations (see e.g., \citealt{Mondal2016}). However, due to the non-Gaussianity present in the CD-EoR signal, the SAPS does not fully quantify the statistical information. For this, we need higher-order statistics, such as the bispectrum (see e.g., \citealt{Peebles1980}).

The bispectrum is the Fourier transform of the three-point correlation function. It depends on three $\mathbf{k}$ vectors that form a closed triangle. \citet{Bharadwaj2005b} first studied the non-Gaussianity of the EoR 21-cm signal using an analytical model consisting of spherical ionized regions. They showed that the bispectrum can be both positive and negative. \citet{Watkinson2017} and \citet{Majumdar2018} independently confirmed these findings using two different bispectrum estimators.

The spherically averaged bispectrum (SABS) can be characterized by the size (scale factor) and the shape of the triangles (\citealt{Bharadwaj2020}; see Section~\ref{sec:uni_tri} for more details). Using this framework, many studies (see e.g., \citealt{Majumdar2020, Kamran2022}) have shown that the 21-cm CD-EoR SABS can be used to distinguish between different sources of non-Gaussianity in the signal, and to learn more about the physical processes that are responsible for the formation of the first stars and galaxies. Notably, \citet{Mondal2021} demonstrated that the squeezed limit bispectrum, among all possible triangle shapes, offers the highest signal-to-noise ratio for detection with the SKA, making it a focal point for observational efforts. However, to fully capture the diversity of non-Gaussian signatures and their physical implications, it is valuable to explore a range of triangle configurations beyond just the squeezed limit. Therefore, while this work emphasizes the squeezed limit bispectrum due to its detectability, it also investigates other triangle configurations to provide a more complete picture of the statistical properties of the 21-cm signal.


The EDGES experiment has claimed to detect a global 21-cm signal \citep{bowman18} in the frequency range 50 to 100 MHz that is significantly stronger than predicted from standard astrophysical assumptions. This anomalous absorption trough is centered at redshift 17 (corresponding to a frequency of 78 MHz) and has an amplitude of $T_{21} = -500^{+200}_{-500}$ mK. Various non-standard theoretical models and new physics have been proposed to explain this signal. Although the SARAS3 experiment has disputed the EDGES signal with 95\% confidence \citep{SARAS3}, the debate over the origin of this anomalous absorption trough is ongoing, with some analyses disfavoring an excess radio background as an explanation for EDGES \citep{Cang2025}. Future experiments may be able to resolve the issue. Therefore, in this paper we consider scenarios that could give rise to an EDGES-like signal and study their implications for the CD-EoR 21-cm SABS.


The ARCADE2 \citep{fixsen11, seiffert11} experiment made a noteworthy discovery of excess radio background over the cosmic microwave background (CMB) at low frequencies. This excess was subsequently corroborated by the LWA1 \citep{dowell18} experiment in the frequency range of 40-80 MHz. The origin of this radio excess is still unclear, but it could potentially be attributed to extragalactic sources. However, the exact fraction of this observed excess that stems from galactic sources compared to extragalactic sources remains unclear \citep[e.g.,][]{Subrahmanyan:2013}. Various models of the excess radio background have been proposed, and their effects on the 21-cm signal have been studied in recent years. \citet{fialkov19} explored the effect of a uniform radio background, not directly related to any astrophysical sources, on the 21-cm signal, focusing on the context of the EDGES low-band observation. In this work, we examine whether such a background introduces additional non-Gaussianity by computing the bispectrum within this model. Also, inhomogeneous galactic radio background models have been proposed by \citet{Reis2020, Sikder2023}. These models predict significant enhancement of the 21-cm power spectrum from the CD-EoR. This non-uniform radio background is also expected to introduce significant non-Gaussianity in the 21-cm signal. However, the impact of this non-uniform excess radio background originating from high-redshift galaxies \citep{Condon2012, Ewall_Wice_2018, mirocha19, Mebane2020} on the 21-cm bispectrum has not yet been explored. In this study, we investigate the 21-cm bispectrum within the context of an excess galactic radio background. We also examine how radio fluctuations along the line of sight from these galaxies impact the 21-cm bispectrum.

The paper is structured as follows: We briefly describe our semi-numerical CD simulation in section \ref{sec:simulation}. In section \ref{sec:excess_radio_model}, we summarize the non-standard excess radio models and their implications for the 21-cm power spectrum and global signal. We briefly present the theoretical background and algorithm that has been used to generate the 21-cm bispectrum from a simulated 21-cm signal for all possible triangle configurations in section \ref{sec:bispectrum}. We present our main results in section \ref{sec:result}. Finally, in section \ref{sec:conclusion}, we conclude the paper with a summary. Throughout this study, we use the $\Lambda$CDM cosmology with cosmological parameters from the \citet{Planck2016}.


\section{Semi-Numerical Simulation}
\label{sec:simulation}

To generate the 21-cm signal over a wide range of redshifts, we use our 21-cm Semi-numerical Predictions Across Cosmic Epochs (\texttt{21cmSPACE}) simulation code \citep[see e.g.,][]{visbal12, fialkov14, cohen17, fialkov19}. The outputs of the simulation are 21-cm brightness temperature cubes in a $[384\,{\rm Mpc}]^3$ comoving cosmological volume with a resolution of 3 comoving Mpc over a wide range of redshifts (6 to 35). We calculate the global signal, power spectrum, bispectrum, etc. using those simulation boxes at each redshift.

We use the publicly available code CAMB\footnote{\url{https://camb.info/}} \citep{camb} to calculate the power spectra of initial density fields. We then obtain Gaussian random realizations of density fluctuations and the relative velocity between dark matter and baryon \citep{tseliakhovich10}. The population of collapsed dark matter halos inside each cell is modeled analytically following a hybrid prescription by \citet{barkana04}, which is based on the Press-Schechter formalism \citep{press74} and the Sheth-Tormen mass function \citep{sheth99}. This approach has been slightly modified to account for the effect of the relative baryon-dark matter velocity ($v_{\rm{bc}}$) on the number of halos in each pixel, following the framework of \citet{Fialkov2012} \citep[see also, ][]{Tseliakhovich2011}.

We assume that galaxies only form within halos where the circular velocity exceeds a specific threshold, denoted as $V_{\rm{c}}$. This threshold is a proxy for the minimum mass of the star-forming halos, which is required for radiative cooling of the infalling gas. In this study, we consider star formation set by atomic cooling by choosing $V_{\rm{c}} = 16.5$ km s$^{-1}$ which corresponds to $M_{\rm{min}} \approx 3\times10^7 M_{\odot}$ at $z=20$. Another important parameter is the star formation efficiency, $f_{\star}$, which quantifies the fractional gas mass in star-forming halos that is converted into stellar mass. While the detailed mechanisms of star formation are observationally unconstrained at high redshift, we fix $f_{\star} = 0.1$ for our primary analysis \citep{Mirocha2017, munoz2022}. Throughout the main paper, we utilize a Pop II-only star formation prescription \citep{visbal12, fialkov2013}. However, to ensure that our predicted non-Gaussian signatures are not too dependent on the specific parameters assumed for star formation, we present supplementary simulations in Appendix~\ref{sec:appendixA} using a substantially different model that incorporates a realistic Pop III to Pop II transition \citep{Gessey_Jones2022}, demonstrating the robustness of our results. The simulation also takes into account the effects of the dark matter-baryon relative velocity, Lyman-Werner feedback on molecular cooling halos \citep{haiman97,fialkov2013}, and photoheating feedback \citep{rees86,sobacchi13,cohen16} on the suppression of star formation.

The formation of a population of galaxies is followed by the computation of the radiation fields emitted by these galaxies. The primary radiation fields that have a significant impact on the 21-cm signal are Ly-$\alpha$ radiation, X-ray radiation, and ionizing radiation. Ly-$\alpha$ photons are responsible for the Wouthuysen-Field coupling and these photons also heat up the IGM. To determine the intensity of the Ly-$\alpha$ radiation field, we assume that galaxies consist of population II stars and that for a given galaxy, Ly-$\alpha$ radiation is proportional to its star formation rate (SFR). For a given overdensity, baryon-dark matter velocity, Lyman-Werner intensity and photoheating feedback, the star formation prescription evaluates the averaged SFR in each cell; specifically, the gas density available for star formation is determined by analytical prescriptions that are based on numerical simulation results, and the SFR is determined by the time derivative of this gas density multiplied by the assumed star formation efficiency \citep[see][for more details]{visbal12, fialkov2013, fialkov14a}. The X-ray luminosity is also assumed to scale with the SFR as follows \citep{fialkov14a}:
\begin{equation}\label{eqn:Lx_SFR_relation}
    \frac{L_{\rm{X}}}{\rm{SFR}} = 3\times10^{40} f_{\rm{X}} \ \rm{erg\ s^{-1} M^{-1}_\odot yr} \ ,
\end{equation}
where the additional normalization factor, $f_{\rm X}$ is the X-ray radiation efficiency of the sources. $f_{\rm X} = 1$ corresponds to the typical observed X-ray luminosity of low-metallicity galaxies at low redshifts \citep[see e.g.,][]{Grimm, Gilfanov, Mineo:2012, fragos13, fialkov14a, Pacucci:2014}. We choose $f_{\rm X} = 1$ for all the simulation runs shown in this work. In addition to the X-ray luminosity, the shape of the X-ray spectral energy density (SED) also affects the 21-cm signal. We assume a power law shape for the X-ray SED, where the slope ($\alpha$) and the low energy cutoff ($E_{\rm min}$) determine the shape of the power law. Both $\alpha$ and $E_{\rm min}$ are two free parameters in our simulation. For this work, we set $\alpha = 1.5$ and $E_{\rm min} = 1$ keV for the X-ray SED.

The 21-cm signal during reionization is primarily governed by ionization fluctuations, rather than by other factors such as heating, Lyman-alpha coupling, and density fluctuations. The process of reionization is modeled using the excursion set formalism, as described by \citet{furlanetto04}. In this formalism, a specific region is considered to be ionized if the fraction of collapsed matter surpasses a threshold of $\zeta^{-1}$, where $\zeta$ represents the overall efficiency of ionizing sources. The course of reionization also depends on the maximum distance traveled by ionizing photons, i.e., the mean free path of ionizing photons, $R_{\rm{mfp}}$, another free parameter in the simulation (see e.g., \citealt{greig15}). $R_{\rm{mfp}}$ represents an effective upper limit for the size of ionization bubbles formed during the EoR. We used a fixed value of $\zeta = 30$ \citep{FURLANETTO2006181} and $R_{\rm{mfp}} = 30$\, cMpc \citep{Wyithe_loeb2004} for all the cases analyzed in this study. Note that the reionization model in this simulation has some limitations compared to simulations that only focus on the EoR (see e.g., \citealt{Iliev2007, Mondal2017}). However, our main concern is to study the 21-cm bispectrum during the CD, so these limitations are not a major drawback for our work.


\section{Non-standard radio background models}
\label{sec:excess_radio_model}
The 21-cm brightness temperature, $T_{\rm{21}}$, is proportional to the difference between the spin temperature, $T_{\rm S}$, of neutral hydrogen and the background radiation temperature, $T_{\rm{rad}}$. From the solution of the radiative transfer equation in an expanding universe, $T_{\rm{21}}$ can be written as
\begin{equation}
T_{21} = \frac{T_{\rm S} - T_{\rm rad}}{1+z}(1 - e^{-\tau_{21}})\ .
\label{eq:T21}
\end{equation}
In the standard astrophysical scenario, we assume the background radiation to be the CMB at redshift $z$, i.e., $T_{\rm{rad}} = T_{\rm{CMB}} = 2.725(1+z)$\,K. However, in the case of a non-standard excess radio model, the background radiation temperature can be written as
\begin{equation}
T_{\rm{rad}} = T_{\rm{Radio}} + T_{\rm{CMB}} \ , 
\label{Eq:rad}
\end{equation}
where $T_{\rm{Radio}}$ is the brightness temperature of the radio excess. The mathematical equation for $T_{\rm{Radio}}$ pertains exclusively to the modeling of the excess radio background. In simulations with an excess radio background, the astrophysical parameters ($f_*$, $V_{\rm{C}}$, $f_{\rm{X}}$, $\alpha$, $E_{\rm{min}}$, $\zeta$, $R_{\rm{mfp}}$) retain the same values and roles as in the standard CMB background case (Section~\ref{sec:simulation}). Recent studies \citep{acharya23, cyr24} suggest that soft photon heating can modify the shape and amplitude of the 21-cm signal in the presence of a radio background during the CD. This effect, which depends on the low-frequency tail of the radio spectrum, is not considered in the present work. Below, we describe the non-standard excess radio models considered in this study.

\subsection{Exotic uniform radio background}
A simple, homogeneous excess radio background that might not be directly related to astrophysical sources was proposed by \citet{Feng_Holder2018}. This background could arise from exotic processes such as dark matter decay (see, e.g., \citealt{Fraser:2018, Pospelov:2018}) and super conducting cosmic strings (see e.g., \citealt{Brandenberger:2019}). Therefore, its intensity is not dependent on the star formation history. \citet{fialkov19} first explored its impact on 21-cm fluctuations. The brightness temperature of this exotic uniform radio excess at the 21-cm rest frame frequency at $z$ can be written as
\begin{equation}
T_{\rm Radio} = 2.725 (1+z)\, A_{\rm r}\, \left[\frac{\nu_{\rm{obs}}}{78 \ \mathrm{MHz}}\right]^{\beta} ~{\rm K}\ ,
\label{Eq:ar}
\end{equation}
where $2.725$ K is the present day CMB temperature. Here $\beta$ is the spectral index of the synchrotron spectrum, which is set to $-2.6$ to match the observed spectrum \citep{fixsen11, seiffert11, dowell18} of the extragalactic radio background, $\nu_{\rm{obs}}$ is the observed frequency and $A_{\rm{r}}$ is the amplitude of the radio background relative to the CMB at the EDGES peak frequency of 78 MHz. To investigate its effect on the 21-cm bispectrum, we set the amplitude $A_{\rm{r}}$, which regulates the strength of the radio background, at $0.0945$, a value chosen to ensure that the global signal trough in our excess radio model deviates only modestly, by less than $15\%$, from that of the standard astrophysical scenario. By keeping the global signal close to the standard scenario, we can test how even a modest excess can impact higher-order statistics, particularly the 21-cm bispectrum. We note that throughout the paper we refer to the model from this subsection as the uniform radio model.

\subsection{Radio fluctuations from high redshift galaxies}
In contrast to the phenomenological description of the excess radio background with a synchrotron spectrum, high redshift radio-loud sources such as star-forming galaxies or Active Galactic Nuclei\,(AGN) could produce an excess radio background over the CMB. The modeling of this astrophysical radio excess and its impact on the global 21-cm signal has been investigated by \citet{Ewall_Wice_2018} and \citet{mirocha19}. \citet{Reis2020} first incorporated this non-uniform excess radio background from high redshift galaxies into a semi-numerical simulation code for the CD, and explored the effect on the 21-cm power spectrum. In this galactic radio model, the radio luminosity of galaxies is assumed to be proportional to the SFR, as \citep[based on the empirical relation of][]{gurkan18}
\begin{equation}
    L_{\rm Radio} (\nu, z ) = f_{\rm Radio} 10^{22} \left(\frac{\nu}{150\, {\rm MHz}}\right)^{-\alpha_{\rm Radio}} \left( \frac{\rm SFR}{M_{\odot}\, \rm{yr}^{-1} }\right)\ \ \mathrm{\frac{W}{Hz}} \,,
    \label{eq:fRadio}
\end{equation}
where $\alpha_{\rm{Radio}}$ is the spectral index in the radio band which is set to 0.7 as in \citet{mirocha19} and \citet{gurkan18}. $f_{\rm{Radio}}$ is the normalization of the radio emissivity so that for present day star-forming galaxies, the value of $f_{\rm{Radio}}$ is 1. In the case of a simplified model with the approximation of an isotropic galactic radio background \citep{Reis2020}, the radio background brightness temperature at 21-cm frequency at redshift $z$ is determined by adding up the contribution from all the galaxies within the past light-cone: 
\begin{multline}
\label{eq:Tgal}
    T_{\rm Radio} (\nu_{21}, z) = \frac{\lambda_{21}^2}{2 k_{\rm B}} \frac{c (1+z)^3}{4 \pi} \times \\ \int\epsilon_{\rm Radio}\left(\nu_{21} \frac{1 + z_{\rm em}}{1 + z}, z_{\rm em}\right) (1 + z_{\rm em})^{-1} H(z_{\rm em})^{-1} dz_{\rm em}\ ,
\end{multline}
where $z_{\rm{em}}$ is the redshift of photon emission, and the comoving radio emissivity $\epsilon_{\rm{Radio}}$ is the radio luminosity per unit frequency per unit comoving volume, that is, $L_{\rm Radio}/V_{\rm cell}$, averaged over radial shells in this spherical integral, where $V_{\rm cell}$ is the volume of a simulation cell.


The computation of $T_{\rm{Radio}}$ follows a similar approach to that of the Ly-$\alpha$ and X-ray radiation fields, assuming an isotropically averaged intensity at each pixel in the simulation box. However, the assumption of isotropy of the radio background is only
approximate. In the next section, we present a more precise and realistic model of the excess radio background from galaxies at high redshift.

\subsection{LoS radio fluctuation model}
Since the 21-cm absorption occurs along the line-of-sight (LoS), the radiative transfer calculation along the line of sight depends solely on the radio intensity from sources located behind the hydrogen cloud, along the LoS. However, the isotropically averaged radio intensity is the relevant quantity for physical effects such as the calculation of Lyman-$\alpha$ coupling coefficients. \citet{Sikder2023} illustrated the effect of these LoS radio fluctuations on the 21-cm signal. The brightness temperature of the radio background from sources that contribute along the LoS ($T_{\rm{Radio, LoS}}$) is calculated similarly to $T_{\rm{Radio}}$. However, unlike $T_{\rm{Radio}}$, where a source emitting at $z_{\rm{em}}$ contributes a spherical shell around the source, $T_{\rm{Radio, los}}$ receives contributions from the same source only along a single direction, i.e., the LoS direction. In contrast to previous non-standard cases mentioned above (eqs. \ref{eq:T21}, \ref{Eq:ar} and \ref{eq:fRadio}), the 21-cm brightness temperature for this case can be written as \citep[for details, see][]{Sikder2023}
\begin{equation}
T_{21} = \frac { 
 T_S - \left( T_{\rm{Radio, LoS}} + T_{\rm CMB} \right) } {1+z}\ \left( 1 - e^{- \tau_{21}} \right)\ .
 \label{eq:T21new}
\end{equation}
For both isotropic and line-of-sight considerations in non-uniform galactic radio background models, the parameter $f_{\rm{Radio}}$ quantifies the strength of the radio background. In this work, unless otherwise specified, we adopt $f_{\rm{Radio}}=60$ for the examples presented. This moderate value is chosen so that the global signal trough in this excess radio model lies within $15\%$ of that of the standard CMB-only scenario. This choice allows us to investigate how a mild excess background from early galaxies influences 21-cm fluctuations, particularly the 21-cm bispectrum. We also investigate an $f_{\rm{Radio}}=3000$ model in order to see the impact of a high radio source model; this is roughly the highest value allowed for a galaxy population at $z \sim 20$ by clustering of the radio background, which sets the strongest current observational constraint on such models \citep{2024ApJ...970L..25S}. For convenience, we list in Table~\ref{tab:parameters_list} the parameters along with their values, including those in the excess radio models. 

We emphasize that the LoS radio fluctuation model should be considered the main, most interesting model for comparison with the standard case (CMB only, i.e., no radio background). In comparison, the exotic radio background model is an ad-hoc model, while the others have an astrophysical motivation as they place the radio emission in galaxies (though with an unusually enhanced amplitude of radio emission). The radio fluctuation model with isotropy is only an approximate model that removes some of the realism of the LoS model and thus allows us to isolate the effect of the LoS anisotropy on the results. In summary, the LoS galactic radio fluctuation model is both astrophysically motivated and accurately calculated. 

\begin{table*}
\centering
\begin{tabular}{lcc} 
\hline
Parameter    & Values  & Description \\ 
\hline\hline
$f_{\star}$  & $0.1$     & Star formation efficiency \\
$V_{\rm{c}}$  & $16.5$ km s$^{-1}$    & Minimum circular velocity \\
$f_{\rm{X}}$  & $1$     & X-ray production efficiency \\
$\alpha$  & $1.5$      & Slope of X-ray SED  \\
$E_{\rm{min}}$  & $1$ keV   & X-ray SED low energy cutoff\\
$\zeta$  & $30$     &  the overall efficiency of ionizing sources \\
$R_{\rm{mfp}}$  & $30$ Mpc  & Mean free path for ionizing photons \\
$f_{\rm{Radio}}$  & 60 or 3000    & Radio production efficiency (radio from early galaxies)\\
$\alpha_{\rm{Radio}}$ & 0.7 & Spectral index in the radio band (radio from early galaxies) \\
$A_{\rm{r}}$  &  $0.0945$   & Amplitude of uniform radio background (radio from exotic processes)\\
$\beta$ & $-2.6$ & Spectral index in the synchrotron spectrum (radio from exotic processes)\\

\hline
\end{tabular}
\caption{The parameters and their values used in this work. The model labeled ``Radio fluctuations'' assumes the isotropic approximation, and always has moderate efficiency ($f_{\rm{Radio}}=60$). Once we add the LoS effect, we assume moderate efficiency by default, unless we indicate high efficiency ($f_{\rm{Radio}}=3000$).}\label{tab:parameters_list}

\end{table*}


\section{21-cm Bispectrum}
\label{sec:bispectrum}
The bispectrum is a statistical measure of three-point correlations of a field in Fourier space. Mathematically, the bispectrum of the 21-cm brightness temperature field can be written as
\begin{equation}
\langle \Delta(\mathbf{k_1}) \Delta(\mathbf{k_2}) \Delta(\mathbf{k_3}) \rangle  = B(\mathbf{k_1, k_2, k_3}) V \delta_{\rm D}(\mathbf{k_1+k_2+ k_3}) \ ,
\end{equation}
where $\Delta(\mathbf{k})$ is the Fourier transform of the brightness temperature fluctuations (in units of mK$\times$Mpc$^3$), $V$ is the comoving volume under consideration, and the angular brackets denote an ensemble average over different realizations of the field. The Dirac delta function $\delta_{\rm D}$, which equals 0 unless $\mathbf{k_1 + k_2 +k_3} = 0$, ensures that the Fourier modes form a closed triangle. 

\subsection{Unique triangle configurations in Fourier space}
\label{sec:uni_tri}

We use the parameterization from \cite{Bharadwaj2020} to identify all possible unique triangle configurations in Fourier space. This formalism characterizes the size and shape of a triangle by designating the largest side as $\bf k_1$ and the second-largest side as $\bf k_2$ which implies $k_1 \geq k_2 \geq k_3$, where $k$ represents the amplitude of the vector $\bf k$. The following two parameters relate $k_1$ and $k_2$ in a way that quantifies the size and shape of the triangle:
\begin{equation}
{\rm Size\,parameter\colon}~    n = \frac{k_2}{k_1} \ ,
\label{eqn:n_k_2}
\end{equation}
\begin{equation}
{\rm Shape\,parameter\colon}~    \cos{\theta} = -\frac{\mathbf{k_1}\cdot\mathbf{k_2}}{k_1 k_2} \ ,
\label{eqn:cost}
\end{equation}
where $\theta$ is the angle between $-\mathbf{k_1}$ and $\mathbf{k_2}$. The value of $n$ is restricted to the range $0.5 \leq n \leq 1$, and $\cos{\theta}$ is limited by  $n\cos{\theta} \geq 0.5$.

It is common to study specific triangle configurations to interpret the bispectrum, such as the equilateral triangle, isosceles triangle, linear triangle, and squeezed triangle. Each of these configurations quantifies distinct non-Gaussian properties of the field. 


\subsection{The Bispectrum estimator}
The spherically averaged binned bispectrum estimator for the $i$-th bin is defined as
\begin{equation}
\hat{B}_i(k_1,n,\cos{\theta}) \equiv \hat{B}_i(k_1,k_2,k_3) = \frac{1}{N_{\rm t}V} \sum_t \Delta({\mathbf k}_1) \Delta({\mathbf k}_2) \Delta({\mathbf k}_3),
\end{equation}
where $t$ indexes triangles in the $i$-th bin and $\sum_t$ is over the number $N_{\rm t}$ of closed triangles within the $i$-th bin. The bins are three-dimensional (3D) voxels of volume $[\Delta k_1\,\Delta k_2\,\Delta k_3]$, which we map to the $(k_1,n,\cos{\theta})$ space using eqs.~(\ref{eqn:n_k_2}) and (\ref{eqn:cost}). The ensemble average of the estimator is the bin-averaged spherically averaged bispectrum (SABS): $\langle \hat{B} (k_1, n, \cos{\theta}) \rangle = \bar{B} (k_1, n, \cos{\theta})$.

The bispectrum is a measure of the statistical dependence of three wave vectors. To calculate the bispectrum, we need to include the condition $\mathbf{k}_1+\mathbf{k}_2 = -\mathbf{k}_3$. The direct estimation method is the most accurate method for calculating the bispectrum, but it is also the most computationally expensive. A conventional direct estimation method requires six nested \texttt{for} loops, three of which are used to find ${\mathbf k}_1$ and the other three to find ${\mathbf k}_2$. This results in an order of $N_{\rm G}^6$ operations, where $N_{\rm G}$ is the number of grid points along each direction.

To reduce computational cost, \citet{Mondal2021} proposed some optimization and parallelization methods. We use their publicly available bispectrum estimation code {\texttt DviSukta}\footnote{\url{https://github.com/rajeshmondal18/DviSukta}}. This code employs the direct estimation method and has been optimized for speed and efficiency. Here we provide a brief summary of the algorithm used in {\texttt DviSukta}. For a more detailed description, the reader is referred to Section 2.2 of \citet{Mondal2021}.

The algorithm begins by partitioning the Fourier space into a set of non-overlapping cells. The bispectrum is then calculated for each cell using the direct estimation method. The code first searches for all possible ${\mathbf k}_1$ and bins them using equally spaced spherical logarithmic binning. Inside the ${\mathbf k}_1$ loop, the code searches for all possible ${\mathbf k}_2$ using eq.~\ref{eqn:n_k_2} and partial use of eq.~\ref{eqn:cost}. The bispectrum values from the (${\mathbf k}_2,\,{\mathbf k}_3$) space are then mapped to the ($n,\,\cos \theta$) space and binned using equally spaced linear bins. The results from the individual cells are then combined and the bin averaging is performed to produce $\bar{B} (k_1, n, \cos{\theta})$.


To show our results, we use the scale-independent spherically averaged bispectrum, which is defined as
\begin{equation}
\Delta^3(k_1, n, \cos\theta) \equiv \frac{k_1^6n^3 B(k_1, n, \cos\theta)}{(2\pi^2)^2} \ .
\end{equation}

\begin{figure}
    \centering
    \includegraphics[width=0.48\textwidth]{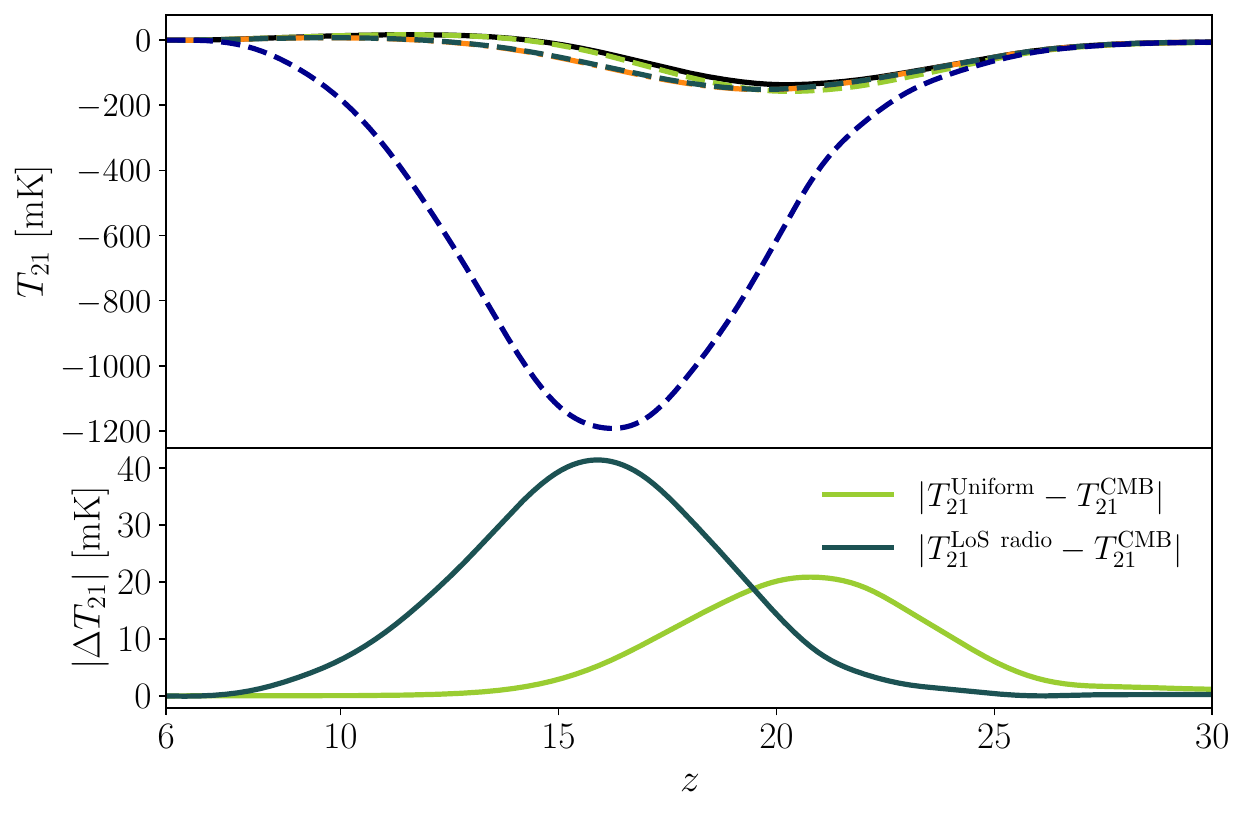}
    \includegraphics[width=0.48\textwidth]{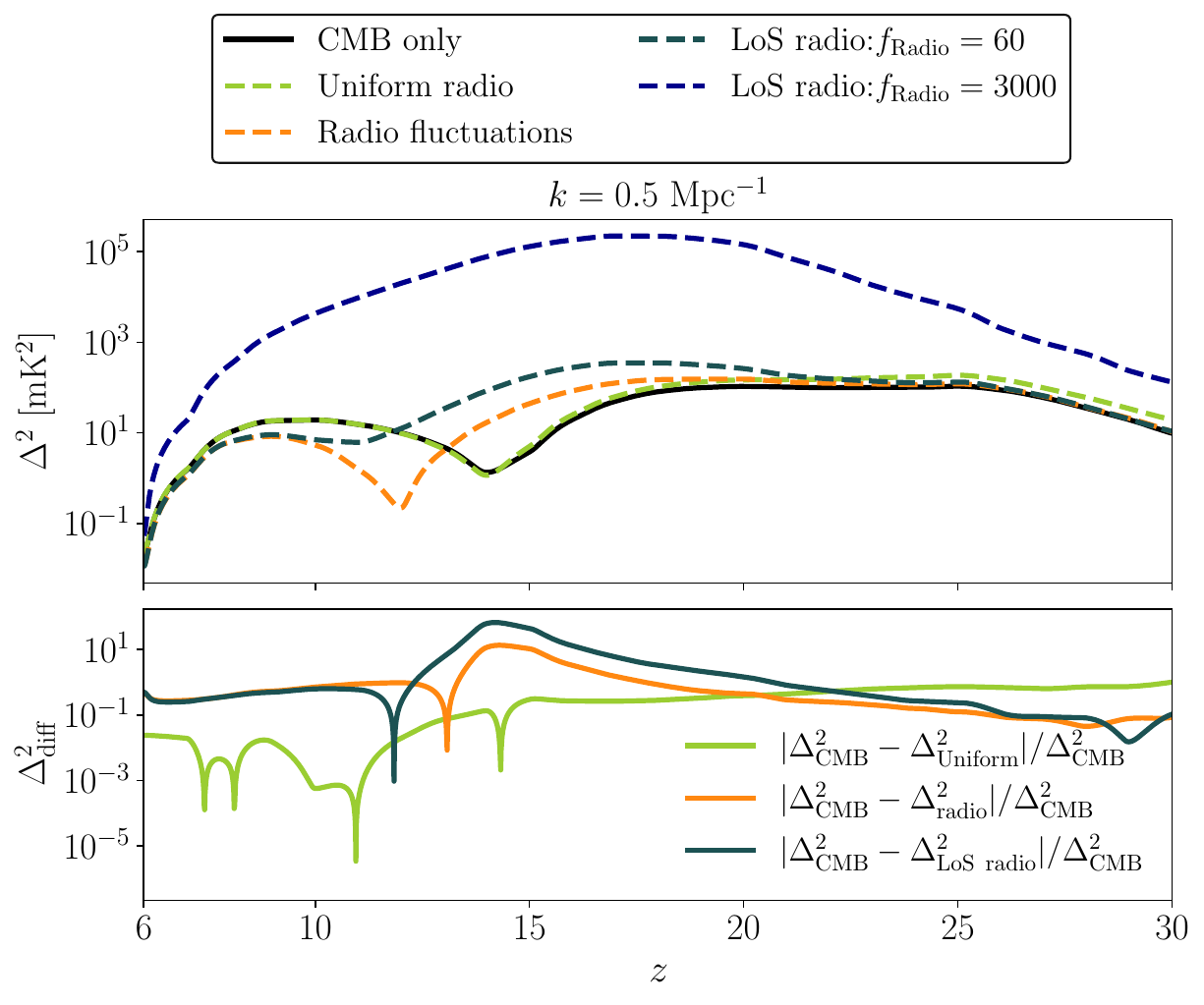}
    \caption{\textbf{Top panel:} Global signal as a function of redshift for the astrophysical models considered in this work. We also show the absolute differences in the global signal for the uniform (green solid line) and moderate LoS radio fluctuation model (teal solid line) compared to the standard astrophysical model. \textbf{Bottom panel:} The 21-cm power spectrum at $k=0.5 \ \rm{Mpc^{-1}}$ as a function of redshift for the five different models. The relative differences of three excess radio models with respect to the standard astrophysical model are also shown.}
    \label{fig:global_signal_power_spectrum}
\end{figure}

\begin{figure}
    \centering
    \includegraphics[width=0.48\textwidth]{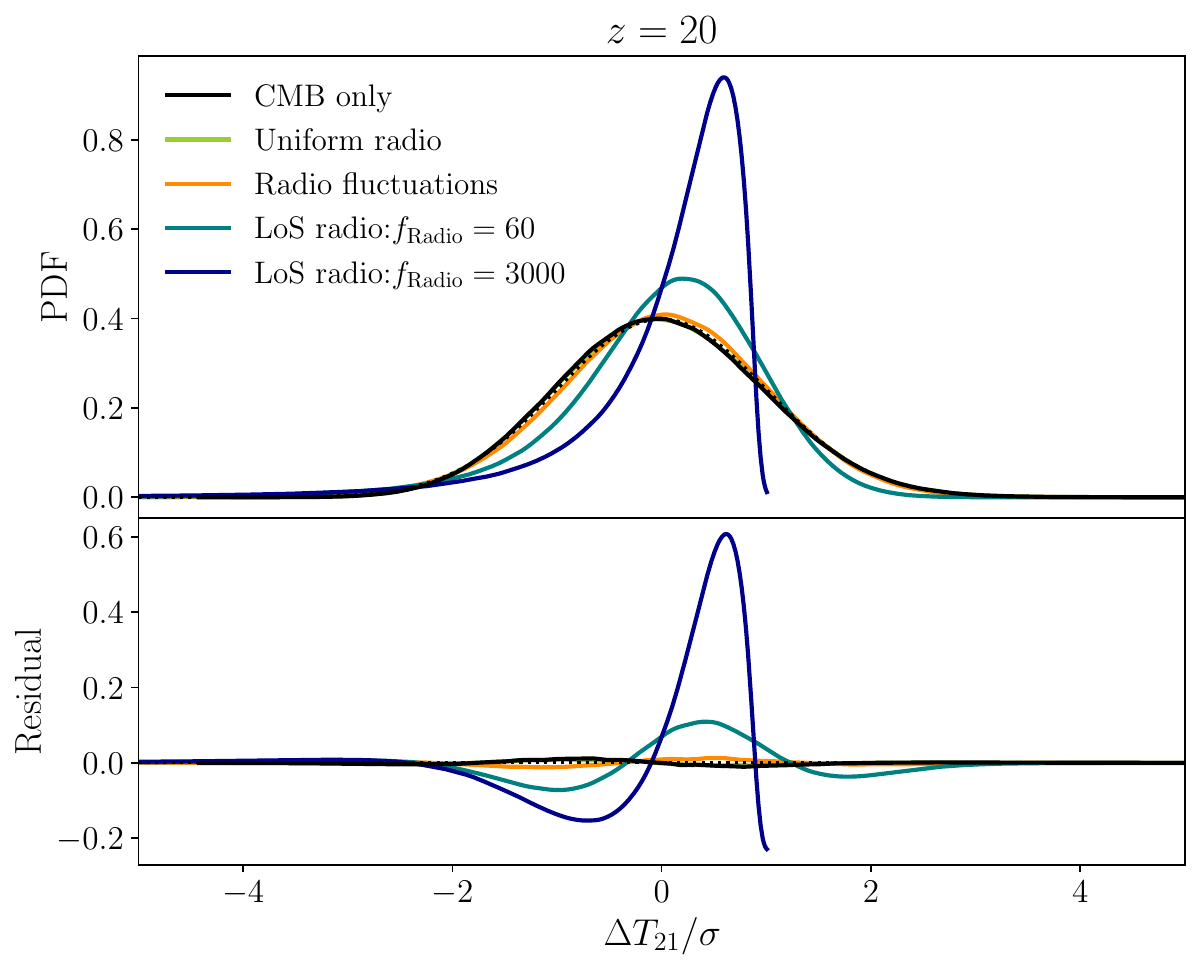}
    \caption{\textbf{Top panel:} Probability density functions (PDFs) of the 21-cm brightness temperature, measured relative to the mean temperature and divided by the standard deviation, for various models at $z = 20$: standard astrophysical model (black solid line), uniform radio background model (green solid line), isotropic galactic radio fluctuation model (orange solid line), and LoS galactic radio fluctuation models with $f_{\rm{Radio}}=60$ (teal solid line) or $f_{\rm{Radio}}=3000$ (blue solid line). Note that the PDFs are normalized to a total area of unity, and the standard deviation for these five cases is $\sigma= 15.47, 18.23, 18.73, 24.89$ and 611.0 mK, respectively. We also compare to a Gaussian PDF (black dotted curve), which has mean 0 and $\sigma=1$ in these normalized units. \textbf{Bottom panel:} We show the difference between each PDF and the Gaussian PDF, for the various models. This quantifies one kind of non-Gaussianity in the 21-cm brightness temperature. Also shown for comparison is a zero difference (black dotted line).}
    \label{fig:histogram_1D_pdf}
\end{figure}

\section{Results}
\label{sec:result}

\subsection{The 21-cm global signal, power spectrum, and probability distribution function}

We first examine the one-point and two-point statistics, namely the global signal and power spectrum, for the same set of astrophysical parameters (see Table \ref{tab:parameters_list}), to get a baseline overview. This is helpful for understanding the next higher statistic, the bispectrum.

Figure \ref{fig:global_signal_power_spectrum} shows the all-sky-averaged global signal (top panel) and 21-cm power spectrum at $k=0.5$ Mpc$^{-1}$ (bottom panel) as a function of redshift for five different models. In the top panel, the maximum absorption for the expected signal from the standard astrophysical scenario (CMB only, black solid line) occurs with an amplitude of $-136.9$ mK at $z=20.2$, corresponding to $\nu \sim 67$ MHz. We also consider three moderate excess radio models: uniform radio excess, isotropic radio fluctuation, and a line-of-sight (LoS) radio fluctuation model, for which the maximum signal amplitudes are $-157$ (green dashed line), $-152$ (dark orange dot-dashed line) and $-151$ (teal dashed line) mK, occurring at $z=20.3, 19.7$ and $19.7$, respectively. These models have moderate values of the radio production efficiency parameter, i.e., $f_{\rm{Radio}}=60$ for the galactic radio background models and $A_{\rm{r}} = 0.0945$ for the radio background of the uniform model, such that the global signal minima of these excess radio models are within $15\%$ of that of the standard astrophysical model (CMB-only case). This allows us to probe the 21-cm bispectrum signatures associated with moderate excess radio backgrounds, even when their global signals are not easily distinguished from the standard (no radio) case. Separately we also show the high radio excess model ($f_{\rm{Radio}}=3000$, for the full galactic radio background model with the line-of-sight effect), which has an order of magnitude deeper trough. 

In a separate frame within the top panel we show the absolute differences between the global signals ($T_{21}^{\rm{Uniform}}, T_{21}^{\rm{LoS \ radio}}$) of two excess radio models and that of the CMB-only case ($T_{21}^{\rm{CMB}}$). For the LoS galactic radio background model, the maximum absolute difference is $\sim 40$~mK, occurring at $z \sim 16$, while for the uniform radio model, the absolute difference peaks at $z \sim 20$ with an amplitude of $\sim 20$~mK. We do not show the absolute difference for two other models: the isotropic radio fluctuation model yields a global signal essentially identical to that of the LoS radio fluctuation model (due to the averaging that is inherent in the global signal); and for the high radio excess model, the difference with the other models is clear.  

The bottom panel of Figure~\ref{fig:global_signal_power_spectrum} displays the 21-cm power spectrum for the same set of models discussed earlier, highlighting the effects of various excess radio background scenarios. Compared to the standard astrophysical case, the two galactic radio models exhibit significant boosts in power. The line-of-sight (LoS) model achieves a maximum relative difference of a factor of $\sim 63$ at $z\sim 14$, while the isotropic model shows a more modest increase of $\sim 17$. These differences reflect the amplified 21-cm signal driven by the spatially varying radio background, with the stronger effect of the LoS model resulting from the additional contribution to fluctuations of the directional dependence along the line of sight. In contrast, the uniform  radio background has a far less pronounced impact on the power spectrum, particularly at low redshifts. For instance, at $z=15$, the relative difference due to the uniform radio background is only 0.39, rising to 0.73 at $z=25$. The redshift dependence - where the effect of uniform radio excess weakens from $z=25$ to $z=15$ - reflects the fact that this smooth uniform background effectively declines with cosmic time. In contrast, the galactic radio backgrounds grow stronger over time, tracing the growth of galaxy formation.

To show the presence of non-Gaussianity in the 21-cm signal, and to quantify one measure of it, we first compare in Figure~\ref{fig:histogram_1D_pdf} the probability distribution functions (PDFs) of the 21-cm brightness temperature \citep{Ciardi_2003, Mellema2006, Ichikawa2010} with a Gaussian distribution. At $z=20$ for each model, we use the variable $\Delta T_{21} = (T_{21} - \bar{T}_{21})$ divided by $\sigma$, i.e., the brightness temperature normalized to zero mean (i.e., measured relative to the mean temperature) and unit variance (i.e., divided by the standard deviation of $\Delta T_{21}$). To assess the deviation from a Gaussian distribution, each PDF is compared to a Gaussian distribution, and the difference is also shown for each of the models (in the bottom panel). The standard (CMB only) case, the uniform radio model, and the radio fluctuation model with the isotropic approximation, all show only a barely noticeable non-Gaussianity in the PDF at this redshift. However, the line of sight effect of bright radio sources, which we have previously shown to enhance the 21-cm power spectrum and its anisotropy \citep{Sikder2023}, also strongly enhances the non-Gaussianity of the 21-cm PDF. Even the moderate case ($f_{\rm{Radio}}=60$) shows a clearly significant deviation from a normal distribution, but the high case ($f_{\rm{Radio}}=3000$) has a highly non-Gaussian PDF; the latter case shows a cutoff on the right-hand side, which corresponds to pixels with no bright radio source behind them (since bright radio sources cause absorption, i.e., a negative 21-cm brightness temperature, and these absorptions are responsible for most of the fluctuations in this case).

In addition to comparing the full PDF with a Gaussian distribution, the non-Gaussianity can be summarized using higher-order moments such as the skewness (a measure of asymmetry) and kurtosis (a measure of the difference between the symmetrically-averaged distribution and a Gaussian), which are important features of the non-Gaussian PDF of the 21-cm signal \citep{Wyithe2007, Harker2009, Watkinson2014, Watkinson2015, Kubota2016, Kittiwisit2017}. In Table~\ref{tab:table1}, we list the skewness and kurtosis values at $z=20$ for the PDFs shown in Figure~\ref{fig:histogram_1D_pdf}. This quantifies the trends that were qualitatively apparent in Figure~\ref{fig:histogram_1D_pdf}. During the epoch of the first stars, brightness temperature fluctuations are primarily driven by density and Lyman-$\alpha$ fluctuations, during the time that the spin temperature becomes coupled to the gas temperature through Lyman-$\alpha$ coupling. Since the gas temperature cools more rapidly than the background radiation temperature (which, in the standard astrophysical scenario, is the CMB), the mean spin temperature falls below the background radiation temperature. In the presence of an excess radio background, the contrast between the spin temperature and background radiation temperature increases, depending on the strength of the radio excess relative to the CMB. When the excess radio background originates from early radio galaxies, we find that the skewness becomes negative. However, only the line-of-sight model substantially increases the size of the non-Gaussian moments. For our moderate LoS radio fluctuation model, the skewness increases $\times 8.4$ and the kurtosis $\times 26$ compared to the standard astrophysical scenario, while the high LoS model further increases the skewness $\times 2.5$ and the kurtosis $\times 3.1$. LoS radio fluctuations enhance the absorption in regions surrounding early radio galaxies and along the same line of sight. When the mean number of significant sources is of order one, there are strong Poisson fluctuations and the 21-cm brightness temperature has a highly non-Gaussianity distribution.

\begin{figure}
    \centering
    \includegraphics[width=0.48\textwidth]{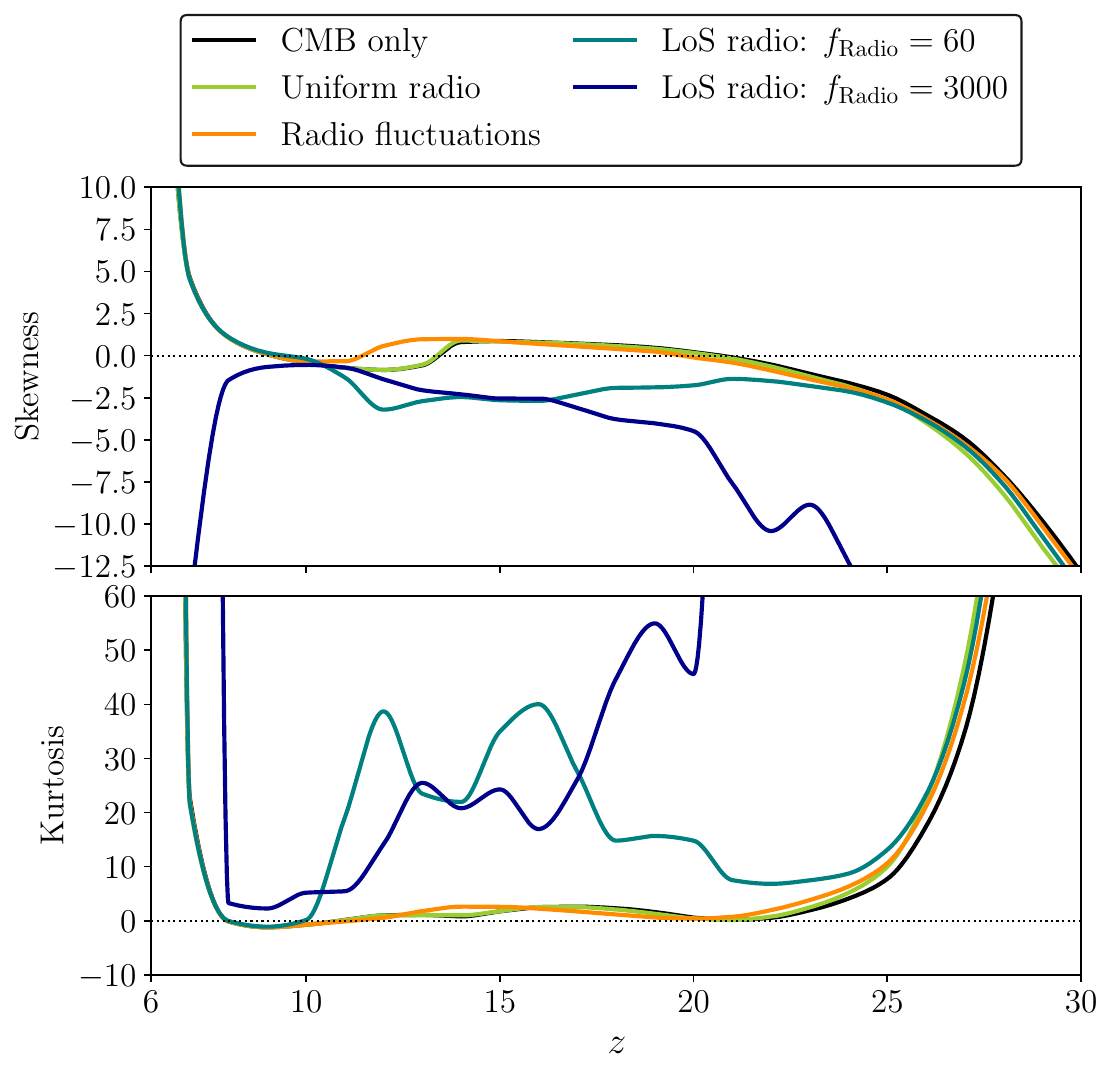}
    \caption{Skewness and kurtosis as a function of redshift, for the same models as shown in Figure~\ref{fig:histogram_1D_pdf}.}
    \label{fig:skew_kurt_z}
\end{figure}

With the help of the moments, we can more easily look at the non-Gaussianity as a function of redshift, in order to characterize the properties of each model over a broad range of cosmic history. Figure~\ref{fig:skew_kurt_z} shows the skewness and kurtosis over a wide redshift range. The non-Gaussianity is in general large towards the end of reionization (large positive skewness and kurtosis, except for the high $f_{\rm{Radio}}$ case that has a strongly negative skewness) and towards the Dark Ages (negative skewness and positive kurtosis). During the CD (which is our focus here), the CMB only and uniform radio cases have nearly identical moments, which are typically of order unity (i.e., not as small as was suggested by the values at $z=20$). The LoS radio model has significantly larger non-Gaussianity, with a negative skewness and positive kurtosis at all $z \gtrsim 11$. The moderate and high LoS models actually have comparable values of these moments, except that the high model takes off to higher magnitudes at $z \gtrsim 18$. The LoS effect is critical here, as can be seen from the fact that the radio fluctuation model without the LoS effect behaves more similarly to the CMB and uniform radio models than to the LoS radio models.

The presence of non-Gaussianity in the 21-cm signal indicates that the power spectrum, which can only fully quantify the statistical properties of a Gaussian random field, is insufficient to capture all relevant information. In a purely Gaussian signal, different Fourier modes are independent, meaning that correlations exist only at the two-point level (power spectrum) and all higher-order statistics are zero. However, for the 21-cm signal from the CD, the bispectrum (three point correlation function) can capture nonlinear and non-Gaussian interactions between different Fourier modes, complementing the power spectrum analysis. Motivated by this, we next investigate how different excess radio models impact the CD 21-cm bispectrum. 

\begin{table}
\centering
\begin{tabular}{lcc} 
\hline
Model             & Skewness & Kurtosis  \\ 
\hline\hline

Standard (CMB only)    & 0.210  & 0.577  \\
Uniform radio         & 0.164  & 0.433  \\
Radio fluctuations     & -0.113  & 0.395  \\
LoS radio: $f_{\rm{Radio}} = 60$ & -1.77  & 14.8  \\
LoS radio: $f_{\rm{Radio}} = 3000$ & -4.48  & 45.6 \\

\hline
\end{tabular}
\caption{Skewness and kurtosis at $z=20$ for our various models considered in this study.}
\label{tab:table1}
\end{table}

\subsection{The effect of an excess radio background on the 21-cm bispectrum}

\subsubsection{Moderate excess radio backgrounds at redshift 20}

With the bispectrum, there is a quite large number of parameters to consider. In order to produce insight rather than confusion, we try to minimize the number of parameters and focus on some regions of the parameter space. As before, we keep the astrophysical model parameters fixed (see Table~\ref{tab:parameters_list}), and consider several distinct excess radio models, alongside the standard astrophysical scenario without any excess radio background (the CMB-only case). We do wish to explore how the various effects vary across cosmic epochs and scales. For simplicity, we begin by only considering in this subsection the radio models with moderate efficiency (i.e., without the high $f_{\rm{Radio}}$ model), and only at a redshift near the global signal absorption trough for these models, specifically $z=20$. 

\begin{figure*}
\centering
\includegraphics[width=0.95\textwidth]{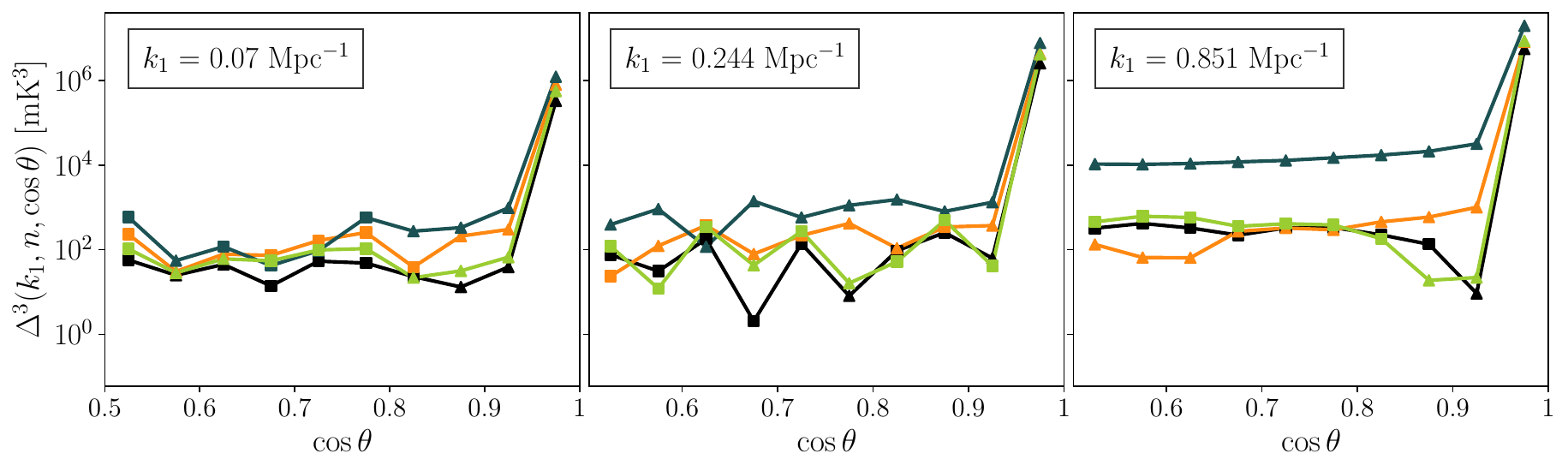}
\caption{The 21-cm SABS at $z=20$ for the limiting L-isosceles triangle ($n \rightarrow 1$,  i.e.,  $k_2 \rightarrow k_1$) versus $\cos \theta$ at $k_{\rm{1}} = 0.07,  0.244$ and $0.851$ Mpc$^{-1}$.   The square and triangle markers represent positive and negative values of the bispectrum, respectively. The colored lines in each panel show various models: CMB-only case (black), uniform background (green), isotropic (orange) and LoS galactic radio fluctuations (teal).}
\label{fig:bispectrum_isosceles_triangles}
\end{figure*}

\begin{figure*}
\centering
\includegraphics[width=0.95\textwidth]{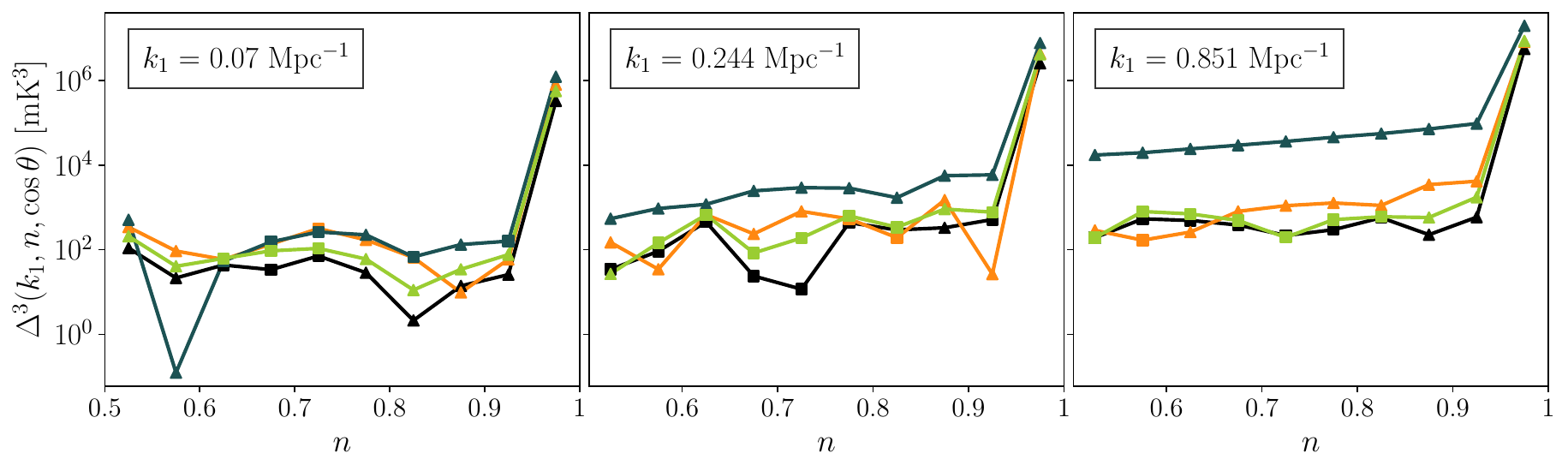}
\caption{The 21-cm SABS at $z=20$ for the limiting linear triangles ($\cos\theta \rightarrow 1$) versus $n$ at $k_{\rm{1}} = 0.07,  0.244$ and $0.851$ Mpc$^{-1}$. The 
markers and colors are as in Figure~\ref{fig:bispectrum_isosceles_triangles}.} \label{fig:bispectrum_linear_triangles}
\end{figure*}

Among several simplified triangle configurations, we first consider the case of limiting L-isosceles triangles ($n \rightarrow 1$, i.e., $k_2 \rightarrow k_1$). In Figure~\ref{fig:bispectrum_isosceles_triangles}, we show the SABS for L-isosceles triangles as a function of $\cos{\theta}$ at three different $k_{\rm{1}}$ values (0.07,  0.244 and 0.851 Mpc$^{-1}$) for the four different models. The black, green, dark orange, and teal lines indicate the simulation runs without any excess radio background (CMB only), with a uniform  radio background, with isotropic radio fluctuations, and the LoS radio fluctuation model, respectively. In contrast to the power spectrum, the SABS can be positive or negative. We show positive and negative values of the SABS using square and triangle markers, respectively. 

The magnitude of the SABS is maximum for all values of $k_1$ in the $\cos{\theta} = 0.975$ bin  (which is the squeezed limit) and then falls sharply for smaller values of $\cos{\theta}$. We note that the length of the smallest arm (and hence the area) of the triangles decreases with increasing $\cos{\theta}$. Now, the squeezed limit ($\cos{\theta} \rightarrow 1$) corresponds to a triangle configuration where $k_1 \approx k_2 \gg k_3$. This means that the large-scale mode (associated with $k_3$) modulates the amplitude of small-scale fluctuations. This effect is strongest and most coherent when the modes are aligned ($\cos\theta \rightarrow 1$), leading to the largest amplitude of the SABS. As $\cos\theta$ decreases, the triangle configuration moves away from the squeezed limit towards more equilateral shapes, the strength of the mode coupling weakens, and thus the amplitude of the SABS drops. We also note that for lower values of $\cos\theta$, the SABS values are nearly independent of $\cos\theta$. The flat behavior of the SABS at lower $\cos\theta$ corresponds to triangle configurations where all three wavevectors ($k_1, k_2, k_3$) are of similar magnitude, resembling near-equilateral triangles. In this regime, the SABS measures the intrinsic non-Gaussianity arising from local, non-linear processes in the 21-cm signal, such as couplings between density and temperature fluctuations, occurring on comparable scales. Unlike the squeezed limit, where a large-scale mode strongly modulates smaller scales, these local interactions are relatively insensitive to the exact triangle shape. As a result, the SABS remains nearly constant across this range of $\cos\theta$, reflecting a consistent level of non-Gaussianity that does not vary much with the angle between the modes.

It is evident that the overall amplitude of the SABS increases with increasing wavenumber, though random variations occur across different values of $\cos{\theta}$. On large scales (left and middle panel of Figure~\ref{fig:bispectrum_isosceles_triangles}), the SABS values randomly oscillate between positive and negative in most models. However, on small scales (i.e., larger $k_1$, shown in the right panel of Figure~\ref{fig:bispectrum_isosceles_triangles}), the SABS exhibits a distinct bias: it is predominantly positive for the standard astrophysical model (CMB-only case) and the uniform radio background model, but takes mostly negative values for both galactic radio models. In these models, the spatially varying $T_{\rm{Radio}}$ (for the isotropic model) or $T_{\rm{Radio, \ LoS}}$ (for the line-of-sight model) introduces an additional source of fluctuations linked to the density field. By enhancing absorption in overdense regions, $T_{\rm{Radio}}$ or $T_{\rm{Radio, \ LoS}}$ amplifies the anti-correlation between density and 21-cm brightness temperature fluctuations. This additional coupling modifies the phase relationships among Fourier modes, resulting in a bispectrum that is biased towards negative values on small scales. Moreover, the LoS effect of radio fluctuations significantly boosts the SABS, increasing its amplitude by approximately two orders of magnitude compared to the standard case for $\cos{\theta} \leq 0.925$. Thus, on small scales the SABS is particularly sensitive to the LoS radio fluctuations from early radio galaxies. These fluctuations enhance small-scale non-Gaussianity due to the combination of non-Gaussian Poisson fluctuations, strong anisotropy, and non-linear 21-cm fluctuations.

Next, we present the results for linear triangle configurations ($\cos{\theta} \rightarrow 1$, $k_1 \rightarrow k_2 + k_3$). Figure~\ref{fig:bispectrum_linear_triangles} illustrates how the SABS varies with the length of the second largest arm $k_2$ of triangles, for the same models and $k_{\rm{1}}$ values as in the previous figure. We vary $k_2$ by adjusting $n$ for a given $k_1$ as in Eq.~\ref{eqn:n_k_2}. The SABS magnitudes are comparable to those of L-isosceles triangles with the same $k_1$ values, although there are some random variations between different models. Similarly to the results for limiting L-isosceles triangles, on small scales the LoS effect of radio fluctuations significantly enhances the SABS amplitude by approximately two orders of magnitude compared to the standard astrophysical case and the other two excess radio background models. In the squeezed limit ($n \rightarrow 1$), the SABS amplitudes increase by factors of 3.7, 3.0, and 3.5 relative to the standard astrophysical scenario at $k_{\rm{1}} = 0.07,  0.244$ and $0.851$ Mpc$^{-1}$, respectively. As seen in the L-isosceles triangle results, the SABS values for the LoS radio fluctuation model remain negative across all values of $n$. These results suggest that the negative SABS values on small scales over the whole range of $\cos{\theta}$ (limiting L-isosceles triangles) and $n$ (limiting linear triangles) could serve as a potential observational signature of the LoS effect of radio sources from the CD. We note, though, that these results are based on theoretical simulations, while the practical prospects depend on also accounting for observational errors and artifacts. 



\begin{figure*}
\centering
\includegraphics[width=0.9\textwidth]{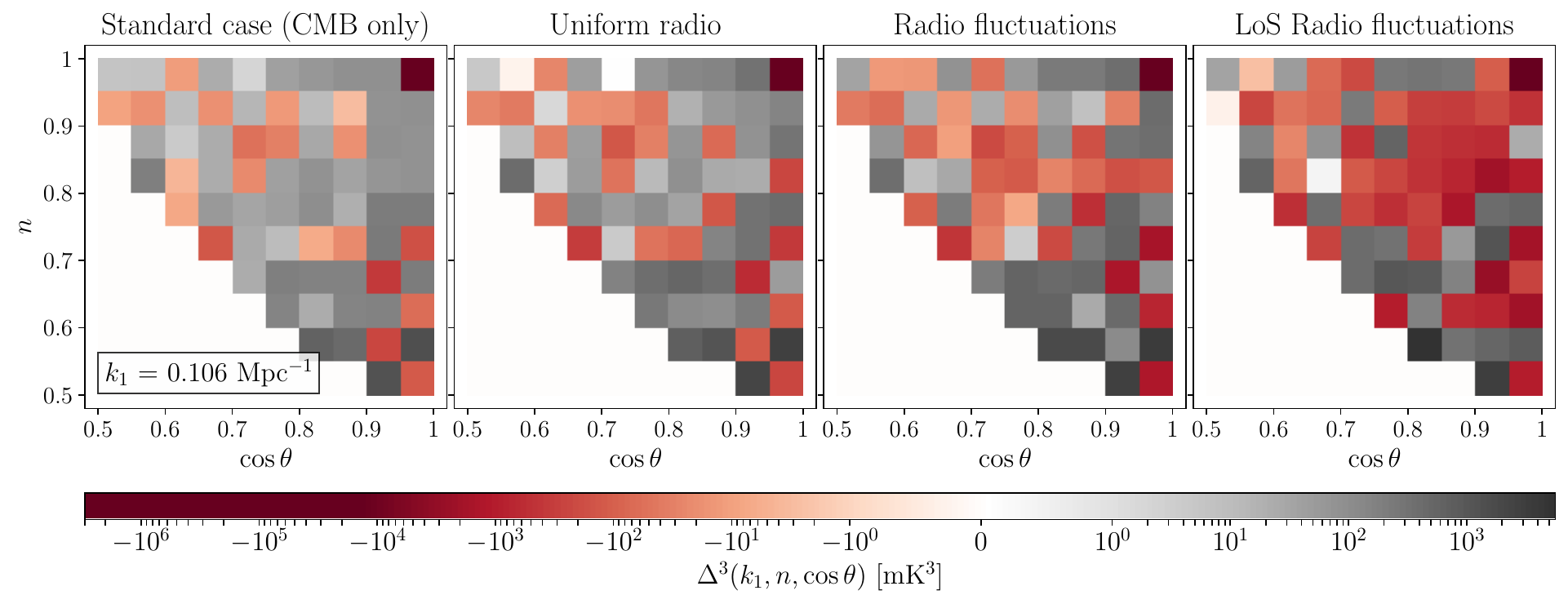}
\includegraphics[width=0.9\textwidth]{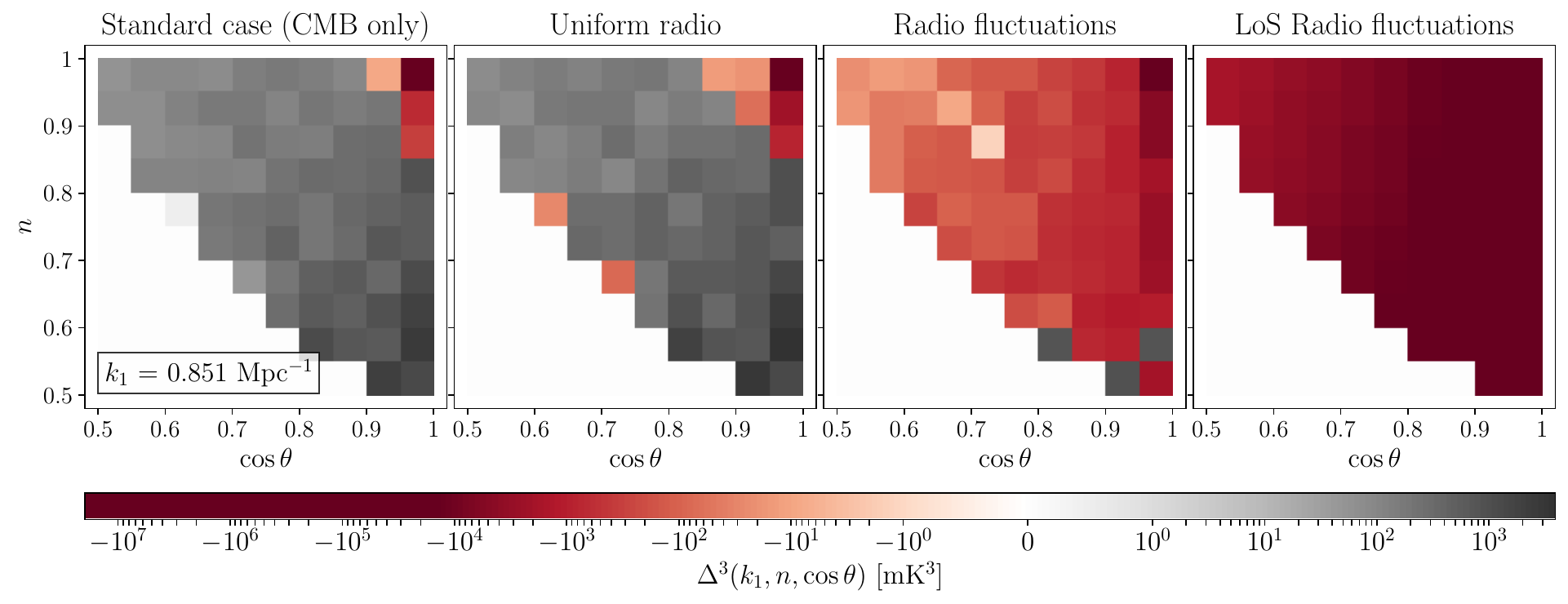}
\caption{The 21-cm SABS at $z=20$ for all unique angle configurations for the four models considered in this section: standard astrophysical model (left panel), uniform radio background (center left), galactic radio background with isotropy (center right) and full galactic radio background with the LoS effect (right), at $k_1 = 0.106$ and $0.851$ Mpc$^{-1}$.  The simulations assume the same astrophysical model parameters as discussed in section \ref{sec:simulation} and presented in Table \ref{tab:parameters_list}.}
\label{fig:bispectrum_all_triangles}
\end{figure*}

\begin{figure*}
\centering
\includegraphics[width=0.9\textwidth]{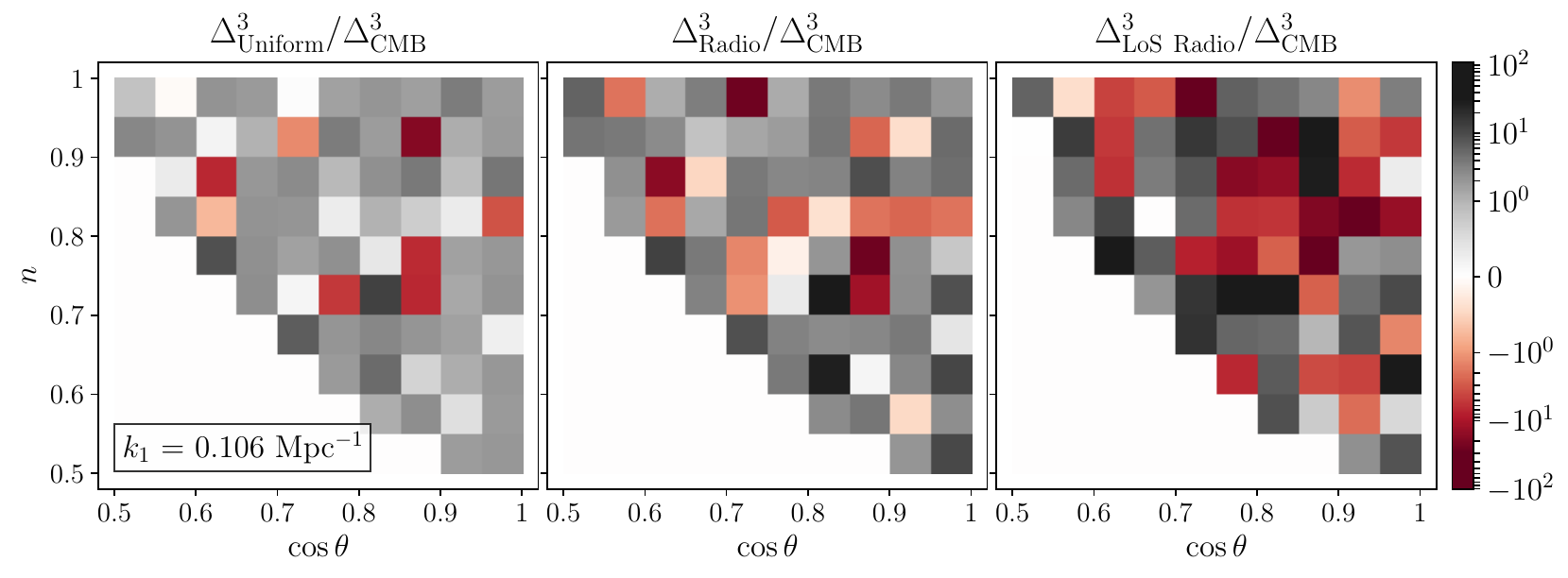}
\includegraphics[width=0.9\textwidth]{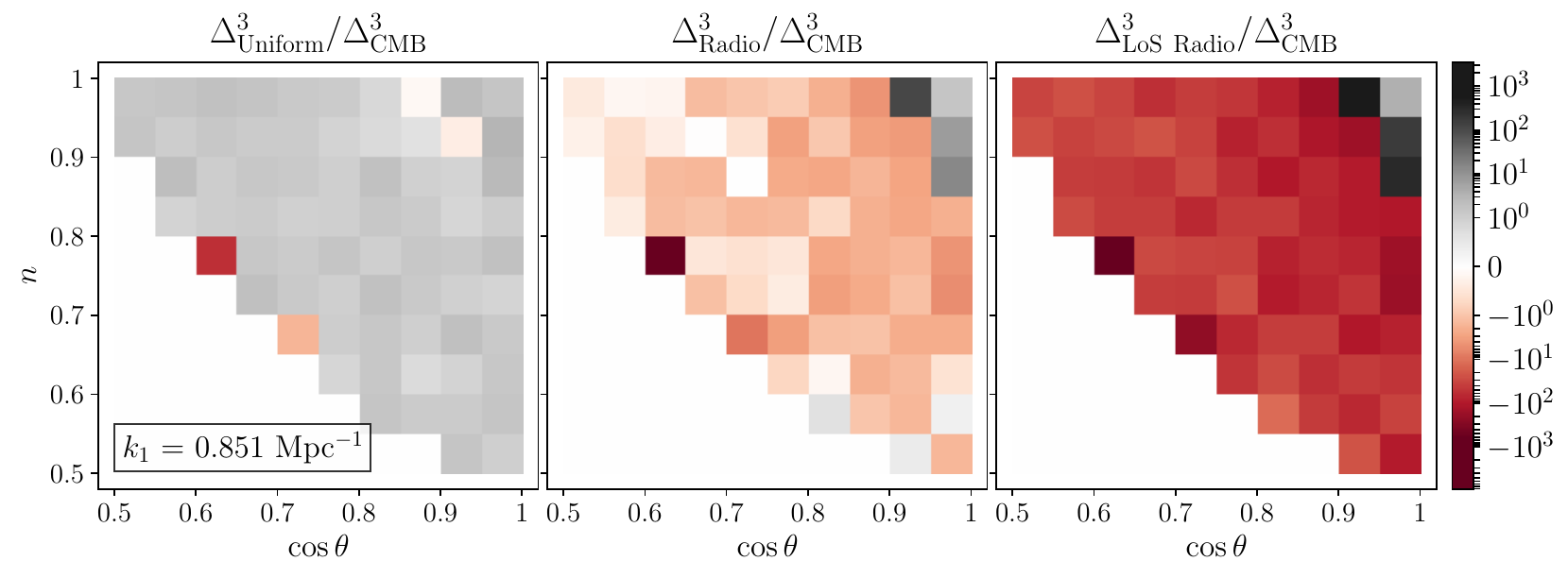}
\caption{The ratio between the 21-cm SABS at $z=20$ in excess radio models and that in the standard astrophysical model, shown in the $n- \cos{\theta}$ space, at $k_1 = 0.106$ and $0.851$ Mpc$^{-1}$. The models shown are: uniform radio background (left), and galactic radio background with isotropy (center) or with the LoS effect (right).}\label{fig:bispectrum_all_triangles_diff}
\end{figure*}

In Figure~\ref{fig:bispectrum_all_triangles} we present the CD 21-cm SABS for all unique triangle configurations in the $n-\cos{\theta}$ space for our four different models at $k_1 = 0.106$ and $0.851$ Mpc$^{-1}$. The figure shows that the SABS is significantly non-zero for almost the entire parameter space of unique triangle configurations, indicating that the 21-cm signal is considerably non-Gaussian during the CD. In all cases, the SABS amplitude in the squeezed limit configuration (top-right corner of each panel) at $z = 20$ is large and negative. Since the bispectrum for the squeezed limit triangle configuration has the maximum magnitude, this has the highest probability of being detected in upcoming 21-cm experiments \citep{Mondal2021}.

The SABS amplitudes are generally lower in magnitude on large scales ($k_1 = 0.106$ Mpc$^{-1}$) than small scales ($k_1 = 0.851$ Mpc$^{-1}$), across all configurations. The power spectrum is often larger on small scales as well, depending on the dominant source of fluctuations. However, in the SABS the difference is very large, with SABS amplitudes that are orders of magnitude higher on small scales, suggesting that non-Gaussian effects are more pronounced on small scales, particularly when radio fluctuations are present. The LoS radio fluctuation case (right panels) shows the strongest enhancement on small scales, consistent with the idea that small-scale fluctuations are amplified by the Poisson character and strong anisotropy of the galactic radio background.

For the standard astrophysical model (CMB-only case, left panels of Figure~\ref{fig:bispectrum_all_triangles}) and the uniform radio background model (center-left panel), the SABS is positive for most of the triangle configurations on small scales ($k_1 = 0.851$ Mpc$^{-1}$). In the standard astrophysical model, the Lyman-$\alpha$ fluctuations dominate the 21-cm brightness temperature fluctuations at $z = 20$. These fluctuations drive mode coupling in a way that keeps the SABS positive, particularly at small scales. For equilateral and isosceles triangles, the dominant interaction occurs between modes of similar size. However, in the squeezed limit ($k_1 \approx k_2 \gg k_3$), a large-scale mode ($k_3$) interacts with two small-scale modes ($k_1, k_2$). We find a negative SABS in the squeezed-limit case. The uniform radio model applies a homogeneous excess radio background across all regions; in this model, the dominant fluctuations are still driven by Lyman-$\alpha$ coupling, resulting in a bispectrum that is only mildly altered compared with the CMB-only case.


Unlike the standard and uniform radio models, which rely mainly on Lyman-$\alpha$ fluctuations to drive the bispectrum, the radio fluctuation models introduce additional structure through spatial variations in the radio background. If radio fluctuations simply amplified other 21-cm fluctuations uniformly throughout space, the SABS would change at most by an overall scaling. However, in reality, radio fluctuations behave quite differently from Lyman-$\alpha$ fluctuations, and modulate the 21-cm fluctuations in a scale-dependent manner. The result that we find is that the SABS is slightly more negative on large scales, but far more on small scales. The LoS effect introduces additional non-Gaussianity and anisotropy in the 21-cm fluctuations, resulting in a SABS that is even more negative compared to the isotropic radio fluctuation model, especially on small scales. 

In order to focus on the effect of the excess radio models, in Figure~\ref{fig:bispectrum_all_triangles_diff} we show the ratio between the SABS in each excess radio model and the SABS in the standard astrophysical model, throughout the $n-\cos{\theta}$ space. On large scales (small $k_{1}$, top panel), the typical enhancement of the SABS increases as we go from the uniform radio model, to the isotropic galactic radio model, to the full LoS galactic radio model; however, no clear pattern emerges in the sign of the SABS ratio, which varies significantly (although it becomes more negative especially for the LoS model). On small scales (large $k_{1}$, bottom panel), the LoS radio model exhibits large ratios, reaching factors of 10 or more, with highly negative values (and positive in a few cases) across many different triangle configurations. In contrast, the uniform radio model shows relatively small deviations from the standard model, while the isotropic model is intermediate. This again confirms that a uniform radio background does not introduce significant new structure but instead acts as a simple scaling factor. Indeed, the SABS usually does not invert its sign in this case. The SABS in the radio fluctuation models has the opposite sign of the standard model, on this small scale. The LoS radio fluctuations enhance SABS more strongly, leading to higher negative ratios. Since the SABS for the squeezed-limit triangle configuration is negative at the CD even in the standard case, the SABS ratios for all the excess radio models turn positive in that specific limit. Our results show that the magnitude and sign of the SABS ratio on small scales across various triangle configurations in the $n-\cos{\theta}$ space can, in principle, distinguish the type of excess radio background present in the 21-cm signal (i.e., galactic or uniform). In particular, the SABS provides a strong observational signature of the LoS effect. 

\subsubsection{More general results}

\begin{figure*}
    \centering
    \includegraphics[width=1\textwidth]{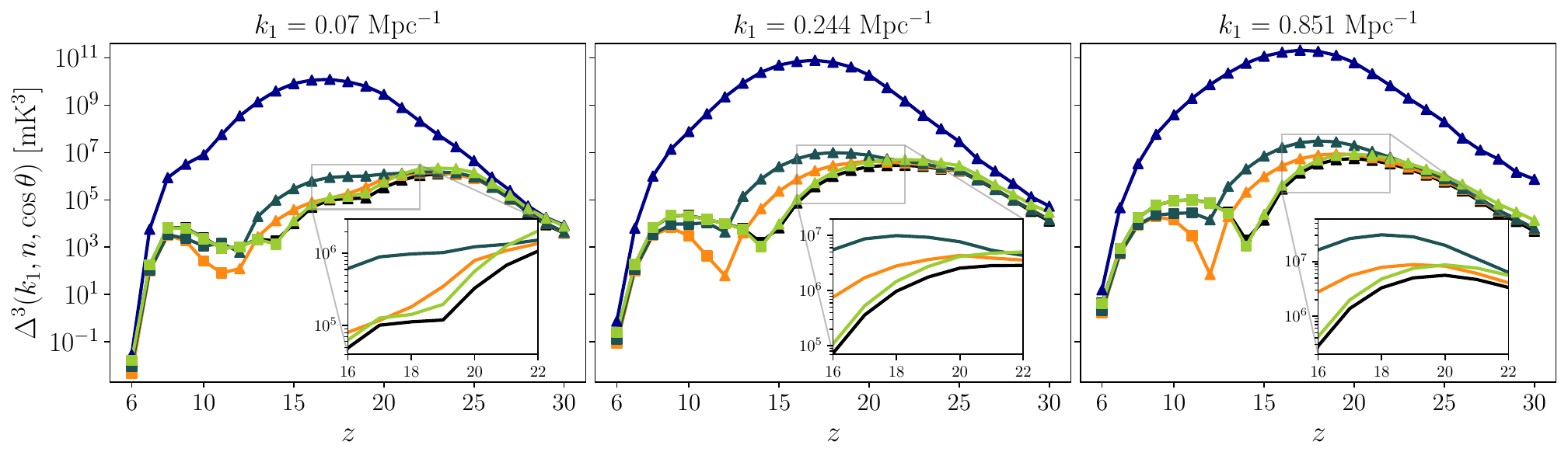}
    \caption{The evolution of the 21-cm SABS as a function of redshift for three different wavenumbers: $k = 0.07$ (left panel), $0.244$ (middle panel) and $0.851$ Mpc$^{-1}$ (right panel). We show the SABS in the squeezed limit ($\cos{\theta} \rightarrow  1$, i.e., $k_{3} \rightarrow 0$, and $n \rightarrow 1$, i.e., $k_{1} \rightarrow k_{2}$). We show our five models as in Figure~\ref{fig:global_signal_power_spectrum}: the standard astrophysical model (black line), uniform radio excess (green line), isotropic galactic radio fluctuations (orange line), and LoS galactic radio fluctuation model with $f_{\rm{Radio}}=60$ (teal line) or $f_{\rm{Radio}}=3000$ (blue line). The inset plot in each panel zooms in to a small redshift range showing the differences in SABS values for the standard and moderate radio models. The square and triangle markers represent positive and negative values of the SABS, respectively.}
    \label{fig:bispec_z_squeezed_limit}
\end{figure*}

\begin{figure*}
    \centering
    \includegraphics[width=1\textwidth]{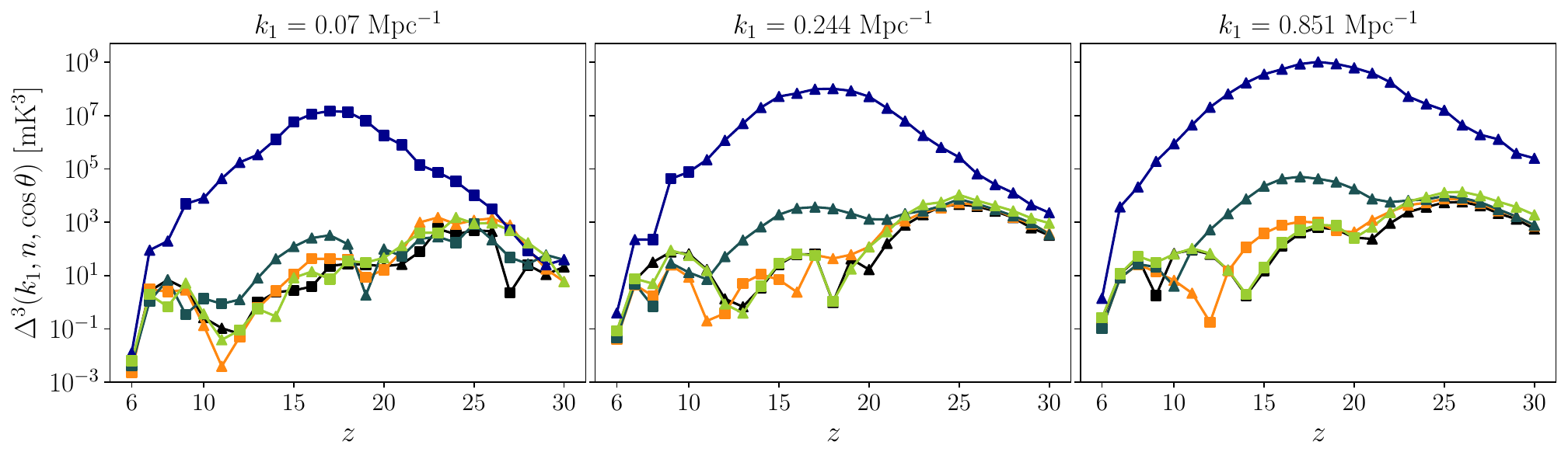}
    \caption{Same as Fig.~\ref{fig:bispec_z_squeezed_limit} except that here we show the SABS for $\cos{\theta} = 0.875$ and $n = 0.875$.}\label{fig:bispec_z_n875_cos875}
\end{figure*}

So far in this section, we have examined the impact of moderate excess radio models on the 21-cm SABS at redshift 20, the approximate location of the global signal absorption trough in our four models. 
In this section, we take a broader look at how excess radio background models affect the 21-cm SABS over the widest redshift range (6 to 30), and compare also to the high galactic radio excess model (with LoS effect). 

In Figure~\ref{fig:bispec_z_squeezed_limit}, we show the evolution of the SABS in the squeezed-triangle limit as a function of redshift, at $k = 0.07, 0.244$ and $0.851$ Mpc$^{-1}$. The inset plot in each panel zooms in to a redshift range corresponding to the CD. Here, we see several notable features. First, the SABS evolves significantly with redshift, exhibiting notable variations across different cosmic epochs. At early redshifts ($z>25$), the SABS values are relatively small, as the Lyman-$\alpha$ fluctuations have not yet emerged, and density and temperature fluctuations dominate. During this epoch, the IGM is cold and neutral. As the first sources begin to form, Lyman-$\alpha$ fluctuations from the earliest stars and galaxies become the dominant source of 21-cm fluctuations, resulting in a strong absorption signal. The prominent peak in the SABS that occurs around $z\sim 15-25$ (depending on the model) reflects the intense 21-cm absorption due to Lyman-$\alpha$ coupling. This characteristic is qualitatively similar to the power spectrum (illustrated in the bottom panel of Figure~\ref{fig:global_signal_power_spectrum}). A key feature is that the SABS for the squeezed-limit triangle configuration remains negative throughout the entire CD redshift range, in all the models. 


At lower redshifts ($z<15$), the SABS drops, likely due to the onset of the heating transition. As the Universe evolves to lower redshifts, X-ray heating from early galaxies becomes significant, altering the dominant source of 21-cm fluctuations. In regions of high galactic density, the gas is strongly heated, and the spin temperature ($T_{\rm{s}}$) increases, reducing the contrast between $T_{\rm{s}}$ and the background radiation temperature. This, in turn, weakens the 21-cm absorption signal. This reduces the correlation between galactic density fluctuations and 21-cm fluctuations. Since the SABS measures how fluctuations at different scales interact, this changes the mode coupling structure observed during the Lyman-$\alpha$ coupling era. In the squeezed limit case, where a long-wavelength mode modulates short-wavelength fluctuations, this new interaction ultimately flips the sign of the SABS from negative to positive as the universe transitions into the heating era. The drop in SABS continues even at lower redshifts ($z<10$), as reionization further reduces small-scale fluctuations. However, the high LoS galactic radio model is different from the others; the strong LoS radio fluctuations dominate and produce a very high, and consistently negative, SABS, down to the end of reionization.

As before, the uniform radio excess only slightly shifts the SABS amplitudes without altering the overall shape, reinforcing the idea that a uniform background does not change the non-Gaussianities in the signal much. For instance, at $k_1 = 0.244$ Mpc$^{-1}$, the uniform radio background enhances the SABS by factors of 1.47, 1.55 and 1.76 at $z = 16, 19$ and 22, respectively. In particular, this boosting factor increases with $z$, as the spatially uniform radio background is more intense during early cosmic epochs before the CD, gradually weakening over cosmic time. In contrast, the excess radio background from early galaxies is non-uniform, and its intensity increases with cosmic time as it traces the formation and growth of galaxies (given our assumption that the radio production efficiency of these galaxies remains constant). For example, at $k_1 = 0.244$ Mpc$^{-1}$, radio fluctuations enhance the SABS (compared to CMB-only) by factors of 10.4, 2.07 and 1.25 at $z=16, 19$ and 22, respectively. Since the LoS radio effect introduces additional anisotropy in the 21-cm fluctuations, it further amplifies the SABS. In this case (with a still moderate $f_{\rm{Radio}}$), it increases it (compared to CMB-only) by factors of 75.4, 5.27 and 1.52 at $z=16, 19$ and 22, respectively. The differences between the isotropic (dark orange) and LoS (teal) radio cases demonstrate that an anisotropic radio background imprints a distinctive non-Gaussian signature on the 21-cm SABS. Finally, the high $f_{\rm{Radio}}$ LoS model produces a much higher SABS, up to four orders of magnitude higher than even the moderate $f_{\rm{Radio}}$ LoS model (at $z \sim 10 - 15$).

It is important to note the sign change in the squeezed-limit SABS, which reflects a fundamental shift in the dominant physical processes shaping the 21-cm fluctuations. In the CMB-only and uniform radio background models, the sign transition (from negative to positive) in the squeezed-limit SABS occurs at $z=14$, marking the beginning of the heating era. However, when an excess galactic radio background over the CMB is present, the coupling transition is delayed because the effective Lyman-$\alpha$ coupling coefficient decreases with the brightness temperature of the radiation background. Furthermore, the heating transition (defined relative to the mean 21-cm background temperature) is also delayed because, in the presence of radio excess, the kinetic temperature needs to reach a higher value, which is the mean $T_{\rm{CMB}} + T_{\rm{Radio}}$ for the isotropic radio fluctuation model or the mean $T_{\rm{CMB}} + T_{\rm{Radio, \ los}}$ for the LoS radio fluctuation model. These mean values are the same (i.e., the LoS model includes the LoS fluctuations of the radio intensity, but the overall mean radio intensity is the same). As a result, in the galactic radio background models, the sign change in the squeezed limit SABS (negative to positive) occurs at a lower redshift compared to the uniform radio background model and the standard astrophysical model. This delayed heating transition, observed as a sign change at lower redshifts, could serve as a smoking-gun signature of the presence of an excess radio background originating from high-redshift radio-loud sources. This sign-change redshift depends on the strength of the galactic radio background, so that it is $z \sim 11$ for $f_{\rm{Radio}}=60$, while the high $f_{\rm{Radio}}=3000$ pushes it beyond the end of reionization.

During the EoR, the primary driver of 21-cm fluctuations is the formation of ionized bubbles (HII regions) around UV-emitting sources. These bubbles introduce large-scale fluctuations in the 21-cm brightness temperature by creating regions where the signal is suppressed due to ionization. As the number of sources increases, excess radio fluctuations gradually diminish across all scales, causing the power spectrum to converge toward that of the standard astrophysical model. The squeezed-limit SABS follows a similar trend, mirroring the evolution of the power spectrum. Figure~\ref{fig:bispec_z_n875_cos875} shows another example of the evolution of the SABS as a function of redshift, for a triangle configuration that is slightly away from the squeezed limit, with $\cos{\theta} = 0.875$ and $n=0.875$, again for $k = 0.07, 0.244$ and $0.851$ Mpc$^{-1}$. The overall behavior of the SABS is similar (as a function of redshift and in terms of the differences among the various models), but the magnitude of the SABS is around two orders of magnitude lower than in the squeezed limit. Also, there are more sign changes so that this feature is more complex to interpret.

\begin{figure}
\includegraphics[width=0.48\textwidth]{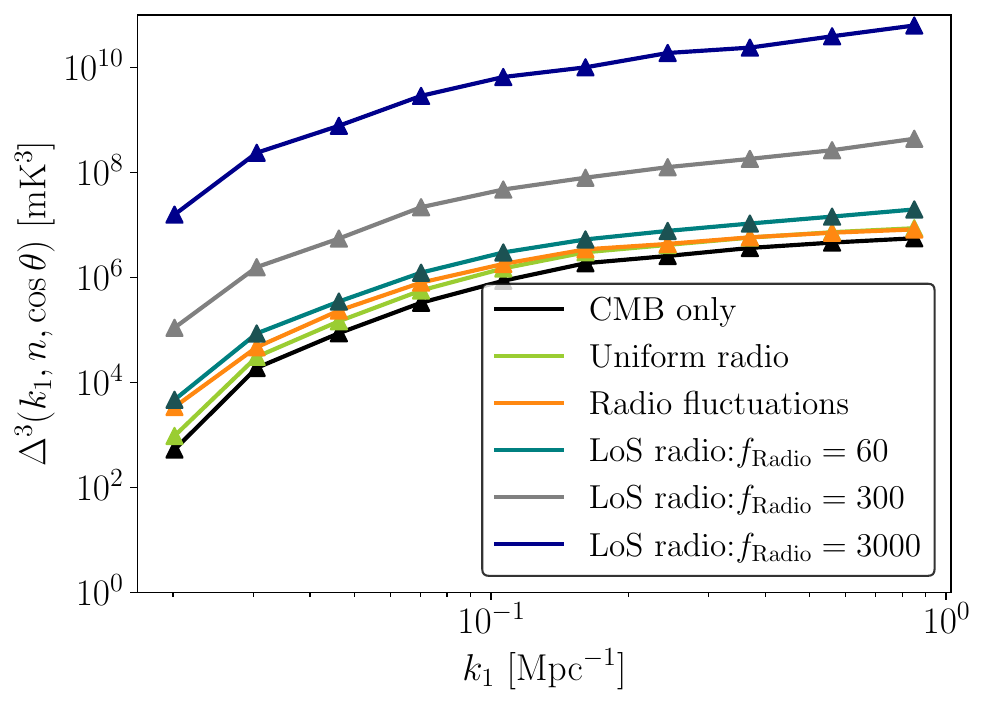}
\caption{The 21-cm SABS for squeezed-limit triangles ($\cos\theta \rightarrow 1$, i.e., $k_3 \rightarrow 0$, and $n \rightarrow 1$,  i.e.,  $k_1 \rightarrow k_2$) as a function of $k_1$ at $z=20$. The triangle markers represent negative values of the bispectrum. The colored lines show various simulation runs: standard astrophysical case (black), uniform radio (green), isotropic galactic radio (orange) and LoS galactic radio fluctuations (shown for three values of $f_{\rm{Radio}}$). }\label{fig:bispectrum_squeezed_triangles}
\end{figure}


Focusing again on the squeezed limit, we show the dependence of the SABS at $z = 20$ on $k_1$ (rather than just a few discrete values of $k_1$ as in the other figures) in Figure~\ref{fig:bispectrum_squeezed_triangles}. The SABS magnitude increases with $k_1$, similarly to the typical behavior of the power spectrum. Since these results correspond to $z=20$, where Lyman-$\alpha$-driven fluctuations dominate, all SABS amplitudes are negative. The Lyman-$\alpha$ fluctuations have a natural smoothing, on the typical scale to which the Lyman-$\alpha$ photons travel, and so the increasing curves flatten at high $k$. However, the LoS fluctuations act over a wide range of scales and reduce this flattening, in addition to their overall boosting of the SABS. For example, at $k_1 = 0.02$ Mpc$^{-1}$, the uniform, isotropic and LoS radio backgrounds (with $f_{\rm{Radio}}=60$ for the latter two) enhance the squeezed-limit SABS by factors of 1.82, 6.48, and 8.89, respectively, compared to the standard astrophysical model. At $k_1 = 0.85$ Mpc$^{-1}$, these factors are 1.55, 1.47, and 3.54. The effect of a uniform radio background on the squeezed-limit SABS is fairly scale-independent, as we saw before. In contrast, radio fluctuations have a stronger impact on large scales than on small scales (due to smoothing on the scale of the typical distance to the nearest radio source).

Figure~\ref{fig:bispectrum_squeezed_triangles} also shows the LoS galactic radio fluctuations for the high  $f_{\rm{Radio}}=3000$, and here also for an intermediate case of $f_{\rm{Radio}}=300$ (which, as expected, gives intermediate results for the SABS). Compared to the standard astrophysical case, the squeezed-limit SABS at $k_1=0.02$ Mpc$^{-1}$ is enhanced by factors of 209 and 29900 for simulation runs with $f_{\rm{Radio}} = 300$ and 3000, respectively. At $k_1 = 0.85$ Mpc$^{-1}$, these factors are 78.5 and 11300. We leave it for future work to analyze the detectability of the CD 21-cm SABS in upcoming observations such as those with the SKA-Low.

\section{Conclusions}
\label{sec:conclusion}
In this paper, we have presented the effects of excess radio background models on the 21-cm SABS during the CD. The novelty of this work lies in being the first study to explore the imprint of high-redshift radio fluctuations on the 21-cm SABS during this epoch. We compared three different excess radio background models to the standard astrophysical scenario (CMB-only case), where no excess radio is present. We examined the 21-cm SABS across all unique triangle configurations in the $n-\cos\theta$ space. Since all astrophysical parameters are held fixed, our findings reflect the impact of different excess radio models relative to the standard scenario. For the astrophysical parameters used here, the SABS is significantly non-zero across essentially the entire range of unique triangle configurations, confirming that the 21-cm signal is inherently non-Gaussian during the CD.

In excess radio models, as well as in the standard astrophysical scenario, the value of the squeezed-limit SABS at $z=20$ is negative. This negative squeezed limit SABS appears to be a characteristic signature of Lyman-$\alpha$-driven fluctuations. The uniform radio model introduces a uniform excess radio background, and the mode coupling remains largely similar to that of the standard model. In contrast, radio fluctuations from early galaxies do more than simply amplifying small-scale fluctuations -- they impose large-scale structure through spatial variations, modulating their strength in a scale-dependent manner. As a result, the SABS shifts in some cases from positive to negative. The LoS effect introduces additional non-Gaussianity and anisotropy in the 21-cm fluctuations, and often makes the SABS more strongly negative compared to the (artificially) isotropic radio fluctuation model.

Given the inherent complexity of the bispectrum, we simplified the discussion by focusing on the squeezed-limit SABS and its redshift evolution. The SABS evolves significantly with redshift, exhibiting notable variations across different cosmic epochs. The prominent peak in the squeezed limit SABS occurring around $z\sim 15 - 25$ reflects the intense 21-cm absorption due to the underlying Lyman-$\alpha$ coupling. This behavior is similar to that of the power spectrum, except that the squeezed-limit SABS remains negative throughout the entire CD redshift range. The sign change in the squeezed-limit SABS marks a fundamental shift in the dominant physical process governing 21-cm fluctuations, and is driven by the heating transition. In the uniform radio background model, the sign changes at the same redshift as in the standard astrophysical scenario. In the galactic radio background model, the transition from negative to positive squeezed-limit SABS occurs at a lower redshift compared to the standard astrophysical scenario; this redshift is independent of the LoS effect, but depends on the overall efficiency of the galactic radio emission (set by $f_{\rm{Radio}}$). We note that the sign is a qualitative feature of the SABS that is not available in the global 21-cm signal or the 21-cm power spectrum.

The amplitude of the squeezed-limit SABS increases with wavenumber ($k_1$, corresponding to the largest side of the triangle). The excess radio background generally enhances the SABS across all scales, with the LoS model exhibiting the strongest amplification. Here we have shown that the 21-cm SABS can in principle provide substantial information on the sources of 21-cm fluctuations. We also considered another signature of non-Gaussianity, namely the probability distribution function of the brightness temperature in 21-cm images. In particular, we considered two higher-order moments, the skewness and kurtosis. The non-Gaussianity is generally large at both low and high redshifts. During the CD, the LoS model (for moderate or high radio efficiency) is substantially more non-Gaussian by these measures. We leave for future work a detailed analysis of the detectability of these various measures of non-Gaussianity with upcoming radio interferometers.

\section*{Acknowledgements}
SS and RB acknowledge the support of the Israel Science Foundation (grant No.\ 1078/24). RM is supported by the NITC FRG Seed Grant (NITC/PRJ/PHY/2024-25/FRG/12). RM is grateful for the helpful discussions that Somnath Bharadwaj, Garrelt Mellema, Abinash Kumar Shaw, and Suman Majumdar provided during the development of the Bispectrum estimation code.

This research made use of: \texttt{Numpy} \citep{harris2020array}, \texttt{Scipy} \citep{2020SciPy-NMeth}, \texttt{matplotlib} \citep{Hunter:2007} and the NASA Astrophysics Data System Bibliographic Services.

\section*{Data Availability}
The data underlying this article will be shared on reasonable request to the corresponding author.



\bibliographystyle{mnras}
\bibliography{example} 

@article{bowman18,
	Adsurl = {https://ui.adsabs.harvard.edu/abs/2018Natur.555...67B},
	Archiveprefix = {arXiv},
	Author = {{Bowman}, Judd D. and {Rogers}, Alan E.~E. and {Monsalve}, Raul A. and {Mozdzen}, Thomas J. and {Mahesh}, Nivedita},
	Doi = {10.1038/nature25792},
	Eprint = {1810.05912},
	Journal = {\nat},
	Month = {Mar},
	Number = {7694},
	Pages = {67-70},
	Primaryclass = {astro-ph.CO},
	Title = {{An absorption profile centred at 78 megahertz in the sky-averaged spectrum}},
	Volume = {555},
	Year = {2018}
 }

@ARTICLE{Mondal2015,
       author = {{Mondal}, R. and {Bharadwaj}, S. and {Majumdar}, S. and {Bera}, A. and {Acharyya}, A.},
        title = "{The effect of non-Gaussianity on error predictions for the Epoch of Reionization (EoR) 21-cm power spectrum.}",
      journal = {\mnras},
         year = 2015,
        month = apr,
       volume = {449},
        pages = {L41-L45},
          doi = {10.1093/mnrasl/slv015},
archivePrefix = {arXiv},
       eprint = {1409.4420},
 primaryClass = {astro-ph.CO},
       adsurl = {https://ui.adsabs.harvard.edu/abs/2015MNRAS.449L..41M}
}

@ARTICLE{Bharadwaj2005,
       author = {{Bharadwaj}, Somnath and {Pandey}, Sanjay K.},
        title = "{Probing non-Gaussian features in the HI distribution at the epoch of re-ionization}",
      journal = {\mnras},
         year = 2005,
        month = apr,
       volume = {358},
       number = {3},
        pages = {968-976},
          doi = {10.1111/j.1365-2966.2005.08836.x},
archivePrefix = {arXiv},
       eprint = {astro-ph/0410581},
 primaryClass = {astro-ph},
       adsurl = {https://ui.adsabs.harvard.edu/abs/2005MNRAS.358..968B}
}

@ARTICLE{Mondal2016,
       author = {{Mondal}, Rajesh and {Bharadwaj}, Somnath and {Majumdar}, Suman},
        title = "{Statistics of the epoch of reionization 21-cm signal - I. Power spectrum error-covariance}",
      journal = {\mnras},
         year = 2016,
        month = feb,
       volume = {456},
       number = {2},
        pages = {1936-1947},
          doi = {10.1093/mnras/stv2772},
archivePrefix = {arXiv},
       eprint = {1508.00896},
 primaryClass = {astro-ph.CO},
       adsurl = {https://ui.adsabs.harvard.edu/abs/2016MNRAS.456.1936M}
}

@ARTICLE{Kamran2022,
       author = {{Kamran}, Mohd and {Ghara}, Raghunath and {Majumdar}, Suman and {Mellema}, Garrelt and {Bharadwaj}, Somnath and {Pritchard}, Jonathan R. and {Mondal}, Rajesh and {Iliev}, Ilian T.},
        title = "{Redshifted 21-cm bispectrum: impact of the source models on the signal and the IGM physics from the Cosmic Dawn}",
      journal = {\jcap},
         year = 2022,
        month = nov,
       volume = {2022},
       number = {11},
          eid = {001},
        pages = {001},
          doi = {10.1088/1475-7516/2022/11/001},
archivePrefix = {arXiv},
       eprint = {2207.09128},
 primaryClass = {astro-ph.CO},
       adsurl = {https://ui.adsabs.harvard.edu/abs/2022JCAP...11..001K}
}

@BOOK{Peebles1980,
       author = {{Peebles}, P.~J.~E.},
        title = "{The large-scale structure of the universe}",
         year = 1980,
       adsurl = {https://ui.adsabs.harvard.edu/abs/1980lssu.book.....P}
}

@ARTICLE{Bharadwaj2005b,
       author = {{Bharadwaj}, Somnath and {Ali}, Sk. Saiyad},
        title = "{On using visibility correlations to probe the HI distribution from the dark ages to the present epoch - I. Formalism and the expected signal}",
      journal = {\mnras},
         year = 2005,
        month = feb,
       volume = {356},
       number = {4},
        pages = {1519-1528},
          doi = {10.1111/j.1365-2966.2004.08604.x},
archivePrefix = {arXiv},
       eprint = {astro-ph/0406676},
 primaryClass = {astro-ph},
       adsurl = {https://ui.adsabs.harvard.edu/abs/2005MNRAS.356.1519B}
}

@ARTICLE{Watkinson2017,
       author = {{Watkinson}, Catherine A. and {Majumdar}, Suman and {Pritchard}, Jonathan R. and {Mondal}, Rajesh},
        title = "{A fast estimator for the bispectrum and beyond - a practical method for measuring non-Gaussianity in 21-cm maps}",
      journal = {\mnras},
         year = 2017,
        month = dec,
       volume = {472},
       number = {2},
        pages = {2436-2446},
          doi = {10.1093/mnras/stx2130},
archivePrefix = {arXiv},
       eprint = {1705.06284},
 primaryClass = {astro-ph.CO},
       adsurl = {https://ui.adsabs.harvard.edu/abs/2017MNRAS.472.2436W}
}

@ARTICLE{Majumdar2018,
       author = {{Majumdar}, Suman and {Pritchard}, Jonathan R. and {Mondal}, Rajesh and {Watkinson}, Catherine A. and {Bharadwaj}, Somnath and {Mellema}, Garrelt},
        title = "{Quantifying the non-Gaussianity in the EoR 21-cm signal through bispectrum}",
      journal = {\mnras},
         year = 2018,
        month = may,
       volume = {476},
       number = {3},
        pages = {4007-4024},
          doi = {10.1093/mnras/sty535},
archivePrefix = {arXiv},
       eprint = {1708.08458},
 primaryClass = {astro-ph.CO},
       adsurl = {https://ui.adsabs.harvard.edu/abs/2018MNRAS.476.4007M}
}

@ARTICLE{Majumdar2020,
       author = {{Majumdar}, Suman and {Kamran}, Mohd and {Pritchard}, Jonathan R. and {Mondal}, Rajesh and {Mazumdar}, Arindam and {Bharadwaj}, Somnath and {Mellema}, Garrelt},
        title = "{Redshifted 21-cm bispectrum - I. Impact of the redshift space distortions on the signal from the Epoch of Reionization}",
      journal = {\mnras},
         year = 2020,
        month = dec,
       volume = {499},
       number = {4},
        pages = {5090-5106},
          doi = {10.1093/mnras/staa3168},
archivePrefix = {arXiv},
       eprint = {2007.06584},
 primaryClass = {astro-ph.CO},
       adsurl = {https://ui.adsabs.harvard.edu/abs/2020MNRAS.499.5090M}
}

@ARTICLE{Iliev2007,
       author = {{Iliev}, Ilian T. and {Mellema}, Garrelt and {Shapiro}, Paul R. and {Pen}, Ue-Li},
        title = "{Self-regulated reionization}",
      journal = {\mnras},
         year = 2007,
        month = apr,
       volume = {376},
       number = {2},
        pages = {534-548},
          doi = {10.1111/j.1365-2966.2007.11482.x},
archivePrefix = {arXiv},
       eprint = {astro-ph/0607517},
 primaryClass = {astro-ph},
       adsurl = {https://ui.adsabs.harvard.edu/abs/2007MNRAS.376..534I}
}

@ARTICLE{Mondal2017,
       author = {{Mondal}, Rajesh and {Bharadwaj}, Somnath and {Majumdar}, Suman},
        title = "{Statistics of the epoch of reionization (EoR) 21-cm signal - II. The evolution of the power-spectrum error-covariance}",
      journal = {\mnras},
         year = 2017,
        month = jan,
       volume = {464},
       number = {3},
        pages = {2992-3004},
          doi = {10.1093/mnras/stw2599},
archivePrefix = {arXiv},
       eprint = {1606.03874},
 primaryClass = {astro-ph.CO},
       adsurl = {https://ui.adsabs.harvard.edu/abs/2017MNRAS.464.2992M}
}

@ARTICLE{Mondal2021,
       author = {{Mondal}, Rajesh and {Mellema}, Garrelt and {Shaw}, Abinash Kumar and {Kamran}, Mohd and {Majumdar}, Suman},
        title = "{The Epoch of Reionization 21-cm bispectrum: the impact of light-cone effects and detectability}",
      journal = {\mnras},
         year = 2021,
        month = dec,
       volume = {508},
       number = {3},
        pages = {3848-3859},
          doi = {10.1093/mnras/stab2900},
archivePrefix = {arXiv},
       eprint = {2107.02668},
 primaryClass = {astro-ph.CO},
       adsurl = {https://ui.adsabs.harvard.edu/abs/2021MNRAS.508.3848M}
}

@ARTICLE{swarup91,
   author = {{Swarup}, G. and {Ananthakrishnan}, S. and {Kapahi}, V.~K. and 
	{Rao}, A.~P. and {Subrahmanya}, C.~R. and {Kulkarni}, V.~K.},
    title = "{The Giant Metre-Wave Radio Telescope}",
  journal = {Current Science, Vol.~60, NO.2/JAN25, P.~95, 1991},
     year = 1991,
    month = jan,
   volume = 60,
    pages = {95},
   adsurl = {http://adsabs.harvard.edu/abs/1991CuSc...60...95S}
}

@ARTICLE{haarlem13,
   author = {{van Haarlem}, M.~P. and {Wise}, M.~W. and {Gunst}, A.~W. and 
	{Heald}, G. and {McKean}, J.~P. and {Hessels}, J.~W.~T. and 
	{de Bruyn}, A.~G. and {Nijboer}, R. and {Swinbank}, J. and {Fallows}, R. and 
	{Brentjens}, M. and {Nelles}, A. and {Beck}, R. and {Falcke}, H. and 
	{Fender}, R. and {H{\"o}randel}, J. and {Koopmans}, L.~V.~E. and 
	{Mann}, G. and {Miley}, G. and {R{\"o}ttgering}, H. and {Stappers}, B.~W. and 
	{Wijers}, R.~A.~M.~J. and {Zaroubi}, S. and {van den Akker}, M. and 
	{Alexov}, A. and {Anderson}, J. and {Anderson}, K. and {van Ardenne}, A. and 
	{Arts}, M. and {Asgekar}, A. and {Avruch}, I.~M. and {Batejat}, F. and 
	{B{\"a}hren}, L. and {Bell}, M.~E. and {Bell}, M.~R. and {van Bemmel}, I. and 
	{Bennema}, P. and {Bentum}, M.~J. and {Bernardi}, G. and {Best}, P. and 
	{B{\^i}rzan}, L. and {Bonafede}, A. and {Boonstra}, A.-J. and 
	{Braun}, R. and {Bregman}, J. and {Breitling}, F. and {van de Brink}, R.~H. and 
	{Broderick}, J. and {Broekema}, P.~C. and {Brouw}, W.~N. and 
	{Br{\"u}ggen}, M. and {Butcher}, H.~R. and {van Cappellen}, W. and 
	{Ciardi}, B. and {Coenen}, T. and {Conway}, J. and {Coolen}, A. and 
	{Corstanje}, A. and {Damstra}, S. and {Davies}, O. and {Deller}, A.~T. and 
	{Dettmar}, R.-J. and {van Diepen}, G. and {Dijkstra}, K. and 
	{Donker}, P. and {Doorduin}, A. and {Dromer}, J. and {Drost}, M. and 
	{van Duin}, A. and {Eisl{\"o}ffel}, J. and {van Enst}, J. and 
	{Ferrari}, C. and {Frieswijk}, W. and {Gankema}, H. and {Garrett}, M.~A. and 
	{de Gasperin}, F. and {Gerbers}, M. and {de Geus}, E. and {Grie{\ss}meier}, J.-M. and 
	{Grit}, T. and {Gruppen}, P. and {Hamaker}, J.~P. and {Hassall}, T. and 
	{Hoeft}, M. and {Holties}, H.~A. and {Horneffer}, A. and {van der Horst}, A. and 
	{van Houwelingen}, A. and {Huijgen}, A. and {Iacobelli}, M. and 
	{Intema}, H. and {Jackson}, N. and {Jelic}, V. and {de Jong}, A. and 
	{Juette}, E. and {Kant}, D. and {Karastergiou}, A. and {Koers}, A. and 
	{Kollen}, H. and {Kondratiev}, V.~I. and {Kooistra}, E. and 
	{Koopman}, Y. and {Koster}, A. and {Kuniyoshi}, M. and {Kramer}, M. and 
	{Kuper}, G. and {Lambropoulos}, P. and {Law}, C. and {van Leeuwen}, J. and 
	{Lemaitre}, J. and {Loose}, M. and {Maat}, P. and {Macario}, G. and 
	{Markoff}, S. and {Masters}, J. and {McFadden}, R.~A. and {McKay-Bukowski}, D. and 
	{Meijering}, H. and {Meulman}, H. and {Mevius}, M. and {Middelberg}, E. and 
	{Millenaar}, R. and {Miller-Jones}, J.~C.~A. and {Mohan}, R.~N. and 
	{Mol}, J.~D. and {Morawietz}, J. and {Morganti}, R. and {Mulcahy}, D.~D. and 
	{Mulder}, E. and {Munk}, H. and {Nieuwenhuis}, L. and {van Nieuwpoort}, R. and 
	{Noordam}, J.~E. and {Norden}, M. and {Noutsos}, A. and {Offringa}, A.~R. and 
	{Olofsson}, H. and {Omar}, A. and {Orr{\'u}}, E. and {Overeem}, R. and 
	{Paas}, H. and {Pandey-Pommier}, M. and {Pandey}, V.~N. and 
	{Pizzo}, R. and {Polatidis}, A. and {Rafferty}, D. and {Rawlings}, S. and 
	{Reich}, W. and {de Reijer}, J.-P. and {Reitsma}, J. and {Renting}, G.~A. and 
	{Riemers}, P. and {Rol}, E. and {Romein}, J.~W. and {Roosjen}, J. and 
	{Ruiter}, M. and {Scaife}, A. and {van der Schaaf}, K. and {Scheers}, B. and 
	{Schellart}, P. and {Schoenmakers}, A. and {Schoonderbeek}, G. and 
	{Serylak}, M. and {Shulevski}, A. and {Sluman}, J. and {Smirnov}, O. and 
	{Sobey}, C. and {Spreeuw}, H. and {Steinmetz}, M. and {Sterks}, C.~G.~M. and 
	{Stiepel}, H.-J. and {Stuurwold}, K. and {Tagger}, M. and {Tang}, Y. and 
	{Tasse}, C. and {Thomas}, I. and {Thoudam}, S. and {Toribio}, M.~C. and 
	{van der Tol}, B. and {Usov}, O. and {van Veelen}, M. and {van der Veen}, A.-J. and 
	{ter Veen}, S. and {Verbiest}, J.~P.~W. and {Vermeulen}, R. and 
	{Vermaas}, N. and {Vocks}, C. and {Vogt}, C. and {de Vos}, M. and 
	{van der Wal}, E. and {van Weeren}, R. and {Weggemans}, H. and 
	{Weltevrede}, P. and {White}, S. and {Wijnholds}, S.~J. and 
	{Wilhelmsson}, T. and {Wucknitz}, O. and {Yatawatta}, S. and 
	{Zarka}, P. and {Zensus}, A. and {van Zwieten}, J.},
    title = "{LOFAR: The LOw-Frequency ARray}",
  journal = {\aap},
archivePrefix = "arXiv",
   eprint = {1305.3550},
 primaryClass = "astro-ph.IM",
     year = 2013,
    month = aug,
   volume = 556,
      eid = {A2},
    pages = {A2},
      doi = {10.1051/0004-6361/201220873},
   adsurl = {http://adsabs.harvard.edu/abs/2013A%26A...556A...2V}
}

@ARTICLE{Greig2021,
       author = {{Greig}, Bradley and {Trott}, Cathryn M. and {Barry}, Nichole and {Mutch}, Simon J. and {Pindor}, Bart and {Webster}, Rachel L. and {Wyithe}, J. Stuart B.},
        title = "{Exploring reionization and high-z galaxy observables with recent multiredshift MWA upper limits on the 21-cm signal}",
      journal = {\mnras},
         year = 2021,
        month = jan,
       volume = {500},
       number = {4},
        pages = {5322-5335},
          doi = {10.1093/mnras/staa3494},
archivePrefix = {arXiv},
       eprint = {2008.02639},
 primaryClass = {astro-ph.CO},
       adsurl = {https://ui.adsabs.harvard.edu/abs/2021MNRAS.500.5322G}
}

@ARTICLE{Ghara2020,
       author = {{Ghara}, R. and {Giri}, S.~K. and {Mellema}, G. and {Ciardi}, B. and {Zaroubi}, S. and {Iliev}, I.~T. and {Koopmans}, L.~V.~E. and {Chapman}, E. and {Gazagnes}, S. and {Gehlot}, B.~K. and {Ghosh}, A. and {Jeli{\'c}}, V. and {Mertens}, F.~G. and {Mondal}, R. and {Schaye}, J. and {Silva}, M.~B. and {Asad}, K.~M.~B. and {Kooistra}, R. and {Mevius}, M. and {Offringa}, A.~R. and {Pandey}, V.~N. and {Yatawatta}, S.},
        title = "{Constraining the intergalactic medium at z {\ensuremath{\approx}} 9.1 using LOFAR Epoch of Reionization observations}",
      journal = {\mnras},
         year = 2020,
        month = feb,
       volume = {493},
       number = {4},
        pages = {4728-4747},
          doi = {10.1093/mnras/staa487},
archivePrefix = {arXiv},
       eprint = {2002.07195},
 primaryClass = {astro-ph.CO},
       adsurl = {https://ui.adsabs.harvard.edu/abs/2020MNRAS.493.4728G}
}

@ARTICLE{tingay13,
   author = {{Tingay}, S.~J. and {Goeke}, R. and {Bowman}, J.~D. and {Emrich}, D. and 
	{Ord}, S.~M. and {Mitchell}, D.~A. and {Morales}, M.~F. and 
	{Booler}, T. and {Crosse}, B. and {Wayth}, R.~B. and {Lonsdale}, C.~J. and 
	{Tremblay}, S. and {Pallot}, D. and {Colegate}, T. and {Wicenec}, A. and 
	{Kudryavtseva}, N. and {Arcus}, W. and {Barnes}, D. and {Bernardi}, G. and 
	{Briggs}, F. and {Burns}, S. and {Bunton}, J.~D. and {Cappallo}, R.~J. and 
	{Corey}, B.~E. and {Deshpande}, A. and {Desouza}, L. and {Gaensler}, B.~M. and 
	{Greenhill}, L.~J. and {Hall}, P.~J. and {Hazelton}, B.~J. and 
	{Herne}, D. and {Hewitt}, J.~N. and {Johnston-Hollitt}, M. and 
	{Kaplan}, D.~L. and {Kasper}, J.~C. and {Kincaid}, B.~B. and 
	{Koenig}, R. and {Kratzenberg}, E. and {Lynch}, M.~J. and {Mckinley}, B. and 
	{Mcwhirter}, S.~R. and {Morgan}, E. and {Oberoi}, D. and {Pathikulangara}, J. and 
	{Prabu}, T. and {Remillard}, R.~A. and {Rogers}, A.~E.~E. and 
	{Roshi}, A. and {Salah}, J.~E. and {Sault}, R.~J. and {Udaya-Shankar}, N. and 
	{Schlagenhaufer}, F. and {Srivani}, K.~S. and {Stevens}, J. and 
	{Subrahmanyan}, R. and {Waterson}, M. and {Webster}, R.~L. and 
	{Whitney}, A.~R. and {Williams}, A. and {Williams}, C.~L. and 
	{Wyithe}, J.~S.~B.},
    title = "{The Murchison Widefield Array: The Square Kilometre Array Precursor at Low Radio Frequencies}",
  journal = {\pasa},
archivePrefix = "arXiv",
   eprint = {1206.6945},
 primaryClass = "astro-ph.IM",
     year = 2013,
    month = jan,
   volume = 30,
      eid = {e007},
    pages = {7},
      doi = {10.1017/pasa.2012.007},
   adsurl = {http://adsabs.harvard.edu/abs/2013PASA...30....7T}
}

@ARTICLE{deboer17,
   author = {{DeBoer}, D.~R. and {Parsons}, A.~R. and {Aguirre}, J.~E. and 
	{Alexander}, P. and {Ali}, Z.~S. and {Beardsley}, A.~P. and 
	{Bernardi}, G. and {Bowman}, J.~D. and {Bradley}, R.~F. and 
	{Carilli}, C.~L. and {Cheng}, C. and {de Lera Acedo}, E. and 
	{Dillon}, J.~S. and {Ewall-Wice}, A. and {Fadana}, G. and {Fagnoni}, N. and 
	{Fritz}, R. and {Furlanetto}, S.~R. and {Glendenning}, B. and 
	{Greig}, B. and {Grobbelaar}, J. and {Hazelton}, B.~J. and {Hewitt}, J.~N. and 
	{Hickish}, J. and {Jacobs}, D.~C. and {Julius}, A. and {Kariseb}, M. and 
	{Kohn}, S.~A. and {Lekalake}, T. and {Liu}, A. and {Loots}, A. and 
	{MacMahon}, D. and {Malan}, L. and {Malgas}, C. and {Maree}, M. and 
	{Martinot}, Z. and {Mathison}, N. and {Matsetela}, E. and {Mesinger}, A. and 
	{Morales}, M.~F. and {Neben}, A.~R. and {Patra}, N. and {Pieterse}, S. and 
	{Pober}, J.~C. and {Razavi-Ghods}, N. and {Ringuette}, J. and 
	{Robnett}, J. and {Rosie}, K. and {Sell}, R. and {Smith}, C. and 
	{Syce}, A. and {Tegmark}, M. and {Thyagarajan}, N. and {Williams}, P.~K.~G. and 
	{Zheng}, H.},
    title = "{Hydrogen Epoch of Reionization Array (HERA)}",
  journal = {\pasp},
archivePrefix = "arXiv",
   eprint = {1606.07473},
 primaryClass = "astro-ph.IM",
     year = 2017,
    month = apr,
   volume = 129,
   number = 4,
    pages = {045001},
      doi = {10.1088/1538-3873/129/974/045001},
   adsurl = {http://adsabs.harvard.edu/abs/2017PASP..129d5001D}
}

@ARTICLE{ali08,
   author = {{Ali}, S.~S. and {Bharadwaj}, S. and {Chengalur}, J.~N.},
    title = "{Foregrounds for redshifted 21-cm studies of reionization: Giant Meter Wave Radio Telescope 153-MHz observations}",
  journal = {\mnras},
archivePrefix = "arXiv",
   eprint = {0801.2424},
     year = 2008,
    month = apr,
   volume = 385,
    pages = {2166-2174},
    adsurl = {http://adsabs.harvard.edu/abs/2008MNRAS.385.2166A}
}

@ARTICLE{barry19,
       author = {{Barry}, N. and {Wilensky}, M. and {Trott}, C.~M. and {Pindor}, B. and
         {Beardsley}, A.~P. and {Hazelton}, B.~J. and {Sullivan}, I.~S. and
         {Morales}, M.~F. and {Pober}, J.~C. and {Line}, J. and {Greig}, B. and
         {Byrne}, R. and {Lanman}, A. and {Li}, W. and {Jordan}, C.~H. and
         {Joseph}, R.~C. and {McKinley}, B. and {Rahimi}, M. and {Yoshiura}, S. and
         {Bowman}, J.~D. and {Gaensler}, B.~M. and {Hewitt}, J.~N. and
         {Jacobs}, D.~C. and {Mitchell}, D.~A. and {Udaya Shankar}, N. and
         {Sethi}, S.~K. and {Subrahmanyan}, R. and {Tingay}, S.~J. and
         {Webster}, R.~L. and {Wyithe}, J.~S.~B.},
        title = "{Improving the Epoch of Reionization Power Spectrum Results from Murchison Widefield Array Season 1 Observations}",
      journal = {\apj},
         year = 2019,
        month = oct,
       volume = {884},
       number = {1},
          eid = {1},
        pages = {1},
          doi = {10.3847/1538-4357/ab40a8},
archivePrefix = {arXiv},
       eprint = {1909.00561},
 primaryClass = {astro-ph.IM},
       adsurl = {https://ui.adsabs.harvard.edu/abs/2019ApJ...884....1B}
}

@article{li19,
  title={First Season MWA Phase II Epoch of Reionization Power Spectrum Results at Redshift 7},
  author={Li, W and Pober, JC and Barry, N and Hazelton, BJ and Morales, MF and Trott, CM and Lanman, A and Wilensky, M and Sullivan, I and Beardsley, AP and others},
  journal={\apj},
  volume={887},
  number={2},
  pages={141},
  year={2019},
  publisher={IOP Publishing}
}

@ARTICLE{Mertens2020,
       author = {{Mertens}, F.~G. and {Mevius}, M. and {Koopmans}, L.~V.~E. and
         {Offringa}, A.~R. and {Mellema}, G. and {Zaroubi}, S. and
         {Brentjens}, M.~A. and {Gan}, H. and {Gehlot}, B.~K. and {Pand
        ey}, V.~N. and {Sardarabadi}, A.~M. and {Vedantham}, H.~K. and
         {Yatawatta}, S. and {Asad}, K.~M.~B. and {Ciardi}, B. and
         {Chapman}, E. and {Gazagnes}, S. and {Ghara}, R. and {Ghosh}, A. and
         {Giri}, S.~K. and {Iliev}, I.~T. and {Jeli{\'c}}, V. and
         {Kooistra}, R. and {Mondal}, R. and {Schaye}, J. and {Silva}, M.~B.},
        title = "{Improved upper limits on the 21 cm signal power spectrum of neutral hydrogen at z {\ensuremath{\approx}} 9.1 from LOFAR}",
      journal = {\mnras},
         year = 2020,
        month = apr,
       volume = {493},
       number = {2},
        pages = {1662-1685},
          doi = {10.1093/mnras/staa327},
archivePrefix = {arXiv},
       eprint = {2002.07196},
 primaryClass = {astro-ph.CO},
       adsurl = {https://ui.adsabs.harvard.edu/abs/2020MNRAS.493.1662M}
}

@article{trott20,
    author = {Trott, Cathryn M and Jordan, C H and Midgley, S and Barry, N and Greig, B and Pindor, B and Cook, J H and Sleap, G and Tingay, S J and Ung, D and Hancock, P and Williams, A and Bowman, J and Byrne, R and Chokshi, A and Hazelton, B J and Hasegawa, K and Jacobs, D and Joseph, R C and Li, W and Line, J L B and Lynch, C and McKinley, B and Mitchell, D A and Morales, M F and Ouchi, M and Pober, J C and Rahimi, M and Takahashi, K and Wayth, R B and Webster, R L and Wilensky, M and Wyithe, J S B and Yoshiura, S and Zhang, Z and Zheng, Q},
    title = "{Deep multiredshift limits on Epoch of Reionization 21 cm power spectra from four seasons of Murchison Widefield Array observations}",
    journal = {\mnras},
    volume = {493},
    number = {4},
    pages = {4711-4727},
    year = {2020},
    month = {02},
    issn = {0035-8711},
    doi = {10.1093/mnras/staa414},
    url = {https://doi.org/10.1093/mnras/staa414},
    eprint = {https://academic.oup.com/mnras/article-pdf/493/4/4711/32927265/staa414.pdf},
}

@ARTICLE{koopmans15,
   author = {{Koopmans}, L. and {Pritchard}, J. and {Mellema}, G. and {Aguirre}, J. and 
	{Ahn}, K. and {Barkana}, R. and {van Bemmel}, I. and {Bernardi}, G. and 
	{Bonaldi}, A. and {Briggs}, F. and {de Bruyn}, A.~G. and {Chang}, T.~C. and 
	{Chapman}, E. and {Chen}, X. and {Ciardi}, B. and {Dayal}, P. and 
	{Ferrara}, A. and {Fialkov}, A. and {Fiore}, F. and {Ichiki}, K. and 
	{Illiev}, I.~T. and {Inoue}, S. and {Jelic}, V. and {Jones}, M. and 
	{Lazio}, J. and {Maio}, U. and {Majumdar}, S. and {Mack}, K.~J. and 
	{Mesinger}, A. and {Morales}, M.~F. and {Parsons}, A. and {Pen}, U.~L. and 
	{Santos}, M. and {Schneider}, R. and {Semelin}, B. and {de Souza}, R.~S. and 
	{Subrahmanyan}, R. and {Takeuchi}, T. and {Vedantham}, H. and 
	{Wagg}, J. and {Webster}, R. and {Wyithe}, S. and {Datta}, K.~K. and 
	{Trott}, C.},
    title = "{The Cosmic Dawn and Epoch of Reionisation with SKA}",
  journal = {Advancing Astrophysics with the Square Kilometre Array (AASKA14)},
archivePrefix = "arXiv",
   eprint = {1505.07568},
     year = 2015,
      eid = {1},
    pages = {1},
   adsurl = {http://adsabs.harvard.edu/abs/2015aska.confE...1K}
}

@ARTICLE{HERA_limit22,
       author = {{Abdurashidova}, Zara and {Aguirre}, James E. and {Alexander}, Paul and {Ali}, Zaki S. and {Balfour}, Yanga and {Beardsley}, Adam P. and {Bernardi}, Gianni and {Billings}, Tashalee S. and {Bowman}, Judd D. and {Bradley}, Richard F. and {Bull}, Philip and {Burba}, Jacob and {Carey}, Steve and {Carilli}, Chris L. and {Cheng}, Carina and {DeBoer}, David R. and {Dexter}, Matt and {de Lera Acedo}, Eloy and {Dibblee-Barkman}, Taylor and {Dillon}, Joshua S. and {Ely}, John and {Ewall-Wice}, Aaron and {Fagnoni}, Nicolas and {Fritz}, Randall and {Furlanetto}, Steven R. and {Gale-Sides}, Kingsley and {Glendenning}, Brian and {Gorthi}, Deepthi and {Greig}, Bradley and {Grobbelaar}, Jasper and {Halday}, Ziyaad and {Hazelton}, Bryna J. and {Hewitt}, Jacqueline N. and {Hickish}, Jack and {Jacobs}, Daniel C. and {Julius}, Austin and {Kern}, Nicholas S. and {Kerrigan}, Joshua and {Kittiwisit}, Piyanat and {Kohn}, Saul A. and {Kolopanis}, Matthew and {Lanman}, Adam and {La Plante}, Paul and {Lekalake}, Telalo and {Lewis}, David and {Liu}, Adrian and {MacMahon}, David and {Malan}, Lourence and {Malgas}, Cresshim and {Maree}, Matthys and {Martinot}, Zachary E. and {Matsetela}, Eunice and {Mesinger}, Andrei and {Molewa}, Mathakane and {Morales}, Miguel F. and {Mosiane}, Tshegofalang and {Murray}, Steven G. and {Neben}, Abraham R. and {Nikolic}, Bojan and {Nunhokee}, Chuneeta D. and {Parsons}, Aaron R. and {Patra}, Nipanjana and {Pascua}, Robert and {Pieterse}, Samantha and {Pober}, Jonathan C. and {Razavi-Ghods}, Nima and {Ringuette}, Jon and {Robnett}, James and {Rosie}, Kathryn and {Sims}, Peter and {Singh}, Saurabh and {Smith}, Craig and {Syce}, Angelo and {Thyagarajan}, Nithyanandan and {Williams}, Peter K.~G. and {Zheng}, Haoxuan and {HERA Collaboration}},
        title = "{First Results from HERA Phase I: Upper Limits on the Epoch of Reionization 21 cm Power Spectrum}",
      journal = {\apj},
         year = 2022,
        month = feb,
       volume = {925},
       number = {2},
          eid = {221},
        pages = {221},
          doi = {10.3847/1538-4357/ac1c78},
archivePrefix = {arXiv},
       eprint = {2108.02263},
 primaryClass = {astro-ph.CO},
       adsurl = {https://ui.adsabs.harvard.edu/abs/2022ApJ...925..221A}
}

@ARTICLE{Mondal2020,
       author = {{Mondal}, R. and {Fialkov}, A. and {Fling}, C. and {Iliev}, I.~T. and {Barkana}, R. and {Ciardi}, B. and {Mellema}, G. and {Zaroubi}, S. and {Koopmans}, L.~V.~E. and {Mertens}, F.~G. and {Gehlot}, B.~K. and {Ghara}, R. and {Ghosh}, A. and {Giri}, S.~K. and {Offringa}, A. and {Pandey}, V.~N.},
        title = "{Tight constraints on the excess radio background at z = 9.1 from LOFAR}",
      journal = {\mnras},
         year = 2020,
        month = nov,
       volume = {498},
       number = {3},
        pages = {4178-4191},
          doi = {10.1093/mnras/staa2422},
archivePrefix = {arXiv},
       eprint = {2004.00678},
 primaryClass = {astro-ph.CO},
       adsurl = {https://ui.adsabs.harvard.edu/abs/2020MNRAS.498.4178M}
}

@ARTICLE{Abdurashidova2022,
       author = {{Abdurashidova}, Zara and {Aguirre}, James E. and {Alexander}, Paul and {Ali}, Zaki S. and {Balfour}, Yanga and {Barkana}, Rennan and {Beardsley}, Adam P. and {Bernardi}, Gianni and {Billings}, Tashalee S. and {Bowman}, Judd D. and {Bradley}, Richard F. and {Bull}, Philip and {Burba}, Jacob and {Carey}, Steve and {Carilli}, Chris L. and {Cheng}, Carina and {DeBoer}, David R. and {Dexter}, Matt and {de Lera Acedo}, Eloy and {Dillon}, Joshua S. and {Ely}, John and {Ewall-Wice}, Aaron and {Fagnoni}, Nicolas and {Fialkov}, Anastasia and {Fritz}, Randall and {Furlanetto}, Steven R. and {Gale-Sides}, Kingsley and {Glendenning}, Brian and {Gorthi}, Deepthi and {Greig}, Bradley and {Grobbelaar}, Jasper and {Halday}, Ziyaad and {Hazelton}, Bryna J. and {Heimersheim}, Stefan and {Hewitt}, Jacqueline N. and {Hickish}, Jack and {Jacobs}, Daniel C. and {Julius}, Austin and {Kern}, Nicholas S. and {Kerrigan}, Joshua and {Kittiwisit}, Piyanat and {Kohn}, Saul A. and {Kolopanis}, Matthew and {Lanman}, Adam and {La Plante}, Paul and {Lekalake}, Telalo and {Lewis}, David and {Liu}, Adrian and {Ma}, Yin-Zhe and {MacMahon}, David and {Malan}, Lourence and {Malgas}, Cresshim and {Maree}, Matthys and {Martinot}, Zachary E. and {Matsetela}, Eunice and {Mesinger}, Andrei and {Mirocha}, Jordan and {Molewa}, Mathakane and {Morales}, Miguel F. and {Mosiane}, Tshegofalang and {Mu{\~n}oz}, Julian B. and {Murray}, Steven G. and {Neben}, Abraham R. and {Nikolic}, Bojan and {Nunhokee}, Chuneeta D. and {Parsons}, Aaron R. and {Patra}, Nipanjana and {Pieterse}, Samantha and {Pober}, Jonathan C. and {Qin}, Yuxiang and {Razavi-Ghods}, Nima and {Reis}, Itamar and {Ringuette}, Jon and {Robnett}, James and {Rosie}, Kathryn and {Santos}, Mario G. and {Sikder}, Sudipta and {Sims}, Peter and {Smith}, Craig and {Syce}, Angelo and {Thyagarajan}, Nithyanandan and {Williams}, Peter K.~G. and {Zheng}, Haoxuan},
        title = "{HERA Phase I Limits on the Cosmic 21 cm Signal: Constraints on Astrophysics and Cosmology during the Epoch of Reionization}",
      journal = {\apj},
         year = 2022,
        month = jan,
       volume = {924},
       number = {2},
          eid = {51},
        pages = {51},
          doi = {10.3847/1538-4357/ac2ffc},
archivePrefix = {arXiv},
       eprint = {2108.07282},
 primaryClass = {astro-ph.CO},
       adsurl = {https://ui.adsabs.harvard.edu/abs/2022ApJ...924...51A}
}

@ARTICLE{SARAS3,
       author = {{Singh}, Saurabh and {Jishnu}, Nambissan T. and {Subrahmanyan}, Ravi and {Udaya Shankar}, N. and {Girish}, B.~S. and {Raghunathan}, A. and {Somashekar}, R. and {Srivani}, K.~S. and {Sathyanarayana Rao}, Mayuri},
        title = "{On the detection of a cosmic dawn signal in the radio background}",
      journal = {Nature Astronomy},
         year = 2022,
        month = feb,
       volume = {6},
        pages = {607-617},
          doi = {10.1038/s41550-022-01610-5},
archivePrefix = {arXiv},
       eprint = {2112.06778},
 primaryClass = {astro-ph.CO},
       adsurl = {https://ui.adsabs.harvard.edu/abs/2022NatAs...6..607S}
}

@ARTICLE{Bharadwaj2020,
       author = {{Bharadwaj}, Somnath and {Mazumdar}, Arindam and {Sarkar}, Debanjan},
        title = "{Quantifying the redshift space distortion of the bispectrum I: primordial non-Gaussianity}",
      journal = {\mnras},
         year = 2020,
        month = mar,
       volume = {493},
       number = {1},
        pages = {594-602},
          doi = {10.1093/mnras/staa279},
archivePrefix = {arXiv},
       eprint = {2001.10243},
 primaryClass = {astro-ph.CO},
       adsurl = {https://ui.adsabs.harvard.edu/abs/2020MNRAS.493..594B}
}

@article{visbal12,
	Adsurl = {https://ui.adsabs.harvard.edu/abs/2012Natur.487...70V},
	Archiveprefix = {arXiv},
	Author = {{Visbal}, E. and {Barkana}, R. and {Fialkov}, A. and {Tseliakhovich}, D. and {Hirata}, C.~M.},
	Doi = {10.1038/nature11177},
	Eprint = {1201.1005},
	Journal = {\nat},
	Month = jul,
	Pages = {70-73},
	Title = {{The signature of the first stars in atomic hydrogen at redshift 20}},
	Volume = 487,
	Year = 2012}

@article{fialkov14,
	Adsurl = {https://ui.adsabs.harvard.edu/abs/2014MNRAS.445..213F},
	Archiveprefix = {arXiv},
	Author = {{Fialkov}, Anastasia and {Barkana}, Rennan},
	Doi = {10.1093/mnras/stu1744},
	Eprint = {1409.3992},
	Journal = {\mnras},
	Month = {Nov},
	Number = {1},
	Pages = {213-224},
	Primaryclass = {astro-ph.CO},
	Title = {{The rich complexity of 21-cm fluctuations produced by the first stars}},
	Volume = {445},
	Year = {2014}}

@article{cohen17,
	Adsurl = {https://ui.adsabs.harvard.edu/abs/2017MNRAS.472.1915C},
	Archiveprefix = {arXiv},
	Author = {{Cohen}, Aviad and {Fialkov}, Anastasia and {Barkana}, Rennan and {Lotem}, Matan},
	Doi = {10.1093/mnras/stx2065},
	Eprint = {1609.02312},
	Journal = {\mnras},
	Month = {Dec},
	Number = {2},
	Pages = {1915-1931},
	Primaryclass = {astro-ph.CO},
	Title = {{Charting the parameter space of the global 21-cm signal}},
	Volume = {472},
	Year = {2017}}

@ARTICLE{fialkov19,
       author = {{Fialkov}, Anastasia and {Barkana}, Rennan},
        title = "{Signature of excess radio background in the 21-cm global signal and power spectrum}",
      journal = {\mnras},
         year = 2019,
        month = jun,
       volume = {486},
       number = {2},
        pages = {1763-1773},
          doi = {10.1093/mnras/stz873},
archivePrefix = {arXiv},
       eprint = {1902.02438},
 primaryClass = {astro-ph.CO},
       adsurl = {https://ui.adsabs.harvard.edu/abs/2019MNRAS.486.1763F}
}

@ARTICLE{Reis2020,
       author = {{Reis}, Itamar and {Fialkov}, Anastasia and {Barkana}, Rennan},
        title = "{High-redshift radio galaxies: a potential new source of 21-cm fluctuations}",
      journal = {\mnras},
         year = 2020,
        month = dec,
       volume = {499},
       number = {4},
        pages = {5993-6008},
          doi = {10.1093/mnras/staa3091},
archivePrefix = {arXiv},
       eprint = {2008.04315},
 primaryClass = {astro-ph.CO},
       adsurl = {https://ui.adsabs.harvard.edu/abs/2020MNRAS.499.5993R}
}

@article{Sikder2023,
    author = {Sikder, Sudipta and Barkana, Rennan and Fialkov, Anastasia and Reis, Itamar},
    title = {Strong 21-cm fluctuations and anisotropy due to the line-of-sight effect of radio galaxies at cosmic dawn},
    journal = {Monthly Notices of the Royal Astronomical Society},
    volume = {527},
    number = {4},
    pages = {10975-10985},
    year = {2023},
    month = {12},
    issn = {0035-8711},
    doi = {10.1093/mnras/stad3847},
    url = {https://doi.org/10.1093/mnras/stad3847},
    eprint = {https://academic.oup.com/mnras/article-pdf/527/4/10975/55024639/stad3847.pdf},
}

@ARTICLE{camb,
       author = {{Lewis}, Antony and {Challinor}, Anthony and {Lasenby}, Anthony},
        title = "{Efficient Computation of Cosmic Microwave Background Anisotropies in Closed Friedmann-Robertson-Walker Models}",
      journal = {\apj},
         year = "2000",
        month = "Aug",
       volume = {538},
       number = {2},
        pages = {473-476},
          doi = {10.1086/309179},
archivePrefix = {arXiv},
       eprint = {astro-ph/9911177},
 primaryClass = {astro-ph},
       adsurl = {https://ui.adsabs.harvard.edu/abs/2000ApJ...538..473L}
}

@article{sheth99,
	Adsurl = {https://ui.adsabs.harvard.edu/abs/1999MNRAS.308..119S},
	Author = {{Sheth}, R.~K. and {Tormen}, G.},
	Doi = {10.1046/j.1365-8711.1999.02692.x},
	Eprint = {astro-ph/9901122},
	Journal = {\mnras},
	Month = sep,
	Pages = {119-126},
	Title = {{Large-scale bias and the peak background split}},
	Volume = 308,
	Year = 1999}

@article{barkana04,
	Adsurl = {https://ui.adsabs.harvard.edu/abs/2004ApJ...609..474B},
	Archiveprefix = {arXiv},
	Author = {{Barkana}, Rennan and {Loeb}, Abraham},
	Doi = {10.1086/421079},
	Eprint = {astro-ph/0310338},
	Journal = {\apj},
	Month = {Jul},
	Number = {2},
	Pages = {474-481},
	Primaryclass = {astro-ph},
	Title = {{Unusually Large Fluctuations in the Statistics of Galaxy Formation at High Redshift}},
	Volume = {609},
	Year = {2004}}

@article{press74,
	Adsurl = {https://ui.adsabs.harvard.edu/abs/1974ApJ...187..425P},
	Author = {{Press}, W.~H. and {Schechter}, P.},
	Doi = {10.1086/152650},
	Journal = {\apj},
	Month = feb,
	Pages = {425-438},
	Title = {{Formation of Galaxies and Clusters of Galaxies by Self-Similar Gravitational Condensation}},
	Volume = 187,
	Year = 1974}

@article{tseliakhovich10,
	Adsurl = {https://ui.adsabs.harvard.edu/abs/2010PhRvD..82h3520T},
	Archiveprefix = {arXiv},
	Author = {{Tseliakhovich}, Dmitriy and {Hirata}, Christopher},
	Doi = {10.1103/PhysRevD.82.083520},
	Eid = {083520},
	Eprint = {1005.2416},
	Journal = {\prd},
	Month = {Oct},
	Number = {8},
	Pages = {083520},
	Primaryclass = {astro-ph.CO},
	Title = {{Relative velocity of dark matter and baryonic fluids and the formation of the first structures}},
	Volume = {82},
	Year = {2010}}

@ARTICLE{Pospelov:2018,
       author = {{Pospelov}, Maxim and {Pradler}, Josef and {Ruderman}, Joshua T. and
         {Urbano}, Alfredo},
        title = "{Room for New Physics in the Rayleigh-Jeans Tail of the Cosmic Microwave Background}",
      journal = {\prl},
         year = 2018,
        month = jul,
       volume = {121},
       number = {3},
          eid = {031103},
        pages = {031103},
          doi = {10.1103/PhysRevLett.121.031103},
archivePrefix = {arXiv},
       eprint = {1803.07048},
 primaryClass = {hep-ph},
       adsurl = {https://ui.adsabs.harvard.edu/abs/2018PhRvL.121c1103P}
}

@ARTICLE{Fraser:2018,
       author = {{Fraser}, Sean and {Hektor}, Andi and {H{\"u}tsi}, Gert and
         {Kannike}, Kristjan and {Marzo}, Carlo and {Marzola}, Luca and
         {Racioppi}, Antonio and {Raidal}, Martti and {Spethmann}, Christian and
         {Vaskonen}, Ville and {Veerm{\"a}e}, Hardi},
        title = "{The EDGES 21 cm anomaly and properties of dark matter}",
      journal = {Physics Letters B},
         year = 2018,
        month = oct,
       volume = {785},
        pages = {159-164},
          doi = {10.1016/j.physletb.2018.08.035},
archivePrefix = {arXiv},
       eprint = {1803.03245},
 primaryClass = {hep-ph},
       adsurl = {https://ui.adsabs.harvard.edu/abs/2018PhLB..785..159F}
}

@ARTICLE{Brandenberger:2019,
       author = {{Brandenberger}, Robert and {Cyr}, Bryce and {Shi}, Rui},
        title = "{Constraints on superconducting cosmic strings from the global 21-cm signal before reionization}",
      journal = {\jcap},
         year = 2019,
        month = sep,
       volume = {2019},
       number = {9},
          eid = {009},
        pages = {009},
          doi = {10.1088/1475-7516/2019/09/009},
archivePrefix = {arXiv},
       eprint = {1902.08282},
 primaryClass = {astro-ph.CO},
       adsurl = {https://ui.adsabs.harvard.edu/abs/2019JCAP...09..009B}
}

@ARTICLE{gurkan18,
       author = {{G{\"u}rkan}, G. and {Hardcastle}, M.~J. and {Smith}, D.~J.~B. and
         {Best}, P.~N. and {Bourne}, N. and {Calistro-Rivera}, G. and
         {Heald}, G. and {Jarvis}, M.~J. and {Prandoni}, I. and
         {R{\"o}ttgering}, H.~J.~A. and {Sabater}, J. and {Shimwell}, T. and
         {Tasse}, C. and {Williams}, W.~L.},
        title = "{LOFAR/H-ATLAS: the low-frequency radio luminosity-star formation rate relation}",
      journal = {\mnras},
         year = 2018,
        month = apr,
       volume = {475},
       number = {3},
        pages = {3010-3028},
          doi = {10.1093/mnras/sty016},
archivePrefix = {arXiv},
       eprint = {1801.02629},
 primaryClass = {astro-ph.GA},
       adsurl = {https://ui.adsabs.harvard.edu/abs/2018MNRAS.475.3010G}
}

@ARTICLE{mirocha19,
       author = {{Mirocha}, Jordan and {Furlanetto}, Steven R.},
        title = "{What does the first highly redshifted 21-cm detection tell us about early galaxies?}",
      journal = {\mnras},
         year = 2019,
        month = feb,
       volume = {483},
       number = {2},
        pages = {1980-1992},
          doi = {10.1093/mnras/sty3260},
archivePrefix = {arXiv},
       eprint = {1803.03272},
 primaryClass = {astro-ph.GA},
       adsurl = {https://ui.adsabs.harvard.edu/abs/2019MNRAS.483.1980M}
}

@Article{         harris2020array,
 title         = {Array programming with {NumPy}},
 author        = {Charles R. Harris and K. Jarrod Millman and St{\'{e}}fan J.
                 van der Walt and Ralf Gommers and Pauli Virtanen and David
                 Cournapeau and Eric Wieser and Julian Taylor and Sebastian
                 Berg and Nathaniel J. Smith and Robert Kern and Matti Picus
                 and Stephan Hoyer and Marten H. van Kerkwijk and Matthew
                 Brett and Allan Haldane and Jaime Fern{\'{a}}ndez del
                 R{\'{i}}o and Mark Wiebe and Pearu Peterson and Pierre
                 G{\'{e}}rard-Marchant and Kevin Sheppard and Tyler Reddy and
                 Warren Weckesser and Hameer Abbasi and Christoph Gohlke and
                 Travis E. Oliphant},
 year          = {2020},
 month         = sep,
 journal       = {Nature},
 volume        = {585},
 number        = {7825},
 pages         = {357--362},
 doi           = {10.1038/s41586-020-2649-2},
 publisher     = {Springer Science and Business Media {LLC}},
 url           = {https://doi.org/10.1038/s41586-020-2649-2}
}

@ARTICLE{2020SciPy-NMeth,
  author  = {Virtanen, Pauli and Gommers, Ralf and Oliphant, Travis E. and
            Haberland, Matt and Reddy, Tyler and Cournapeau, David and
            Burovski, Evgeni and Peterson, Pearu and Weckesser, Warren and
            Bright, Jonathan and {van der Walt}, St{\'e}fan J. and
            Brett, Matthew and Wilson, Joshua and Millman, K. Jarrod and
            Mayorov, Nikolay and Nelson, Andrew R. J. and Jones, Eric and
            Kern, Robert and Larson, Eric and Carey, C J and
            Polat, {\.I}lhan and Feng, Yu and Moore, Eric W. and
            {VanderPlas}, Jake and Laxalde, Denis and Perktold, Josef and
            Cimrman, Robert and Henriksen, Ian and Quintero, E. A. and
            Harris, Charles R. and Archibald, Anne M. and
            Ribeiro, Ant{\^o}nio H. and Pedregosa, Fabian and
            {van Mulbregt}, Paul and {SciPy 1.0 Contributors}},
  title   = {{{SciPy} 1.0: Fundamental Algorithms for Scientific
            Computing in Python}},
  journal = {Nature Methods},
  year    = {2020},
  volume  = {17},
  pages   = {261--272},
  adsurl  = {https://rdcu.be/b08Wh},
  doi     = {10.1038/s41592-019-0686-2},
}

@Article{Hunter:2007,
  Author    = {Hunter, J. D.},
  Title     = {Matplotlib: A 2D graphics environment},
  Journal   = {Computing in Science \& Engineering},
  Volume    = {9},
  Number    = {3},
  Pages     = {90--95},
  publisher = {IEEE COMPUTER SOC},
  doi       = {10.1109/MCSE.2007.55},
  year      = 2007
}

@article{haiman97,
	Adsurl = {https://ui.adsabs.harvard.edu/abs/1997ApJ...476..458H},
	Author = {{Haiman}, Z. and {Rees}, M.~J. and {Loeb}, A.},
	Doi = {10.1086/303647},
	Eprint = {astro-ph/9608130},
	Journal = {\apj},
	Month = feb,
	Pages = {458-463},
	Title = {{Destruction of Molecular Hydrogen during Cosmological Reionization}},
	Volume = 476,
	Year = 1997}

@ARTICLE{fialkov2013,
       author = {{Fialkov}, Anastasia and {Barkana}, Rennan and {Visbal}, Eli and
         {Tseliakhovich}, Dmitriy and {Hirata}, Christopher M.},
        title = "{The 21-cm signature of the first stars during the Lyman-Werner feedback era}",
      journal = {\mnras},
         year = 2013,
        month = jul,
       volume = {432},
       number = {4},
        pages = {2909-2916},
          doi = {10.1093/mnras/stt650},
archivePrefix = {arXiv},
       eprint = {1212.0513},
 primaryClass = {astro-ph.CO},
       adsurl = {https://ui.adsabs.harvard.edu/abs/2013MNRAS.432.2909F}
}

@ARTICLE{rees86,
       author = {{Rees}, Martin J.},
        title = "{Baryon concentration in string wakes at Z greater than about 200 - Implications for galaxy formation and large-scale structure}",
      journal = {\mnras},
         year = 1986,
        month = oct,
       volume = {222},
        pages = {27P-32P},
          doi = {10.1093/mnras/222.1.27P},
       adsurl = {https://ui.adsabs.harvard.edu/abs/1986MNRAS.222P..27R}
}

@ARTICLE{sobacchi13,
       author = {{Sobacchi}, Emanuele and {Mesinger}, Andrei},
        title = "{How does radiative feedback from an ultraviolet background impact reionization?}",
      journal = {\mnras},
         year = 2013,
        month = jul,
       volume = {432},
       number = {4},
        pages = {3340-3348},
          doi = {10.1093/mnras/stt693},
archivePrefix = {arXiv},
       eprint = {1301.6781},
 primaryClass = {astro-ph.CO},
       adsurl = {https://ui.adsabs.harvard.edu/abs/2013MNRAS.432.3340S}
}

@article{cohen16,
	Adsurl = {https://ui.adsabs.harvard.edu/abs/2016MNRAS.459L..90C},
	Archiveprefix = {arXiv},
	Author = {{Cohen}, Aviad and {Fialkov}, Anastasia and {Barkana}, Rennan},
	Doi = {10.1093/mnrasl/slw047},
	Eprint = {1508.04138},
	Journal = {\mnras},
	Month = {Jun},
	Number = {1},
	Pages = {L90-L94},
	Primaryclass = {astro-ph.CO},
	Title = {{The 21-cm BAO signature of enriched low-mass galaxies during cosmic reionization}},
	Volume = {459},
	Year = {2016}}

@ARTICLE{Gilfanov,
       author = {{Gilfanov}, M. and {Grimm}, H. -J. and {Sunyaev}, R.},
        title = "{L$_{X}$-SFR relation in star-forming galaxies}",
      journal = {\mnras},
         year = 2004,
        month = jan,
       volume = {347},
       number = {3},
        pages = {L57-L60},
          doi = {10.1111/j.1365-2966.2004.07450.x},
archivePrefix = {arXiv},
       eprint = {astro-ph/0301331},
 primaryClass = {astro-ph},
       adsurl = {https://ui.adsabs.harvard.edu/abs/2004MNRAS.347L..57G}
}

@ARTICLE{Grimm,
       author = {{Grimm}, H. -J. and {Gilfanov}, M. and {Sunyaev}, R.},
        title = "{High-mass X-ray binaries as a star formation rate indicator in distant galaxies}",
      journal = {\mnras},
         year = 2003,
        month = mar,
       volume = {339},
       number = {3},
        pages = {793-809},
          doi = {10.1046/j.1365-8711.2003.06224.x},
archivePrefix = {arXiv},
       eprint = {astro-ph/0205371},
 primaryClass = {astro-ph},
       adsurl = {https://ui.adsabs.harvard.edu/abs/2003MNRAS.339..793G}
}

@ARTICLE{Mineo:2012,
       author = {{Mineo}, S. and {Gilfanov}, M. and {Sunyaev}, R.},
        title = "{X-ray emission from star-forming galaxies - I. High-mass X-ray binaries}",
      journal = {\mnras},
         year = 2012,
        month = jan,
       volume = {419},
       number = {3},
        pages = {2095-2115},
          doi = {10.1111/j.1365-2966.2011.19862.x},
archivePrefix = {arXiv},
       eprint = {1105.4610},
 primaryClass = {astro-ph.HE},
       adsurl = {https://ui.adsabs.harvard.edu/abs/2012MNRAS.419.2095M}
}

@article{fragos13,
	Adsurl = {https://ui.adsabs.harvard.edu/abs/2013ApJ...764...41F},
	Archiveprefix = {arXiv},
	Author = {{Fragos}, T. and {Lehmer}, B. and {Tremmel}, M. and {Tzanavaris}, P. and {Basu-Zych}, A. and {Belczynski}, K. and {Hornschemeier}, A. and {Jenkins}, L. and {Kalogera}, V. and {Ptak}, A. and {Zezas}, A.},
	Doi = {10.1088/0004-637X/764/1/41},
	Eid = {41},
	Eprint = {1206.2395},
	Journal = {\apj},
	Month = {Feb},
	Number = {1},
	Pages = {41},
	Primaryclass = {astro-ph.HE},
	Title = {{X-Ray Binary Evolution Across Cosmic Time}},
	Volume = {764},
	Year = {2013}}

@ARTICLE{Pacucci:2014,
       author = {{Pacucci}, Fabio and {Mesinger}, Andrei and {Mineo}, Stefano and
         {Ferrara}, Andrea},
        title = "{The X-ray spectra of the first galaxies: 21 cm signatures}",
      journal = {\mnras},
         year = 2014,
        month = sep,
       volume = {443},
       number = {1},
        pages = {678-686},
          doi = {10.1093/mnras/stu1240},
archivePrefix = {arXiv},
       eprint = {1403.6125},
 primaryClass = {astro-ph.CO},
       adsurl = {https://ui.adsabs.harvard.edu/abs/2014MNRAS.443..678P}
}

@article{fialkov14a,
	Adsurl = {https://ui.adsabs.harvard.edu/abs/2014Natur.506..197F},
	Archiveprefix = {arXiv},
	Author = {{Fialkov}, Anastasia and {Barkana}, Rennan and {Visbal}, Eli},
	Doi = {10.1038/nature12999},
	Eprint = {1402.0940},
	Journal = {\nat},
	Month = {Feb},
	Number = {7487},
	Pages = {197-199},
	Primaryclass = {astro-ph.CO},
	Title = {{The observable signature of late heating of the Universe during cosmic reionization}},
	Volume = {506},
	Year = {2014}}

@ARTICLE{seiffert11,
       author = {{Seiffert}, M. and {Fixsen}, D.~J. and {Kogut}, A. and {Levin}, S.~M. and
         {Limon}, M. and {Lubin}, P.~M. and {Mirel}, P. and {Singal}, J. and
         {Villela}, T. and {Wollack}, E. and {Wuensche}, C.~A.},
        title = "{Interpretation of the ARCADE 2 Absolute Sky Brightness Measurement}",
      journal = {\apj},
         year = 2011,
        month = jun,
       volume = {734},
       number = {1},
          eid = {6},
        pages = {6},
          doi = {10.1088/0004-637X/734/1/6},
       adsurl = {https://ui.adsabs.harvard.edu/abs/2011ApJ...734....6S}
}

@ARTICLE{fixsen11,
       author = {{Fixsen}, D.~J. and {Kogut}, A. and {Levin}, S. and {Limon}, M. and
         {Lubin}, P. and {Mirel}, P. and {Seiffert}, M. and {Singal}, J. and
         {Wollack}, E. and {Villela}, T. and {Wuensche}, C.~A.},
        title = "{ARCADE 2 Measurement of the Absolute Sky Brightness at 3-90 GHz}",
      journal = {\apj},
         year = 2011,
        month = jun,
       volume = {734},
       number = {1},
          eid = {5},
        pages = {5},
          doi = {10.1088/0004-637X/734/1/5},
archivePrefix = {arXiv},
       eprint = {0901.0555},
 primaryClass = {astro-ph.CO},
       adsurl = {https://ui.adsabs.harvard.edu/abs/2011ApJ...734....5F}
}

@ARTICLE{dowell18,
       author = {{Dowell}, Jayce and {Taylor}, Greg B.},
        title = "{The Radio Background below 100 MHz}",
      journal = {\apjl},
         year = 2018,
        month = may,
       volume = {858},
       number = {1},
          eid = {L9},
        pages = {L9},
          doi = {10.3847/2041-8213/aabf86},
archivePrefix = {arXiv},
       eprint = {1804.08581},
 primaryClass = {astro-ph.CO},
       adsurl = {https://ui.adsabs.harvard.edu/abs/2018ApJ...858L...9D}
}

@ARTICLE{Subrahmanyan:2013,
       author = {{Subrahmanyan}, Ravi and {Cowsik}, Ramanath},
        title = "{Is there an Unaccounted for Excess in the Extragalactic Cosmic Radio Background?}",
      journal = {The Astrophysical Journal},
         year = "2013",
        month = "Oct",
       volume = {776},
       number = {1},
          eid = {42},
        pages = {42},
          doi = {10.1088/0004-637X/776/1/42},
archivePrefix = {arXiv},
       eprint = {1305.7060},
 primaryClass = {astro-ph.CO},
       adsurl = {https://ui.adsabs.harvard.edu/abs/2013ApJ...776...42S}
}

@ARTICLE{furlanetto04,
       author = {{Furlanetto}, Steven R. and {Zaldarriaga}, Matias and {Hernquist}, Lars},
        title = "{Statistical Probes of Reionization with 21 Centimeter Tomography}",
      journal = {\apj},
         year = "2004",
        month = "Sep",
       volume = {613},
       number = {1},
        pages = {16-22},
          doi = {10.1086/423028},
archivePrefix = {arXiv},
       eprint = {astro-ph/0404112},
 primaryClass = {astro-ph},
       adsurl = {https://ui.adsabs.harvard.edu/abs/2004ApJ...613...16F}
}

@article{greig15,
    author = {Greig, Bradley and Mesinger, Andrei},
    title = "{21CMMC: an MCMC analysis tool enabling astrophysical parameter studies of the cosmic 21 cm signal}",
    journal = {Monthly Notices of the Royal Astronomical Society},
    volume = {449},
    number = {4},
    pages = {4246-4263},
    year = {2015},
    month = {04},
    issn = {0035-8711},
    doi = {10.1093/mnras/stv571},
    url = {https://doi.org/10.1093/mnras/stv571},
    eprint = {https://academic.oup.com/mnras/article-pdf/449/4/4246/18509895/stv571.pdf},
}

@ARTICLE{acharya23,
       author = {{Acharya}, Sandeep Kumar and {Cyr}, Bryce and {Chluba}, Jens},
        title = "{The role of soft photon injection and heating in 21 cm cosmology}",
      journal = {\mnras},
         year = 2023,
        month = aug,
       volume = {523},
       number = {2},
        pages = {1908-1918},
          doi = {10.1093/mnras/stad1540},
archivePrefix = {arXiv},
       eprint = {2303.17311},
 primaryClass = {astro-ph.CO},
       adsurl = {https://ui.adsabs.harvard.edu/abs/2023MNRAS.523.1908A}
}

@ARTICLE{cyr24,
       author = {{Cyr}, Bryce and {Acharya}, Sandeep Kumar and {Chluba}, Jens},
        title = "{Soft photon heating: a semi-analytic framework and applications to 21-cm cosmology}",
      journal = {\mnras},
         year = 2024,
        month = oct,
       volume = {534},
       number = {1},
        pages = {738-757},
          doi = {10.1093/mnras/stae2113},
archivePrefix = {arXiv},
       eprint = {2404.11743},
 primaryClass = {astro-ph.CO},
       adsurl = {https://ui.adsabs.harvard.edu/abs/2024MNRAS.534..738C}
}

@article{Ciardi_2003,
doi = {10.1086/377634},
url = {https://dx.doi.org/10.1086/377634},
year = {2003},
month = {oct},
publisher = {},
volume = {596},
number = {1},
pages = {1},
author = {Ciardi, Benedetta and Madau, Piero},
title = {Probing beyond the Epoch of Hydrogen Reionization with 21 Centimeter Radiation},
journal = {The Astrophysical Journal},
}

@article{Mellema2006,
    author = {Mellema, Garrelt and Iliev, Ilian T. and Pen, Ue-Li and Shapiro, Paul R.},
    title = {Simulating cosmic reionization at large scales – II. The 21-cm emission features and statistical signals},
    journal = {Monthly Notices of the Royal Astronomical Society},
    volume = {372},
    number = {2},
    pages = {679-692},
    year = {2006},
    month = {09},
    issn = {0035-8711},
    doi = {10.1111/j.1365-2966.2006.10919.x},
    url = {https://doi.org/10.1111/j.1365-2966.2006.10919.x},
    eprint = {https://academic.oup.com/mnras/article-pdf/372/2/679/2986158/mnras0372-0679.pdf},
}

@article{Ichikawa2010,
    author = {Ichikawa, Kazuhide and Barkana, Rennan and Iliev, Ilian T. and Mellema, Garrelt and Shapiro, Paul R.},
    title = {Measuring the history of cosmic reionization using the 21-cm probability distribution function from simulations},
    journal = {Monthly Notices of the Royal Astronomical Society},
    volume = {406},
    number = {4},
    pages = {2521-2532},
    year = {2010},
    month = {08},
    issn = {0035-8711},
    doi = {10.1111/j.1365-2966.2010.16828.x},
    url = {https://doi.org/10.1111/j.1365-2966.2010.16828.x},
    eprint = {https://academic.oup.com/mnras/article-pdf/406/4/2521/3346632/mnras0406-2521.pdf},
}

@article{Wyithe2007,
    author = {Wyithe, J. Stuart B. and Morales, Miguel F.},
    title = {Biased reionization and non-Gaussianity in redshifted 21-cm intensity maps of the reionization epoch},
    journal = {Monthly Notices of the Royal Astronomical Society},
    volume = {379},
    number = {4},
    pages = {1647-1657},
    year = {2007},
    month = {07},
    doi = {10.1111/j.1365-2966.2007.12048.x},
    url = {https://doi.org/10.1111/j.1365-2966.2007.12048.x},
    eprint = {https://academic.oup.com/mnras/article-pdf/379/4/1647/3201798/mnras0379-1647.pdf},
}

@article{Harker2009,
    author = {Harker, Geraint J. A. and Zaroubi, Saleem and Thomas, Rajat M. and Jelić, Vibor and Labropoulos, Panagiotis and Mellema, Garrelt and Iliev, Ilian T. and Bernardi, Gianni and Brentjens, Michiel A. and De Bruyn, A. G. and Ciardi, Benedetta and Koopmans, Leon V. E. and Pandey, V. N. and Pawlik, Andreas H. and Schaye, Joop and Yatawatta, Sarod},
    title = {Detection and extraction of signals from the epoch of reionization using higher-order one-point statistics},
    journal = {Monthly Notices of the Royal Astronomical Society},
    volume = {393},
    number = {4},
    pages = {1449-1458},
    year = {2009},
    month = {02},
    doi = {10.1111/j.1365-2966.2008.14209.x},
    url = {https://doi.org/10.1111/j.1365-2966.2008.14209.x},
    eprint = {https://academic.oup.com/mnras/article-pdf/393/4/1449/3256067/mnras0393-1449.pdf},
}

@article{Watkinson2014,
    author = {Watkinson, C. A. and Pritchard, J. R.},
    title = {Distinguishing models of reionization using future radio observations of 21-cm 1-point statistics},
    journal = {Monthly Notices of the Royal Astronomical Society},
    volume = {443},
    number = {4},
    pages = {3090-3106},
    year = {2014},
    month = {08},
    doi = {10.1093/mnras/stu1384},
    url = {https://doi.org/10.1093/mnras/stu1384},
    eprint = {https://academic.oup.com/mnras/article-pdf/443/4/3090/6289915/stu1384.pdf},
}

@article{Watkinson2015,
    author = {Watkinson, C. A. and Pritchard, J. R.},
    title = {The impact of spin-temperature fluctuations on the 21-cm moments},
    journal = {Monthly Notices of the Royal Astronomical Society},
    volume = {454},
    number = {2},
    pages = {1416-1431},
    year = {2015},
    month = {10},
    issn = {0035-8711},
    doi = {10.1093/mnras/stv2010},
    url = {https://doi.org/10.1093/mnras/stv2010},
    eprint = {https://academic.oup.com/mnras/article-pdf/454/2/1416/9379588/stv2010.pdf},
}

@article{Shimabukuro2015,
    author = {Shimabukuro, Hayato and Yoshiura, Shintaro and Takahashi, Keitaro and Yokoyama, Shuichiro and Ichiki, Kiyotomo},
    title = {Studying 21cm power spectrum with one-point statistics},
    journal = {Monthly Notices of the Royal Astronomical Society},
    volume = {451},
    number = {1},
    pages = {467-474},
    year = {2015},
    month = {05},
    issn = {0035-8711},
    doi = {10.1093/mnras/stv965},
    url = {https://doi.org/10.1093/mnras/stv965},
    eprint = {https://academic.oup.com/mnras/article-pdf/451/1/467/4179513/stv965.pdf},
}

@article{Kubota2016,
    author = {Kubota, Kenji and Yoshiura, Shintaro and Shimabukuro, Hayato and Takahashi, Keitaro},
    title = {Expected constraints on models of the epoch of reionization with the variance and skewness in redshifted 21 cm-line fluctuations},
    journal = {Publications of the Astronomical Society of Japan},
    volume = {68},
    number = {4},
    pages = {61},
    year = {2016},
    month = {06},
    issn = {0004-6264},
    doi = {10.1093/pasj/psw059},
    url = {https://doi.org/10.1093/pasj/psw059},
    eprint = {https://academic.oup.com/pasj/article-pdf/68/4/61/54680621/pasj\_68\_4\_61.pdf},
}

@article{Kittiwisit2017,
    author = {Kittiwisit, Piyanat and Bowman, Judd D and Jacobs, Daniel C and Beardsley, Adam P and Thyagarajan, Nithyanandan},
    title = {Sensitivity of the Hydrogen Epoch of Reionization Array and its build-out stages to one-point statistics from redshifted 21 cm observations},
    journal = {Monthly Notices of the Royal Astronomical Society},
    volume = {474},
    number = {4},
    pages = {4487-4499},
    year = {2017},
    month = {12},
    issn = {0035-8711},
    doi = {10.1093/mnras/stx3099},
    url = {https://doi.org/10.1093/mnras/stx3099},
    eprint = {https://academic.oup.com/mnras/article-pdf/474/4/4487/22978348/stx3099.pdf},
}

@ARTICLE{Planck2016,
       author = {{Planck Collaboration} and {Ade}, P.~A.~R. and {Aghanim}, N. and {Arnaud}, M. and {Ashdown}, M. and {Aumont}, J. and {Baccigalupi}, C. and {Banday}, A.~J. and {Barreiro}, R.~B. and {Bartlett}, J.~G. and {Bartolo}, N. and {Battaner}, E. and {Battye}, R. and {Benabed}, K. and {Beno{\^\i}t}, A. and {Benoit-L{\'e}vy}, A. and {Bernard}, J. -P. and {Bersanelli}, M. and {Bielewicz}, P. and {Bock}, J.~J. and {Bonaldi}, A. and {Bonavera}, L. and {Bond}, J.~R. and {Borrill}, J. and {Bouchet}, F.~R. and {Boulanger}, F. and {Bucher}, M. and {Burigana}, C. and {Butler}, R.~C. and {Calabrese}, E. and {Cardoso}, J. -F. and {Catalano}, A. and {Challinor}, A. and {Chamballu}, A. and {Chary}, R. -R. and {Chiang}, H.~C. and {Chluba}, J. and {Christensen}, P.~R. and {Church}, S. and {Clements}, D.~L. and {Colombi}, S. and {Colombo}, L.~P.~L. and {Combet}, C. and {Coulais}, A. and {Crill}, B.~P. and {Curto}, A. and {Cuttaia}, F. and {Danese}, L. and {Davies}, R.~D. and {Davis}, R.~J. and {de Bernardis}, P. and {de Rosa}, A. and {de Zotti}, G. and {Delabrouille}, J. and {D{\'e}sert}, F. -X. and {Di Valentino}, E. and {Dickinson}, C. and {Diego}, J.~M. and {Dolag}, K. and {Dole}, H. and {Donzelli}, S. and {Dor{\'e}}, O. and {Douspis}, M. and {Ducout}, A. and {Dunkley}, J. and {Dupac}, X. and {Efstathiou}, G. and {Elsner}, F. and {En{\ss}lin}, T.~A. and {Eriksen}, H.~K. and {Farhang}, M. and {Fergusson}, J. and {Finelli}, F. and {Forni}, O. and {Frailis}, M. and {Fraisse}, A.~A. and {Franceschi}, E. and {Frejsel}, A. and {Galeotta}, S. and {Galli}, S. and {Ganga}, K. and {Gauthier}, C. and {Gerbino}, M. and {Ghosh}, T. and {Giard}, M. and {Giraud-H{\'e}raud}, Y. and {Giusarma}, E. and {Gjerl{\o}w}, E. and {Gonz{\'a}lez-Nuevo}, J. and {G{\'o}rski}, K.~M. and {Gratton}, S. and {Gregorio}, A. and {Gruppuso}, A. and {Gudmundsson}, J.~E. and {Hamann}, J. and {Hansen}, F.~K. and {Hanson}, D. and {Harrison}, D.~L. and {Helou}, G. and {Henrot-Versill{\'e}}, S. and {Hern{\'a}ndez-Monteagudo}, C. and {Herranz}, D. and {Hildebrandt}, S.~R. and {Hivon}, E. and {Hobson}, M. and {Holmes}, W.~A. and {Hornstrup}, A. and {Hovest}, W. and {Huang}, Z. and {Huffenberger}, K.~M. and {Hurier}, G. and {Jaffe}, A.~H. and {Jaffe}, T.~R. and {Jones}, W.~C. and {Juvela}, M. and {Keih{\"a}nen}, E. and {Keskitalo}, R. and {Kisner}, T.~S. and {Kneissl}, R. and {Knoche}, J. and {Knox}, L. and {Kunz}, M. and {Kurki-Suonio}, H. and {Lagache}, G. and {L{\"a}hteenm{\"a}ki}, A. and {Lamarre}, J. -M. and {Lasenby}, A. and {Lattanzi}, M. and {Lawrence}, C.~R. and {Leahy}, J.~P. and {Leonardi}, R. and {Lesgourgues}, J. and {Levrier}, F. and {Lewis}, A. and {Liguori}, M. and {Lilje}, P.~B. and {Linden-V{\o}rnle}, M. and {L{\'o}pez-Caniego}, M. and {Lubin}, P.~M. and {Mac{\'\i}as-P{\'e}rez}, J.~F. and {Maggio}, G. and {Maino}, D. and {Mandolesi}, N. and {Mangilli}, A. and {Marchini}, A. and {Maris}, M. and {Martin}, P.~G. and {Martinelli}, M. and {Mart{\'\i}nez-Gonz{\'a}lez}, E. and {Masi}, S. and {Matarrese}, S. and {McGehee}, P. and {Meinhold}, P.~R. and {Melchiorri}, A. and {Melin}, J. -B. and {Mendes}, L. and {Mennella}, A. and {Migliaccio}, M. and {Millea}, M. and {Mitra}, S. and {Miville-Desch{\^e}nes}, M. -A. and {Moneti}, A. and {Montier}, L. and {Morgante}, G. and {Mortlock}, D. and {Moss}, A. and {Munshi}, D. and {Murphy}, J.~A. and {Naselsky}, P. and {Nati}, F. and {Natoli}, P. and {Netterfield}, C.~B. and {N{\o}rgaard-Nielsen}, H.~U. and {Noviello}, F. and {Novikov}, D. and {Novikov}, I. and {Oxborrow}, C.~A. and {Paci}, F. and {Pagano}, L. and {Pajot}, F. and {Paladini}, R. and {Paoletti}, D. and {Partridge}, B. and {Pasian}, F. and {Patanchon}, G. and {Pearson}, T.~J. and {Perdereau}, O. and {Perotto}, L. and {Perrotta}, F. and {Pettorino}, V. and {Piacentini}, F. and {Piat}, M. and {Pierpaoli}, E. and {Pietrobon}, D. and {Plaszczynski}, S. and {Pointecouteau}, E. and {Polenta}, G. and {Popa}, L. and {Pratt}, G.~W. and {Pr{\'e}zeau}, G.},
        title = "{Planck 2015 results. XIII. Cosmological parameters}",
      journal = {\aap},
         year = 2016,
        month = sep,
       volume = {594},
          eid = {A13},
        pages = {A13},
          doi = {10.1051/0004-6361/201525830},
archivePrefix = {arXiv},
       eprint = {1502.01589},
 primaryClass = {astro-ph.CO},
       adsurl = {https://ui.adsabs.harvard.edu/abs/2016A&A...594A..13P}
}

@ARTICLE{Fialkov2012,
       author = {{Fialkov}, Anastasia and {Barkana}, Rennan and {Tseliakhovich}, Dmitriy and {Hirata}, Christopher M.},
        title = "{Impact of the relative motion between the dark matter and baryons on the first stars: semi-analytical modelling}",
      journal = {\mnras},
         year = 2012,
        month = aug,
       volume = {424},
       number = {2},
        pages = {1335-1345},
          doi = {10.1111/j.1365-2966.2012.21318.x},
archivePrefix = {arXiv},
       eprint = {1110.2111},
 primaryClass = {astro-ph.CO},
       adsurl = {https://ui.adsabs.harvard.edu/abs/2012MNRAS.424.1335F}
}

@ARTICLE{Tseliakhovich2011,
       author = {{Tseliakhovich}, Dmitriy and {Barkana}, Rennan and {Hirata}, Christopher M.},
        title = "{Suppression and spatial variation of early galaxies and minihaloes}",
      journal = {\mnras},
         year = 2011,
        month = dec,
       volume = {418},
       number = {2},
        pages = {906-915},
          doi = {10.1111/j.1365-2966.2011.19541.x},
archivePrefix = {arXiv},
       eprint = {1012.2574},
 primaryClass = {astro-ph.CO},
       adsurl = {https://ui.adsabs.harvard.edu/abs/2011MNRAS.418..906T}
}

@ARTICLE{Gessey_Jones2022,
       author = {{Gessey-Jones}, T. and {Sartorio}, N.~S. and {Fialkov}, A. and {Mirouh}, G.~M. and {Magg}, M. and {Izzard}, R.~G. and {de Lera Acedo}, E. and {Handley}, W.~J. and {Barkana}, R.},
        title = "{Impact of the primordial stellar initial mass function on the 21-cm signal}",
      journal = {\mnras},
         year = 2022,
        month = oct,
       volume = {516},
       number = {1},
        pages = {841-860},
          doi = {10.1093/mnras/stac2049},
archivePrefix = {arXiv},
       eprint = {2202.02099},
 primaryClass = {astro-ph.CO},
       adsurl = {https://ui.adsabs.harvard.edu/abs/2022MNRAS.516..841G}
}

@ARTICLE{Feng_Holder2018,
       author = {{Feng}, Chang and {Holder}, Gilbert},
        title = "{Enhanced Global Signal of Neutral Hydrogen Due to Excess Radiation at Cosmic Dawn}",
      journal = {\apjl},
         year = 2018,
        month = may,
       volume = {858},
       number = {2},
          eid = {L17},
        pages = {L17},
          doi = {10.3847/2041-8213/aac0fe},
archivePrefix = {arXiv},
       eprint = {1802.07432},
 primaryClass = {astro-ph.CO},
       adsurl = {https://ui.adsabs.harvard.edu/abs/2018ApJ...858L..17F}
}

@ARTICLE{Ewall_Wice_2018,
       author = {{Ewall-Wice}, A. and {Chang}, T. -C. and {Lazio}, J. and {Dor{\'e}}, O. and {Seiffert}, M. and {Monsalve}, R.~A.},
        title = "{Modeling the Radio Background from the First Black Holes at Cosmic Dawn: Implications for the 21 cm Absorption Amplitude}",
      journal = {\apj},
         year = 2018,
        month = nov,
       volume = {868},
       number = {1},
          eid = {63},
        pages = {63},
          doi = {10.3847/1538-4357/aae51d},
archivePrefix = {arXiv},
       eprint = {1803.01815},
 primaryClass = {astro-ph.CO},
       adsurl = {https://ui.adsabs.harvard.edu/abs/2018ApJ...868...63E}
}

@ARTICLE{Condon2012,
       author = {{Condon}, J.~J. and {Cotton}, W.~D. and {Fomalont}, E.~B. and {Kellermann}, K.~I. and {Miller}, N. and {Perley}, R.~A. and {Scott}, D. and {Vernstrom}, T. and {Wall}, J.~V.},
        title = "{Resolving the Radio Source Background: Deeper Understanding through Confusion}",
      journal = {\apj},
         year = 2012,
        month = oct,
       volume = {758},
       number = {1},
          eid = {23},
        pages = {23},
          doi = {10.1088/0004-637X/758/1/23},
archivePrefix = {arXiv},
       eprint = {1207.2439},
 primaryClass = {astro-ph.CO},
       adsurl = {https://ui.adsabs.harvard.edu/abs/2012ApJ...758...23C}
}

@ARTICLE{Mebane2020,
       author = {{Mebane}, Richard H. and {Mirocha}, Jordan and {Furlanetto}, Steven R.},
        title = "{The effects of population III radiation backgrounds on the cosmological 21-cm signal}",
      journal = {\mnras},
         year = 2020,
        month = mar,
       volume = {493},
       number = {1},
        pages = {1217-1226},
          doi = {10.1093/mnras/staa280},
archivePrefix = {arXiv},
       eprint = {1910.10171},
 primaryClass = {astro-ph.GA},
       adsurl = {https://ui.adsabs.harvard.edu/abs/2020MNRAS.493.1217M}
}

@ARTICLE{2024ApJ...970L..25S,
       author = {{Sikder}, Sudipta and {Barkana}, Rennan and {Fialkov}, Anastasia},
        title = "{Constraining the Clustering and 21 cm Signature of Radio Galaxies at Cosmic Dawn}",
      journal = {\apjl},
         year = 2024,
        month = aug,
       volume = {970},
       number = {2},
          eid = {L25},
        pages = {L25},
          doi = {10.3847/2041-8213/ad5c5f},
archivePrefix = {arXiv},
       eprint = {2401.05865},
 primaryClass = {astro-ph.CO},
       adsurl = {https://ui.adsabs.harvard.edu/abs/2024ApJ...970L..25S}
}

@ARTICLE{Cang2025,
       author = {{Cang}, Junsong and {Mesinger}, Andrei and {Murray}, Steven G. and {Breitman}, Daniela and {Qin}, Yuxiang and {Trotta}, Roberto},
        title = "{The EDGES measurement disfavors an excess radio background during the cosmic dawn}",
      journal = {Astronomy \& Astrophysics},
         year = 2025,
        month = jun,
       volume = {698},
          eid = {A152},
        pages = {A152},
          doi = {10.1051/0004-6361/202452982},
archivePrefix = {arXiv},
       eprint = {2411.08134},
 primaryClass = {astro-ph.CO},
       adsurl = {https://ui.adsabs.harvard.edu/abs/2025A&A...698A.152C}
}

@ARTICLE{Mirocha2017,
       author = {{Mirocha}, Jordan and {Furlanetto}, Steven R. and {Sun}, Guochao},
        title = "{The global 21-cm signal in the context of the high- z galaxy luminosity function}",
      journal = {\mnras},
         year = 2017,
        month = jan,
       volume = {464},
       number = {2},
        pages = {1365-1379},
          doi = {10.1093/mnras/stw2412},
archivePrefix = {arXiv},
       eprint = {1607.00386},
 primaryClass = {astro-ph.GA},
       adsurl = {https://ui.adsabs.harvard.edu/abs/2017MNRAS.464.1365M}
}

@ARTICLE{munoz2022,
       author = {{Mu{\~n}oz}, Julian B. and {Qin}, Yuxiang and {Mesinger}, Andrei and {Murray}, Steven G. and {Greig}, Bradley and {Mason}, Charlotte},
        title = "{The impact of the first galaxies on cosmic dawn and reionization}",
      journal = {\mnras},
         year = 2022,
        month = apr,
       volume = {511},
       number = {3},
        pages = {3657-3681},
          doi = {10.1093/mnras/stac185},
archivePrefix = {arXiv},
       eprint = {2110.13919},
 primaryClass = {astro-ph.CO},
       adsurl = {https://ui.adsabs.harvard.edu/abs/2022MNRAS.511.3657M}
}

@ARTICLE{Wyithe_loeb2004,
       author = {{Wyithe}, J. Stuart B. and {Loeb}, Abraham},
        title = "{A characteristic size of \raisebox{-0.5ex}\textasciitilde10Mpc for the ionized bubbles at the end of cosmic reionization}",
      journal = {\nat},
         year = 2004,
        month = nov,
       volume = {432},
       number = {7014},
        pages = {194-196},
          doi = {10.1038/nature03033},
archivePrefix = {arXiv},
       eprint = {astro-ph/0409412},
 primaryClass = {astro-ph},
       adsurl = {https://ui.adsabs.harvard.edu/abs/2004Natur.432..194W}
}

@article{FURLANETTO2006181,
title = {Cosmology at low frequencies: The 21cm transition and the high-redshift Universe},
journal = {Physics Reports},
volume = {433},
number = {4},
pages = {181-301},
year = {2006},
issn = {0370-1573},
doi = {https://doi.org/10.1016/j.physrep.2006.08.002},
url = {https://www.sciencedirect.com/science/article/pii/S0370157306002730},
author = {Steven R. Furlanetto and S. {Peng Oh} and Frank H. Briggs}
}




\appendix

\section{Non-Gaussian signatures in a substantially different astrophysical model}\label{sec:appendixA}

In the main text, our analysis was performed using a single, fiducial astrophysical model. To ensure that the predicted non-Gaussian signatures are robust, and not merely an artifact of a specific parameter choice, here we present the results from an alternative astrophysical model, with substantially different parameters for high-redshift star formation. In particular, while our fiducial model was driven exclusively by Population II (Pop II) stars, the alternative model incorporates the contribution of the smallest star-forming halos (driven by molecular hydrogen cooling) as well as Population III (Pop III) stars, and models the transition from Pop III to Pop II star formation \citep{Gessey_Jones2022}. The specific parameter values adopted for these supplementary simulations are detailed in Table~\ref{tab:parameters_list_add}. We note that unlike our fiducial model, the alternative model does not include Poisson fluctuations in the numbers of galactic halos (which only affects the highest redshifts when galaxies were rare).

\begin{table*}
\centering
\begin{tabular}{lcc} 
\hline
Parameter    & Values  & Description \\ 
\hline\hline
$f_{\star, \ \rm{III}}$  & $0.01$     & Pop III star formation efficiency  \\
$f_{\star, \ \rm{II}}$  & $0.1$     & Pop II star formation efficiency \\
$V_{\rm{c}}$  & $4.2$ km s$^{-1}$    & Minimum circular velocity \\
$f_{\rm{X, \ \rm{III}}}$  & $0$     & Pop III X-ray production efficiency \\
$f_{\rm{X, \ \rm{II}}}$  & $1$     & Pop II X-ray production efficiency \\
$\alpha$  & $1.5$      & Slope of X-ray SED  \\
$E_{\rm{min}}$  & $1$ keV   & X-ray SED low energy cutoff\\
$\zeta$  & $30$     &  the overall efficiency of ionizing sources \\
$R_{\rm{mfp}}$  & $30$ Mpc  & Mean free path for ionizing photons \\
$f_{\rm{Radio}}$  & 60 or 3000    & Radio production efficiency (radio from early galaxies)\\
$\alpha_{\rm{Radio}}$ & 0.7 & Spectral index in the radio band (radio from early galaxies) \\
$A_{\rm{r}}$  &  $0.0945$   & Amplitude of uniform radio background (radio from exotic processes)\\
$\beta$ & $-2.6$ & Spectral index in the synchrotron spectrum (radio from exotic processes)\\


\hline
\end{tabular}
\caption{Similar to Table \ref{tab:parameters_list} but for the astrophysical model shown in this Appendix, incorporating a transition from Pop III to Pop II stars. These parameters were used for the supplementary simulations presented in this appendix to demonstrate that our predicted non-Gaussian signatures are robust across different astrophysical regimes.}\label{tab:parameters_list_add}

\end{table*}

Running this different set of astrophysical parameters allows us to test the robustness of our primary conclusions. Figure~\ref{fig:global_signal_power_spectrum_popII_III} is the counterpart of  Figure~\ref{fig:global_signal_power_spectrum}, illustrating the evolution of the global 21-cm signal and the 21-cm power spectrum in the alternative model. Similarly, Figure~\ref{fig:skew_kurt_z_popII_III}
replicates the analysis shown in Figure~\ref{fig:skew_kurt_z} but for the alternative model. Despite the addition of Pop III stars and the altered star formation efficiencies, the qualitative behavior shown in this figures is similar to the fiducial model, though the detailed numbers are somewhat different, particularly at $z>15$. The biggest difference is that at the highest redshifts the skewness and kurtosis become larger (in absolute magnitude); however, these redshifts are at the edge of the CD and will be more difficult to observe. During the heart of the CD ($z \sim 15$), the moderate LoS radio model has a significantly larger kurtosis in the alternative model. Overall, given that our conclusions from these figures were that there is significant non-Gaussianity in the 21-cm signal and it is increased by radio fluctuations, then the results for the alternative model only strengthen these conclusions.

\begin{figure}
    \centering
    \includegraphics[width=0.48\textwidth]{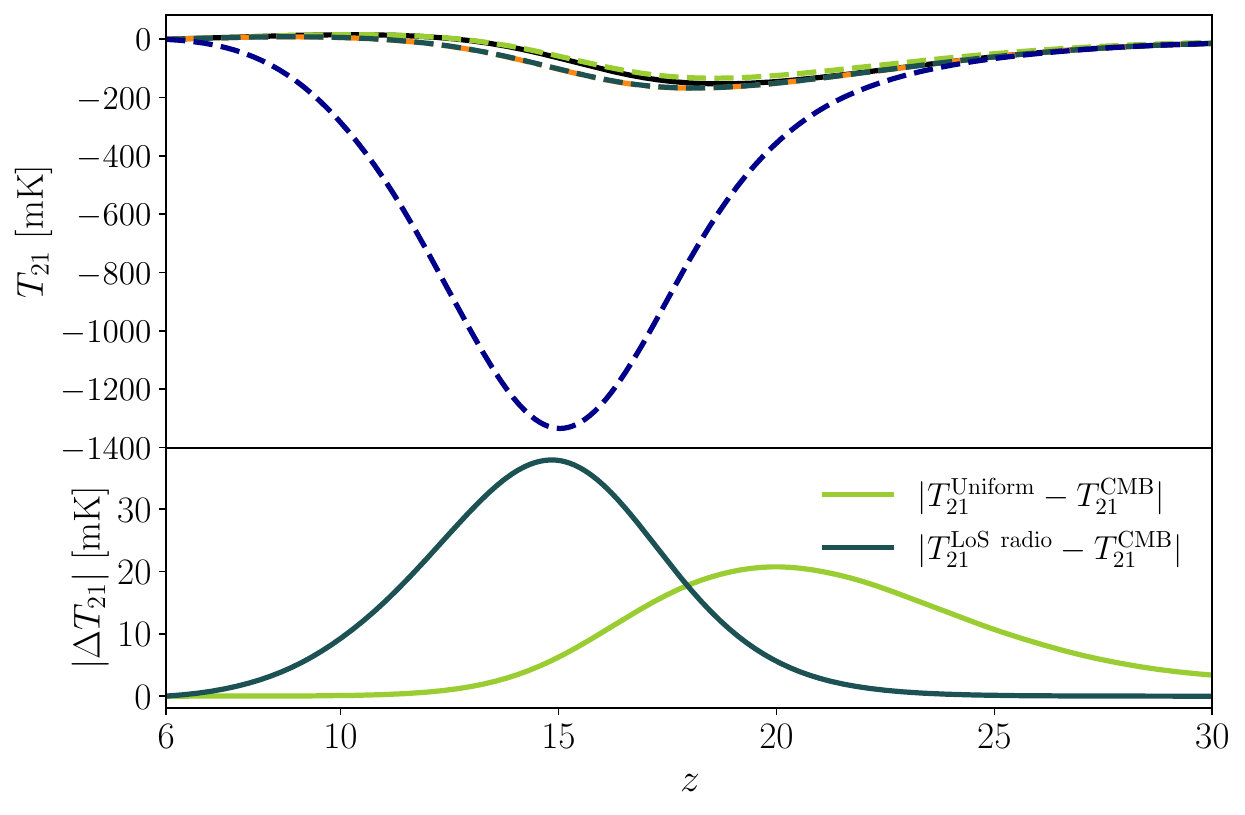}
    \includegraphics[width=0.48\textwidth]{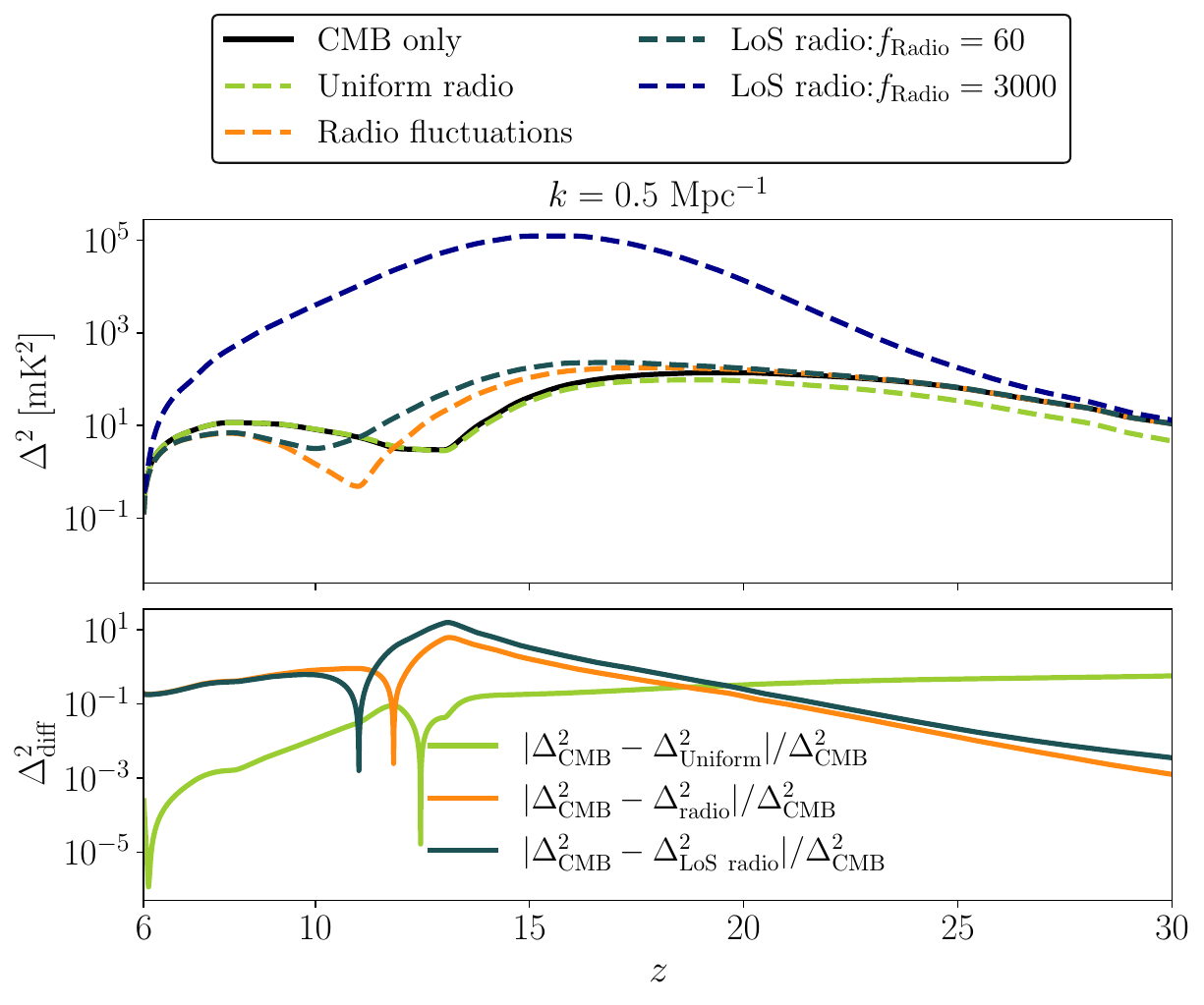}
    \caption{Same as Figure \ref{fig:global_signal_power_spectrum} in the main text, but generated using the extended astrophysical model incorporating a transition from Population III to Population II stars (parameters detailed in Table \ref{tab:parameters_list_add}). This demonstrates that the overall evolution of the global 21-cm signal and the 21-cm power spectrum remains consistent and the predicted signatures are robust under a completely different set of astrophysical assumptions.}
    \label{fig:global_signal_power_spectrum_popII_III}
\end{figure}

\begin{figure}
    \centering
    \includegraphics[width=0.48\textwidth]{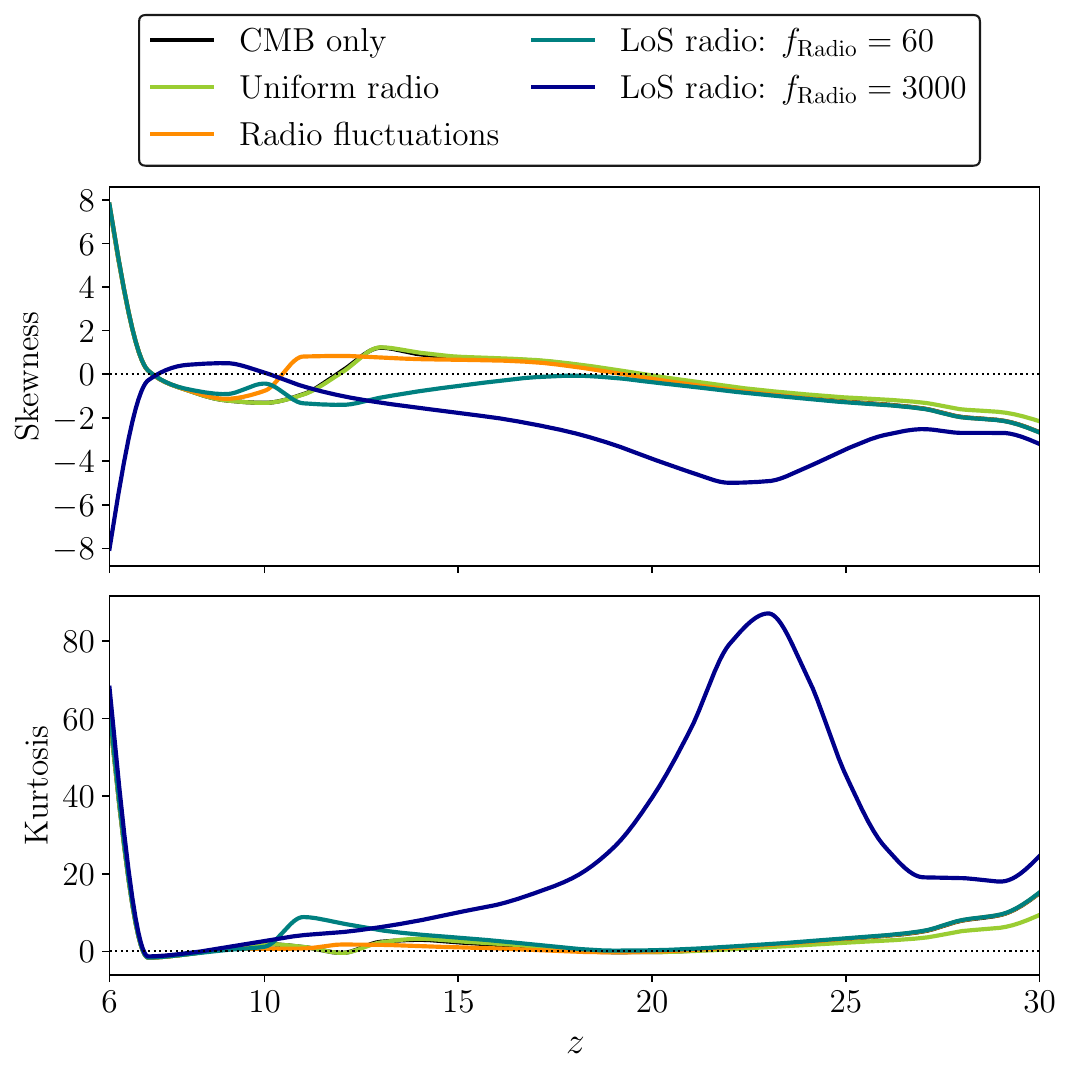}
    \caption{Same as Figure \ref{fig:skew_kurt_z} in the main text, but generated for the alternative Pop III to Pop II transition model detailed in Table~\ref{tab:parameters_list_add}.}
    \label{fig:skew_kurt_z_popII_III}
\end{figure}

Figure~\ref{fig:bispectrum_all_triangles_popII_III} extends the analysis of Figure \ref{fig:bispectrum_all_triangles} from the main text. Alongside the original panels showing the results for two values of $k$ at $z=20$, we have added two lower rows to display the corresponding behavior at $z=19$; the reason is that Figure~\ref{fig:skew_kurt_z} shows that $z=20$ for the fiducial model is roughly where the skewness and kurtosis first reach near zero (coming from high redshift) in the CMB only case, and Figure~\ref{fig:skew_kurt_z_popII_III} shows that $z=19$ is closer to this point for the alternative model. At the high $k$ value, the results are qualitatively similar for $z=20$ and 19 in the alternative model as at $z=20$ in the fiducial model; namely, the sign of the SABS varies quite a bit, but is always strongly negative in the squeezed limit (upper right corner of each panel). At the low $k$ value, the behavior is more complex: The SABS is quantitatively similar for the LoS Radio fluctuations between $z=20$ in the two models, and it remains negative (but quantitatively smaller in absolute value) at $z=19$ in the alternative model. In the other models, there is a switch in sign (for most configurations, but never the squeezed limit) going from $z=20$ (negative) to $z=19$ in the alternative model; this makes $z=19$ a better match to $z=20$ of the fiducial model, for the CMB only and Uniform radio cases. Finally, Figure~\ref{fig:bispec_z_squeezed_limit_popII_III} reproduces the analysis of Figure~\ref{fig:bispec_z_squeezed_limit} for the alternative model. The behavior of the SABS in the squeezed limit versus redshift is remarkably similar in the two models, despite some differences in the detailed numbers. Overall, the results derived from this alternative example of the parameter space confirm that the non-Gaussian signatures discussed in this work are not confined to a narrow set of conditions. Rather, they likely represent a robust outcome of the various models (CMB only or various excess radio backgrounds) during the CD.

\begin{figure*}
\centering
\includegraphics[width=0.8\textwidth]{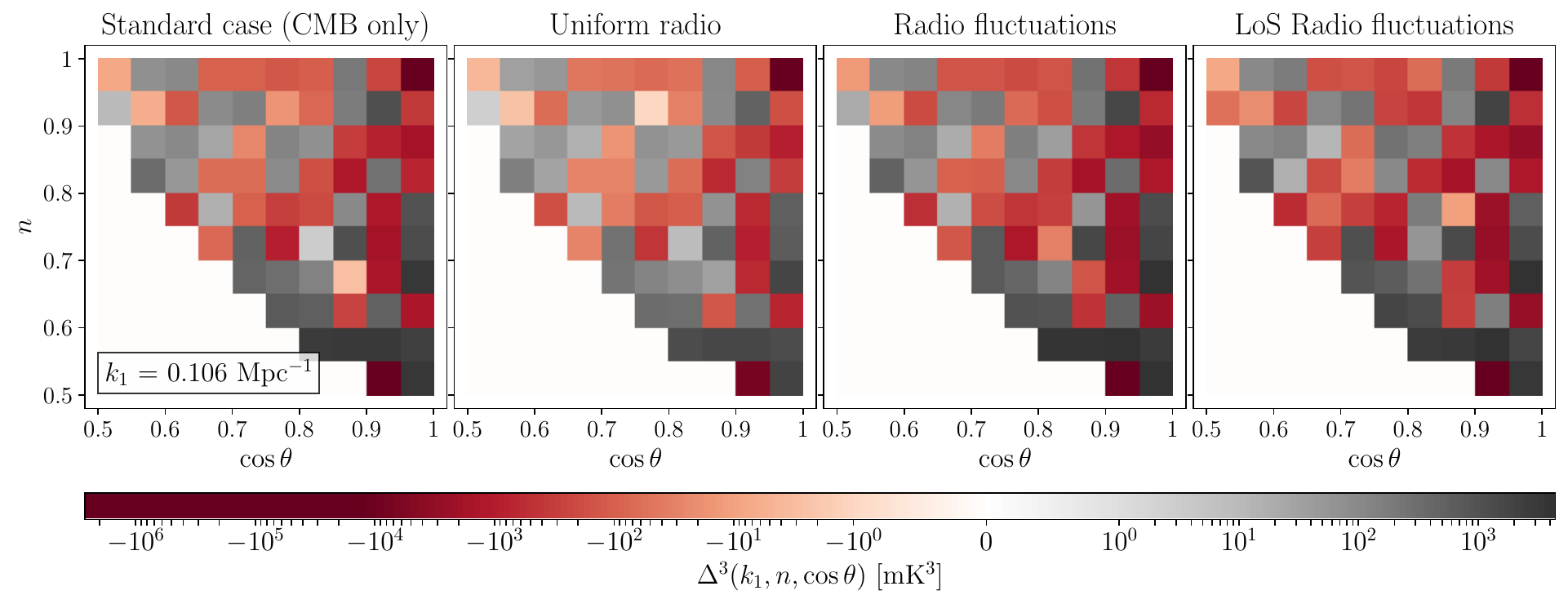}
\includegraphics[width=0.8\textwidth]{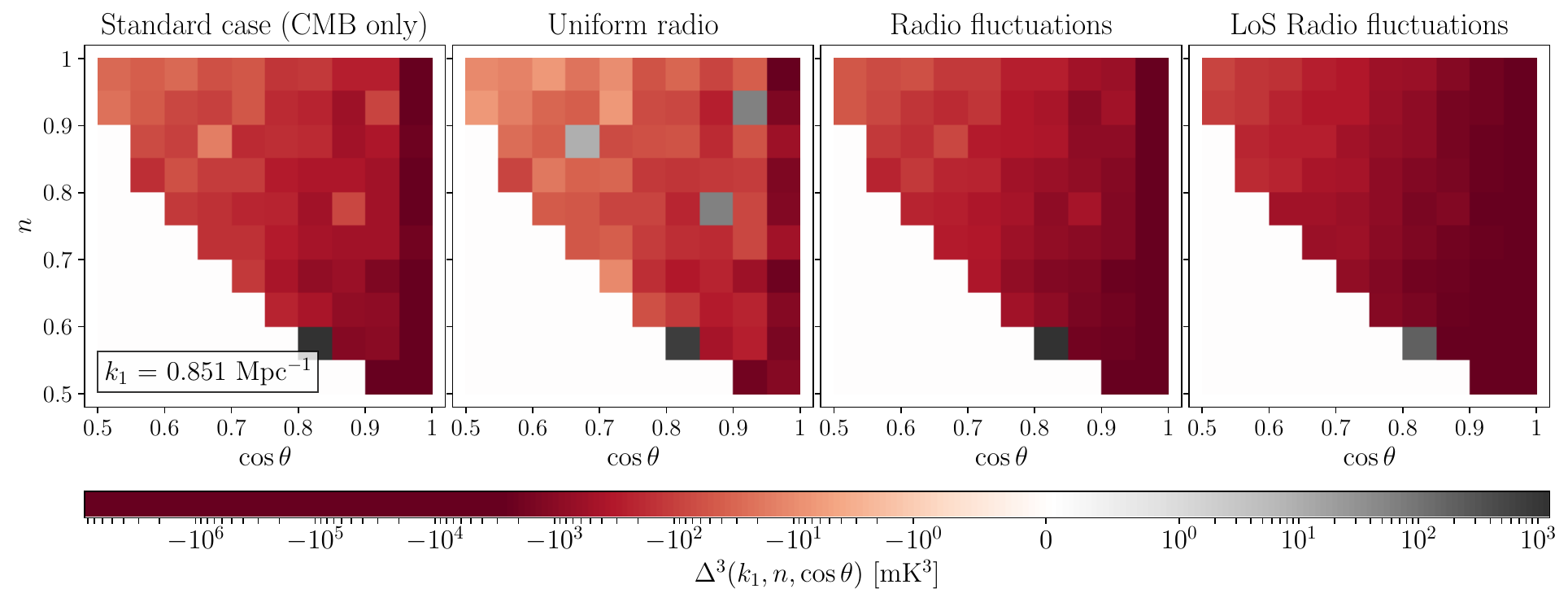}
\includegraphics[width=0.8\textwidth]{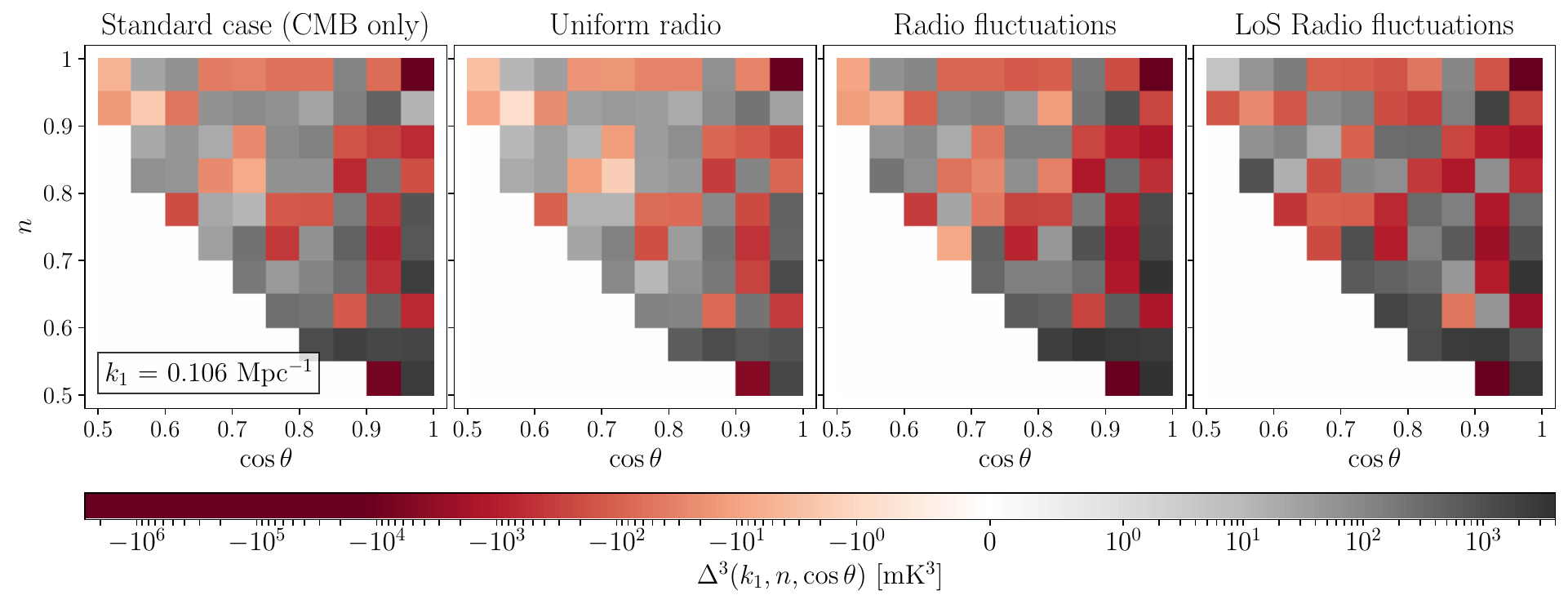}
\includegraphics[width=0.8\textwidth]{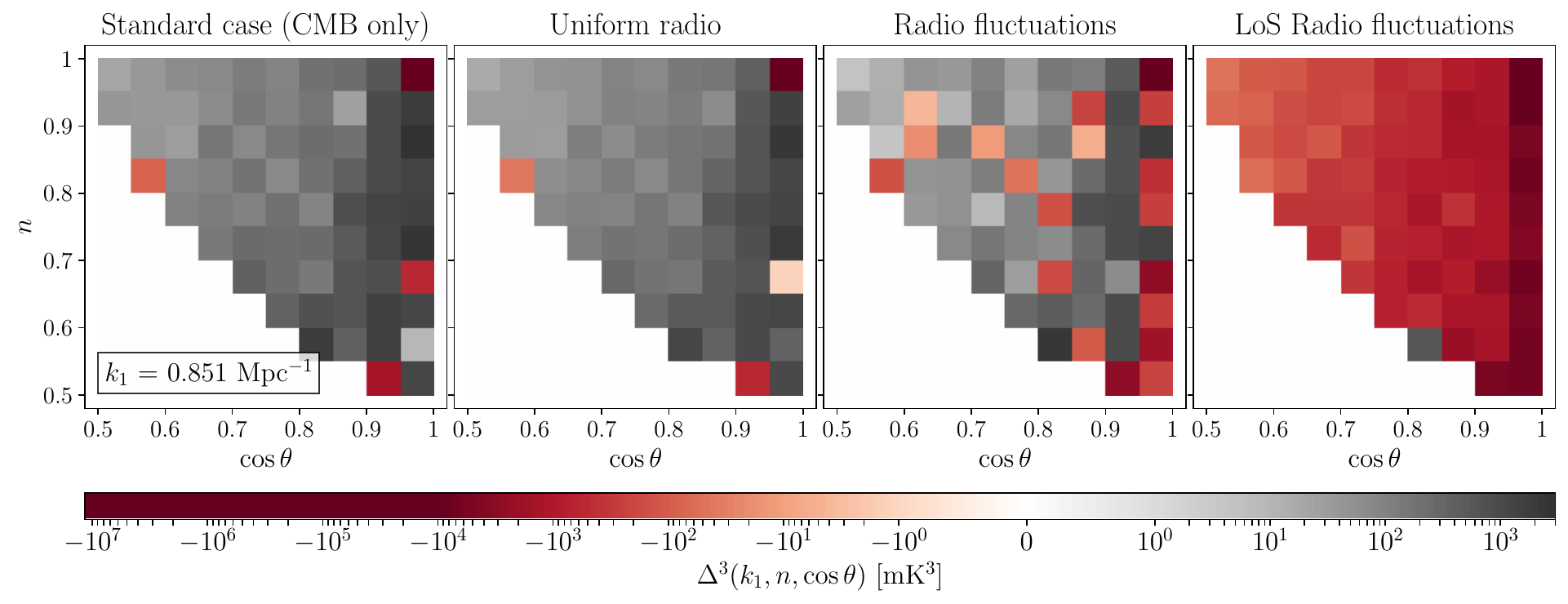}
\caption{Same as Figure \ref{fig:bispectrum_all_triangles} in the main text, but computed using the alternative Pop III to Pop II transition model detailed in Table \ref{tab:parameters_list_add}. In addition to replicating the panels for the two $k$ values at $z=20$, we have included two additional rows displaying the corresponding results for $z=19$.}
\label{fig:bispectrum_all_triangles_popII_III}
\end{figure*}

\begin{figure*}
    \centering
    \includegraphics[width=1\textwidth]{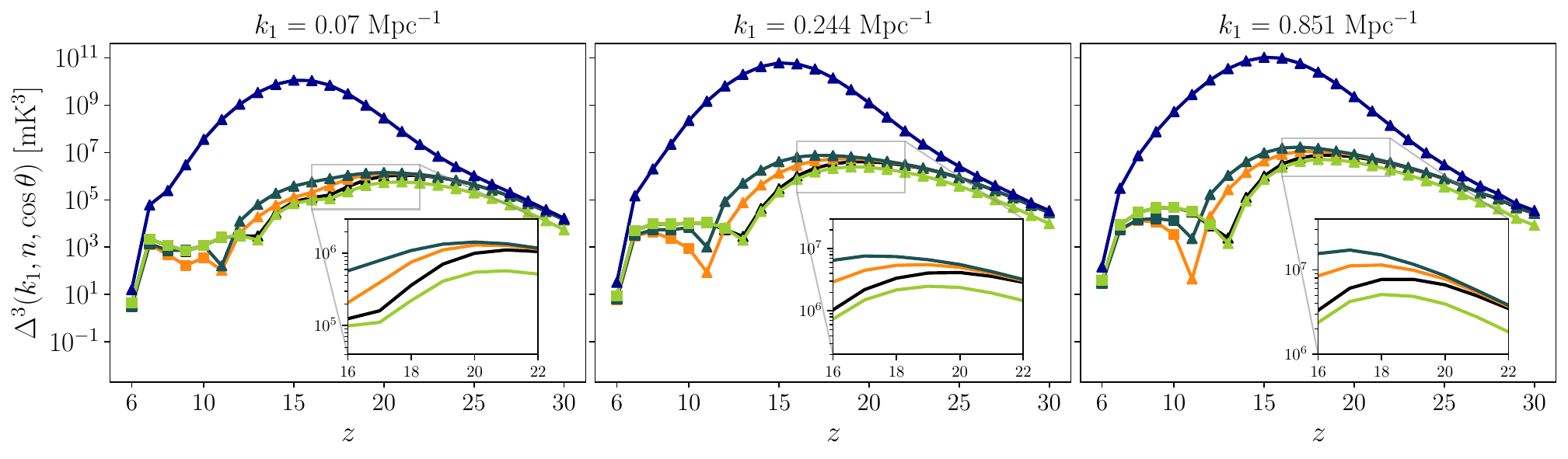}
    \caption{Same as Figure \ref{fig:bispec_z_squeezed_limit} in the main text, but generated using the alternative Pop III to Pop II transition model detailed in Table \ref{tab:parameters_list_add}.}
    \label{fig:bispec_z_squeezed_limit_popII_III}
\end{figure*}


\bsp	
\label{lastpage}
\end{document}